\pdfoutput=1
\documentclass{JINST}
\usepackage{xspace}
\usepackage{multirow}

\def\github {\href{https://github.com/yunjie-yang/TuneMC}{{\sc TuneMC}}}

\def\spearmint {\mbox{\textsc{Spearmint}}\xspace}
\def\skopt {\mbox{\textsc{Scikit-Optimize}}\xspace}
\def\pythia     {\mbox{\textsc{Pythia}}\xspace}

\title{Event generator tuning using Bayesian optimization}

\author{Philip Ilten, Mike Williams, and Yunjie Yang\\
Laboratory for Nuclear Science, Massachusetts Institute of Technology, Cambridge, MA 02139}

\abstract{
Monte Carlo event generators contain a large number of parameters that must be determined by comparing the output of the generator with experimental data.
Generating enough events with a fixed set of parameter values to enable making such a comparison is extremely CPU intensive,
which prohibits performing a simple brute-force grid-based tuning of the parameters.
Bayesian optimization is a powerful method designed for such black-box tuning applications.
In this article, we show that Monte Carlo event generator parameters can be accurately obtained using Bayesian optimization and minimal expert-level physics knowledge.
A tune of the \pythia~8 event generator using $e^+e^-$ events, where 20 parameters are optimized, can be run on a modern laptop in just two days.
Combining the Bayesian optimization approach with expert knowledge should enable producing better tunes in the future, by making it faster and easier to study discrepancies between Monte Carlo and experimental data.
}

\begin{document}

\section{Introduction}

Monte Carlo event generators, which are used to simulate particle collisions, contain a large number of parameters that must be determined (tuned) by comparing the output of the generator with experimental data.
Generating enough events with a fixed set of parameter values to enable making such a comparison is extremely CPU intensive.
For example, it takes $\mathcal{O}({\rm hour})$ on a modern CPU core to generate 1M events for a single set of parameter values using the \pythia~8 event generator~\cite{pythia,Sjostrand:2014zea}.
A full tune of \pythia to $e^+e^-$ data involves optimizing $\approx 20$ parameters, which clearly cannot be performed using a brute-force grid-based approach.
Even a tune of only a small subset of parameters, {\em e.g.}, the 6 parameters that control fragmentation, would take $\mathcal{O}(100)$ CPU years using a coarse 10-bins-per-parameter scheme.

All available tunes provided with the \pythia~8 package were obtained either:
{\em manually}, where an expert chose how to vary the parameters based on extensive knowledge and insight, guided by comparing generated and experimental distributions;
or {\em parametrically}, where the generator response to changes in the parameters was itself parametrized based on a large set of reference generator data sets, which then facilitated optimizing the parameters via minimization of an objective function, {\em e.g.}, a $\chi^2$.
Of course, this characterization is oversimplified since experts have performed parametric tunes, but it is sufficient to motivate this study.
Examples of manual and parametric tunes are described in detail in Refs.~\cite{Skands:2014pea} and \cite{Buckley:2009bj}, respectively.

Each of these approaches has both merit and limitations.
We believe that only a few physicists are capable of performing a large-scale manual tune of \pythia, and even for such an expert it takes considerable effort.
The manual approach does not scale to larger parameter sets, and is not well suited to less-intuitive models or to producing many experiment-specific tunes (or a large number of dedicated tunes in general).
That said, the manual approach is less prone to finding an {\em unphysical} local minimum in the parameter space that happens to provide a decent description of the data distributions being compared to during the tuning process; the expert can intuitively identify such situations.
The parametrization approach is easily parallelized, but requires that the generator response --- including multi-parameter correlations --- is well approximated by the chosen parametric function within the parameter hypercube to be explored.
 Furthermore, the optimal working point must be included in the parameter hypercube, though this can be achieved by first doing a coarse scan of the parameter space.

In this article, we propose treating Monte Carlo event-generator tuning as a black-box optimization problem to be addressed using the framework of Bayesian optimization.
We will show that Monte Carlo generator parameters can be accurately obtained using Bayesian optimization and minimal expert-level physics knowledge.
 Furthermore, a tune of the \pythia~8 event generator using $e^+e^-$ data, where 20 parameters are optimized, can be run on a modern laptop in just two days.
Combining the Bayesian optimization approach with expert-level knowledge should enable producing better tunes in the future, by making it faster and easier to study discrepancies between Monte Carlo and experimental data.
This article is organized as follows:
the Bayesian optimization framework is described in Sec.~\ref{sec:bayes};
its application to a full \pythia $e^+e^-$ tune is presented in Sec.~\ref{sec:tune};
Sec.~\ref{sec:cpu} describes CPU usage;
moving towards a real-world tune is discussed in Sec.~\ref{sec:real}
and a summary
is provided in Sec.~\ref{sec:sum}.

\section{Bayesian Optimization}
\label{sec:bayes}

For each Monte Carlo data sample produced by the event generator (for a given set of parameters), a number of observable distributions can be constructed and compared between Monte Carlo and experimental data.
Such a comparison is done using an objective function, {\em e.g.}, a two-sample $\chi^2$ statistic built from the binned distributions in data and Monte Carlo.\footnote{In practice, a pseudo-$\chi^2$ statistic is typically used, where correlations between the various observable distributions are ignored.  Regardless, a smaller $\chi^2$ is taken to mean a better tune.}
Since the dependence of the objective function on the parameters is unknown, the strategy employed in Bayesian optimization is to treat the $\chi^2$ as a random function over which a prior must be assigned (see, {\em e.g.}, Ref.~\cite{BO}).
We choose to use the Gaussian process prior, which is a common choice as it permits computing marginals and conditionals in closed form.
For an overview of Gaussian processes, see Ref.~\cite{GP}.

Each time a Monte Carlo sample is generated with a different set of parameters, a $\chi^2$ value is computed comparing the Monte Carlo to the experimental data.
From the initial prior and all of the sampled $\chi^2$ values, a posterior over functions is constructed.
The main idea is to use all information available, and not just the local gradient to find the best possible $\chi^2$.
Another choice that must be made in Bayesian optimization is the so-called acquisition function, which is used to determine the next point in parameter space to query.
We choose to focus on maximizing the expected improvement over the current best $\chi^2$ found, as implemented in the \spearmint software package~\cite{spearmint}, and use the default \spearmint settings for balancing exploration versus exploitation.
For a detailed discussion of the algorithms implemented in \spearmint, see Ref.~\cite{NIPS2012_4522}.
{\em N.b.}, working within the Gaussian process framework is not suited to discrete parameters; however, other automated optimization procedures do handle discrete parameters well (see Sec.~\ref{sec:discrete}).
Finally, it is also possible within Bayesian optimization to account for the CPU cost of generating each Monte Carlo data set, and attempt to maximize the expected improvement per unit time~\cite{NIPS2012_4522}.

\section{Tuning \pythia}
\label{sec:tune}

To demonstrate how to apply Bayesian optimization to Monte Carlo event generator tuning --- and to validate its performance --- the following closure test is performed:
\begin{itemize}
\item a 10M-event $e^+e^-$ data sample is generated using \pythia~8 with its default parameter values, collectively referred to as the {\em Monash} tune~\cite{Skands:2014pea};
\item various observable distributions are built from the Monash simulated data sample and treated as experimental data;
\item a set of 20 parameters in \pythia is chosen for tuning;
\item a minimal amount of expert knowledge is input on each parameter, as each is allowed to vary freely within a large pre-defined range (of course, the true Monash values are treated as unknown in the tuning);
\item and, finally, the Bayesian optimization framework is applied using \spearmint to obtain the 20 optimal (tuned) parameter values.
\end{itemize}
Treating the Monash data sample as experimental data permits validating the performance by comparing the Monash parameter values to the optimal ones found by \spearmint.
This treatment ensures that each distribution can be perfectly
modeled by \pythia. 
In reality, Monte Carlo event generators often times do not model parts of the
experimental data well; therefore, it is important that any tuning
method can also handle optimizing imperfect models (see Sec.~\ref{sec:real}).

\subsection{Objective Function}

We define our objective function as a pseudo-$\chi^2$ in a similar way to the one used in producing the Monash tune~\cite{Skands:2014pea}:
\begin{equation}
\label{eq:chi2}
\chi^2 \equiv \sum\limits_{i=1}^{n_{\rm bins}} \frac{({\rm Monash}_i - {\rm MC}_i)^2}{\sigma_{\rm Monash,i}^2 + \sigma^2_{\rm MC,i}},
\end{equation}
where $\sigma$ denote the statistical uncertainties
on the Monash and Monte Carlo values
in the $i^{\rm th}$ bin.
Any $\sigma_i$ value that corresponds to less than a 1\% relative uncertainty is set to be 1\%.
The choice of setting a minimum value avoids having a few bins with large occupancies dominating the tuning.  In practice, the systematic uncertainties on the experimental distributions implicitly accomplish this.
The sample size of each Monte Carlo data set is chosen such that the 1\% value is used in most bins.
We ignore correlations between bins in our definition of $\chi^2$, since this information is often not available for experimental data.
It may be desirable to alter the $\chi^2$ to include weight factors for each bin.
Incorporating correlations or bin weights into the tuning procedure is straightforward, as only the definition of the $\chi^2$ needs to be modified.

\subsection{Parameters \& Observables}

\begin{figure}[t]
  \centering
 \includegraphics[height=0.27\textheight]{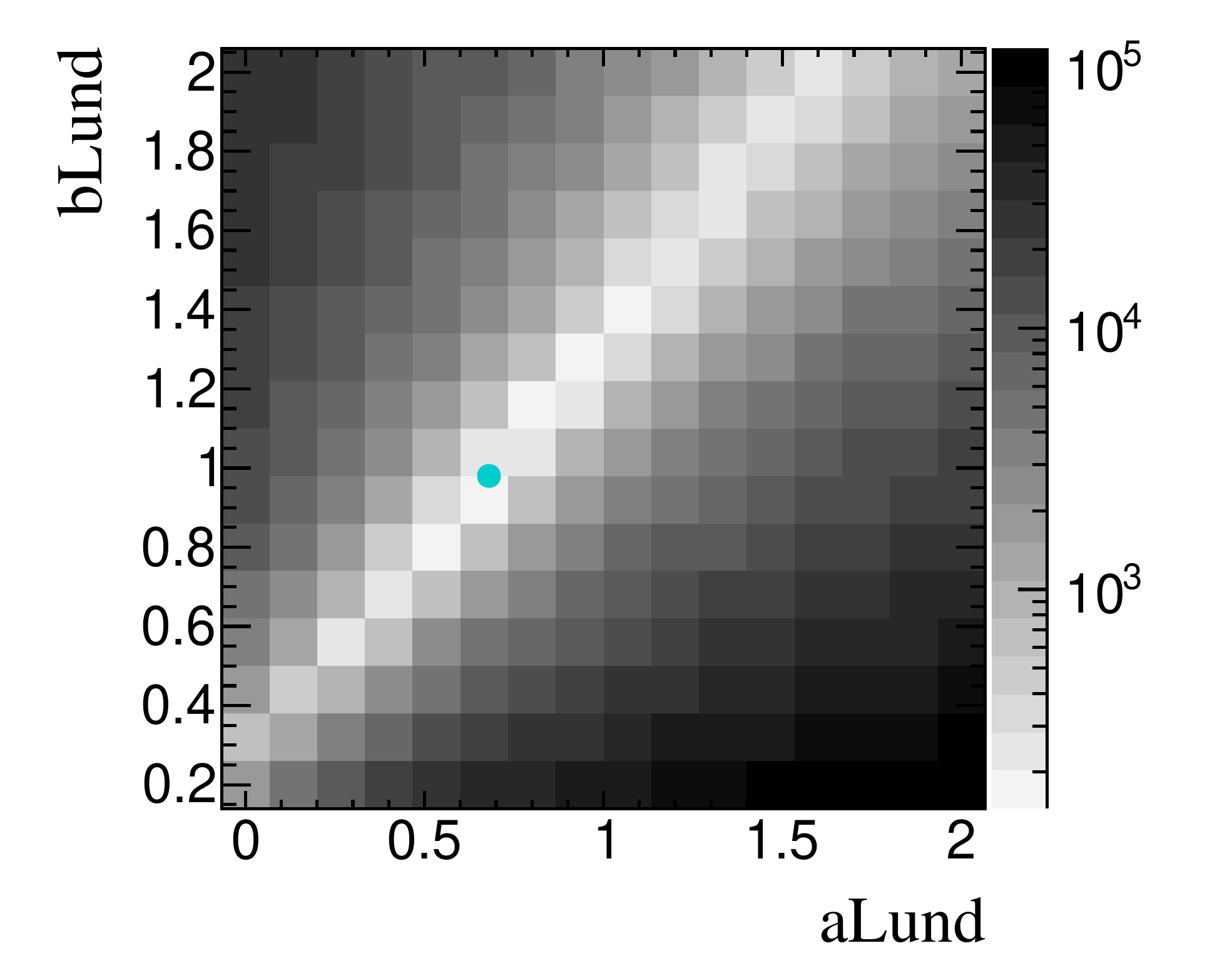}
  \caption{Example of a parameter in \pythia that is not included in our tune.  The 10M-event Monash data sample is compared to samples where the aLund and bLund parameters are varied.
The $z$ axis (greyscale) denotes the $\chi^2$ value, and the cyan marker shows the Monash value of (aLund, bLund)$=$(0.68,0.98).   The parameters aLund and bLund are strongly correlated, and so only bLund is tuned.
}
  \label{fig:parsnotused}
\end{figure}

Since the Gaussian process framework is not suited to tuning discrete parameters, all discrete parameters in \pythia are left at their Monash values.\footnote{See Sec.~\ref{sec:discrete} for discussion on how to tune discrete parameters. We note that no discrete parameters were altered from the default \pythia values in the Monash tune itself.}
We choose to tune a large set of 20 continuous parameters, which roughly corresponds to the full set of \pythia~8 parameters constrained by the observable distributions from $e^+e^-$ data that were used in the Monash tune, and that enter into the $\chi^2$ defined in Eq.~\ref{eq:chi2}. This excludes parameters that when varied have either negligible impact on the $\chi^2$ value, or those that are $\approx 100\%$ correlated with another \pythia parameter. An example of a correlated parameter not included in our tune, aLund,  is shown in Fig.~\ref{fig:parsnotused}.
{\em N.b.}, we chose to include in our tune a few parameters that are not well constrained by the $\chi^2$ to study how \spearmint performs in the presence of such parameters.

The full list of parameters included in the tune is given in Table~\ref{tab:pars}.
The range in which each parameter is allowed to vary is also provided.
We place a uniform prior over each parameter within the specified range, {\em i.e.}, parameters are allowed to vary freely within these ranges.
Expert knowledge could be used here by assigning non-uniform priors to the parameters that capture the physics belief about their behavior (see Sec.~\ref{sec:improve}); however, as our goal in this article is to demonstrate the power of the Bayesian optimization process, we choose to use minimal expert knowledge.
For a detailed discussion on the meaning of each parameter, see Ref.~\cite{Skands:2014pea}.
Table~\ref{tab:dists} provides a full list of the distributions that enter into the $\chi^2$ defined in Eq.~\ref{eq:chi2}.
We choose to use the same set of distributions that was used to produce the Monash tune.
For our purposes, the physical meaning of each distribution is not important, so we omit any detailed description and instead refer the interested reader to Ref.~\cite{Skands:2014pea}.

\begin{table}
  \begin{center}
    \caption{\label{tab:pars}Full list of parameters considered in our tune, categorized into blocks, along with their values in the Monash tune and the interval in which we allow them to vary.}
      \begin{tabular}{cccc}
        \hline
        Block & Parameter & Monash Value & Range Considered \\
        \hline
        \multirow{3}{*}{1} & alphaSvalue & 0.1365 & $[0.06,0.25]$ \\
        & pTmin  & 0.5 & $[0.1,2.0]$ \\
        & pTminChgQ & 0.5 & $[0.1,2.0]$ \\
        \hline
        \multirow{6}{*}{2}         & bLund & 0.98 & $[0.2,2]$ \\
        & sigma & 0.335 & $[0,1]$ \\
        & aExtraSQuark & 0 & $[0,2]$ \\
        & aExtraDiQuark & 0.97 & $[0,2]$ \\
        & rFactC & 1.32 & $[0,2]$ \\
        & rFactB & 0.855 & $[0,2]$ \\
        \hline
        \multirow{11}{*}{3} & probStoUD & 0.217 & $[0,1]$ \\
        & probQQtoQ & 0.081 & $[0,1]$ \\
        & probSQtoQQ & 0.915 & $[0,1]$ \\
        & probQQ1toQQ0 & 0.0275 & $[0,1]$ \\
        & etaSup & 0.6 & $[0,1]$ \\
        & etaPrimeSup & 0.12 & $[0,1]$ \\
        & decupletSup & 1 & $[0,1]$ \\
        & mesonUDvector & 0.5 & $[0,3]$ \\
        & mesonSvector & 0.55 & $[0,3]$ \\
        & mesonCvector & 0.88 & $[0,3]$ \\
        & mesonBvector & 2.2 & $[0,3]$ \\
        \hline
      \end{tabular}
  \end{center}
\end{table}

\begin{table}
  \begin{center}
    \caption{\label{tab:dists}Full list of $e^+e^-$ distributions that contribute to the $\chi^2$ in our tune. The notation ``$2\times$'' refers to distributions that are considered separately for events with and without a $b$ tag.
}
      \begin{tabular}{ccc}
        \hline
         Category & Number & Distributions \\
         \hline
         event shapes & 10 &  $2\times$(thrust, $C$ and $D$ parameter, wide and total jet broadening) \\
         \hline
         \multirow{2}{*}{fragmentation} & \multirow{2}{*}{6} & $2\times$(charged-particle multiplicity and momentum fraction),\\
         & & scaled momentum spectra $x_{D^*,B}\equiv 2p_{D^*,B}/E_{\rm cm}$ for $D^*,B$ hadrons\\
         \hline
         hadron types & 4 & hadron types in $e^+e^-\!\to X$ and $Z\!\to$\,heavy flavor\\
         \hline
      \end{tabular}
  \end{center}
\end{table}

\subsection{Tuning Closure Test}

We consider two approaches to tuning the 20 parameters listed in blocks 1--3 in Table~\ref{tab:pars} using the $\chi^2$ built from the 20 distributions given in Table~\ref{tab:dists}:
\begin{itemize}
\item a block-diagonal strategy, where the parameters in blocks 1, 2, and 3 are tuned using only the event-shape, fragmentation, and hadron-type distributions, respectively, and each block is tuned independently;
\item and a global strategy, where all 20 parameters are tuned simultaneously on all distributions.
\end{itemize}
The pseudo-experimental data distributions are obtained by generating 10M events using the Monash parameters. For each set of parameters that \spearmint chooses to evaluate, a data sample of 1M events is generated.
The number of \spearmint queries and the total CPU time used are discussed in Sec.~\ref{sec:cpu}.
In all tunes, the optimal parameter set is taken to be the one that the internal \spearmint $\chi^2$ model predicts is the best, and not the set for which a \pythia data sample was generated and found to have the best $\chi^2$ (in practice this makes little difference).  Assigning error bars to parameters using the \spearmint $\chi^2$ model is discussed in detail in Appendix~\ref{app:errors}, while more details about the technical aspects of the tuning procedure are given in Appendix~\ref{app:procedure}.

\begin{figure}
  \centering
  \includegraphics[width=0.49\textwidth]{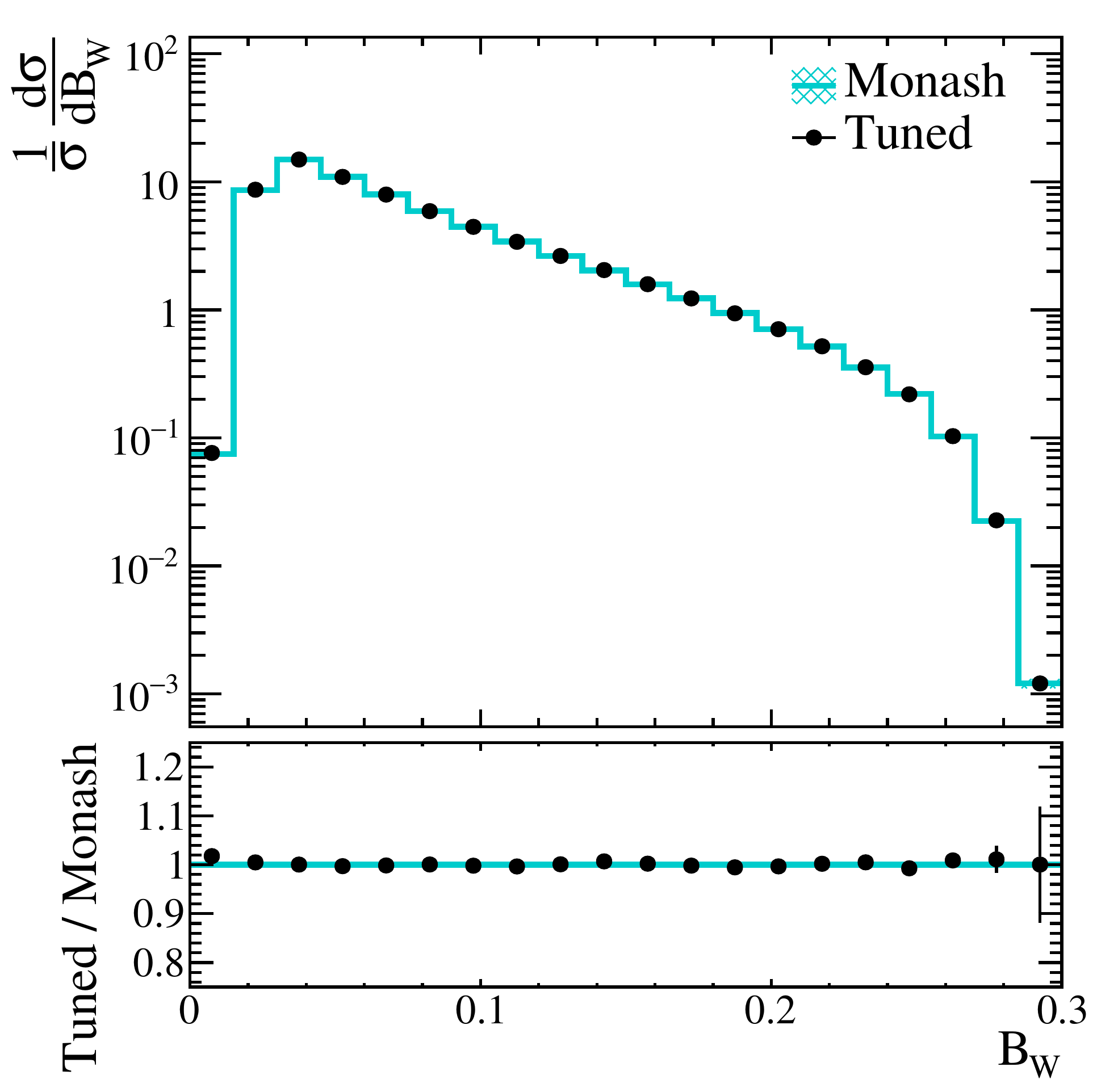}
  \includegraphics[width=0.49\textwidth]{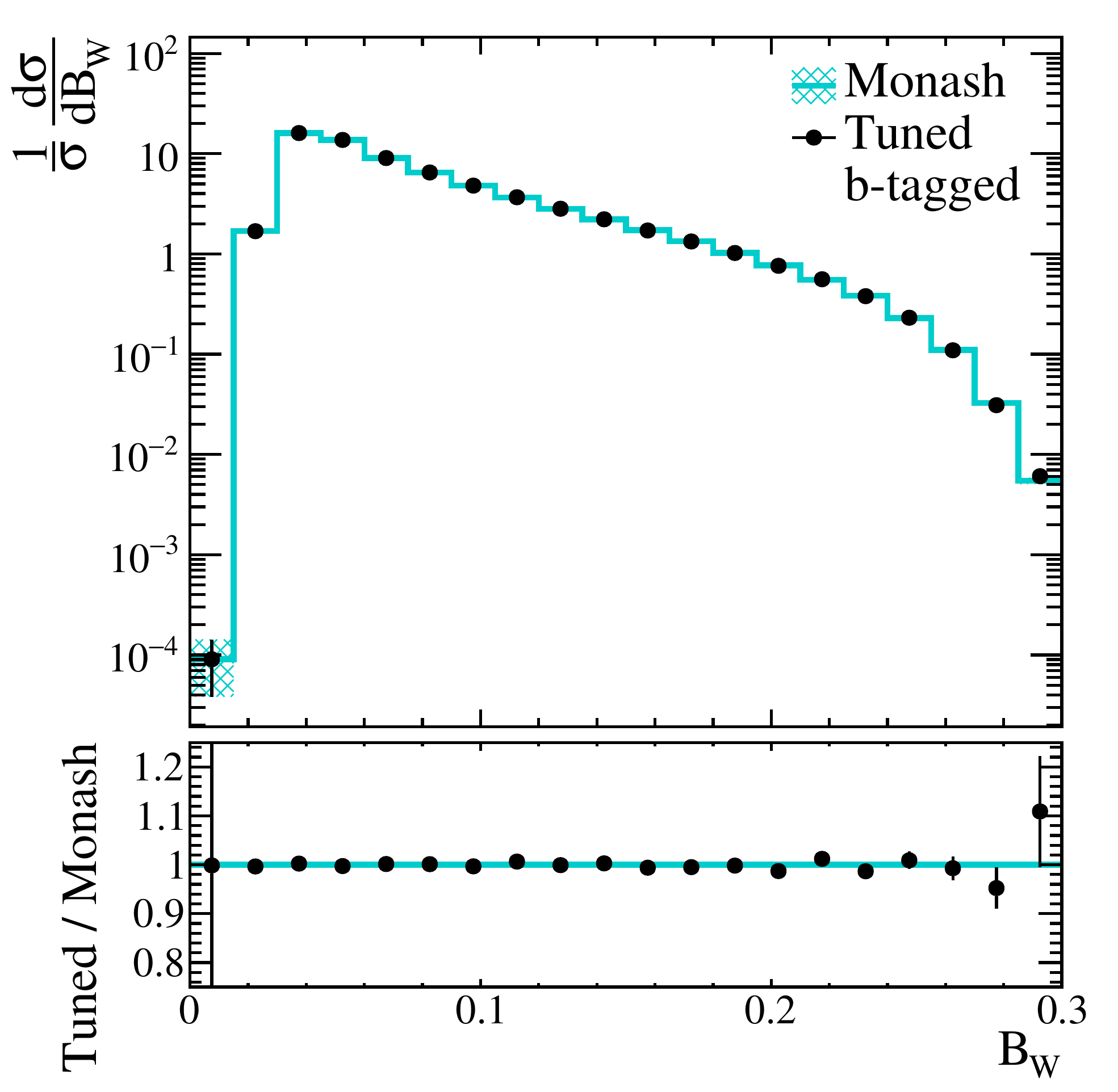}
  \includegraphics[width=0.49\textwidth]{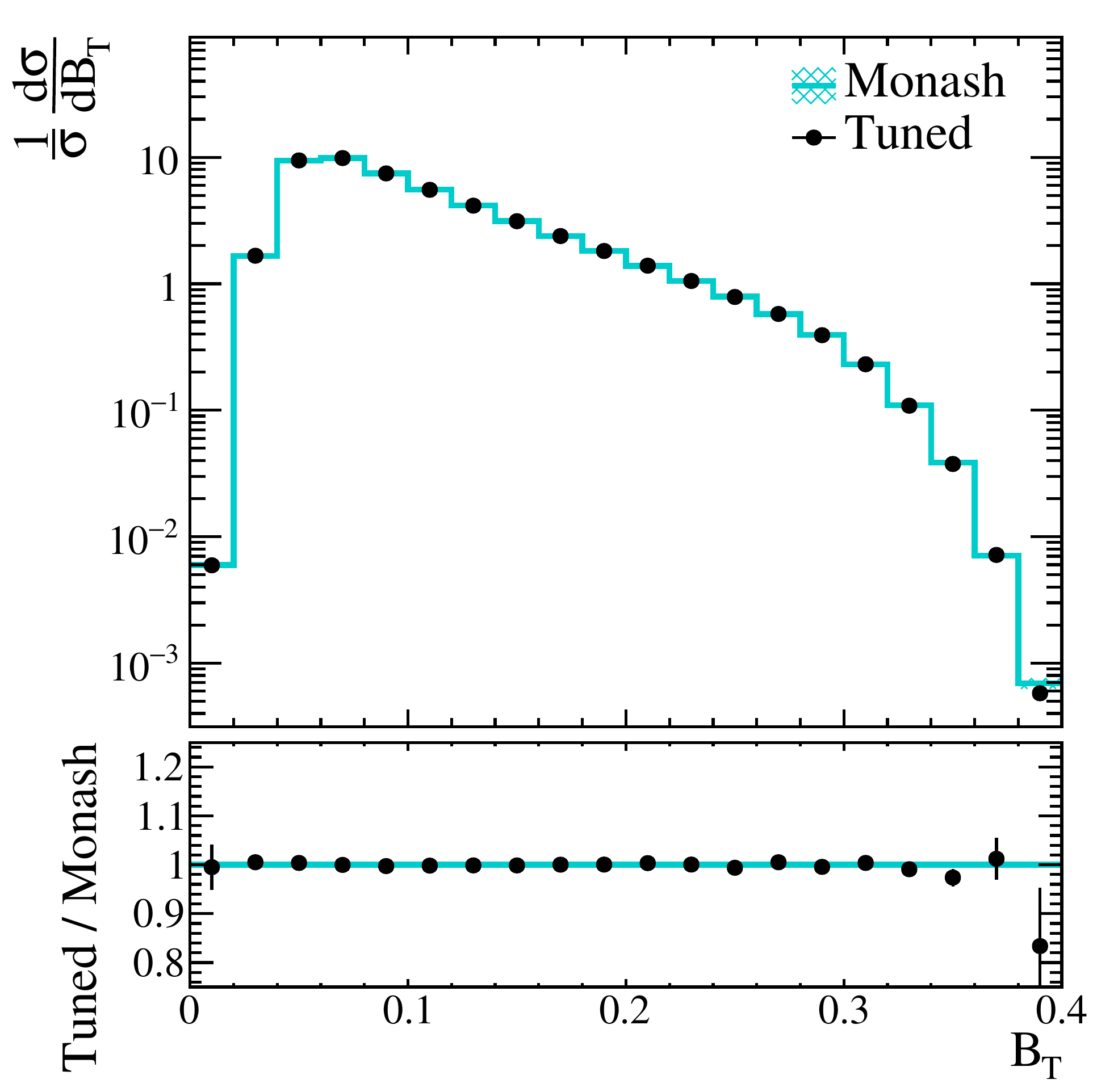}
  \includegraphics[width=0.49\textwidth]{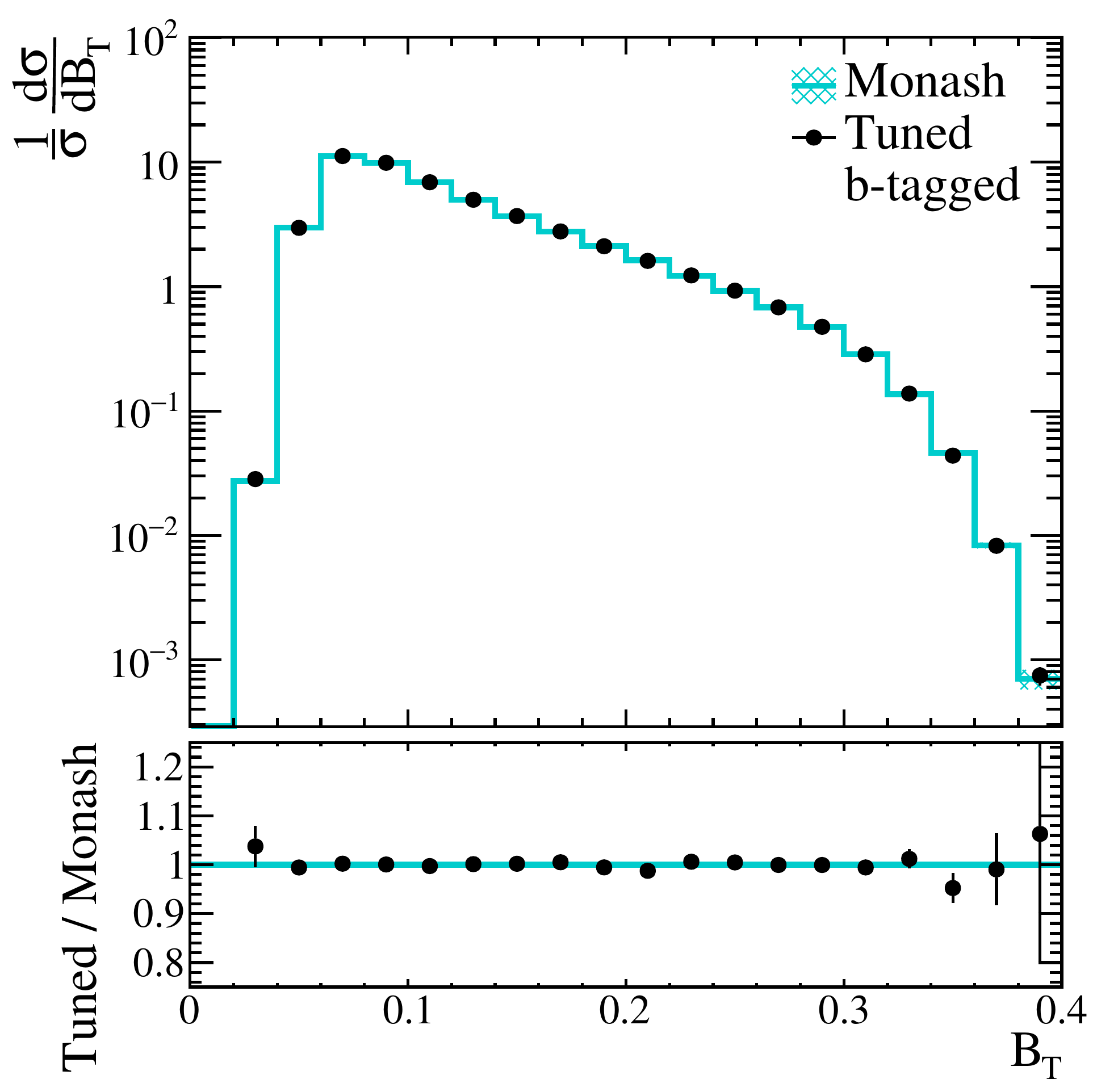}
  \caption{Additional event-shape distributions obtained from the Monash data sample compared to those obtained from our optimal tune of the parameters in block 1. Both samples used here have 10M events.
}
  \label{fig:block1_dists2}
\end{figure}

\begin{figure}
  \centering
  \includegraphics[width=0.99\textwidth]{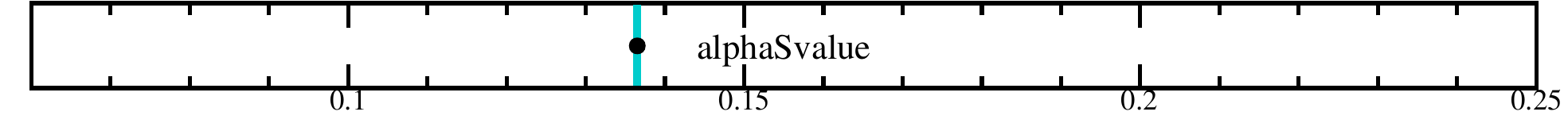}
  \includegraphics[width=0.99\textwidth]{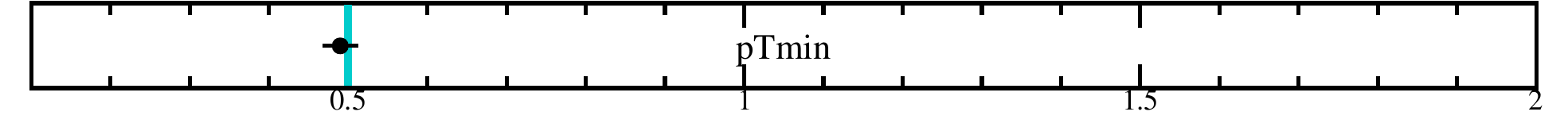}
  \includegraphics[width=0.99\textwidth]{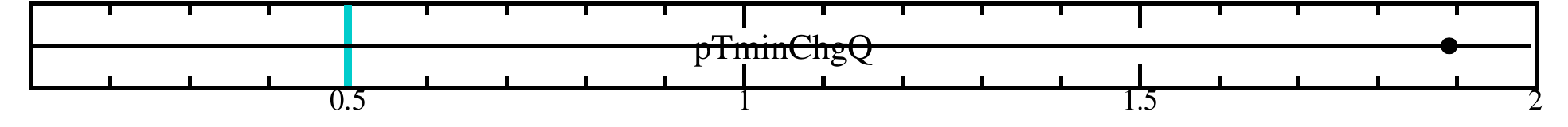}
  \caption{(black points) Block 1 parameters from our optimal tune compared to their (vertical cyan lines) Monash values. The horizontal-axis ranges are the regions considered by \spearmint during tuning.
}
  \label{fig:block1_pars}
\end{figure}

Figures~\ref{fig:block1_dists1}--\ref{fig:block1_pars} show the results of tuning the parameters in block~1 on the event-shape distributions, while the optimal parameter values are listed in Table~\ref{tab:block1_pars}.
The Monash event-shape data distributions are all well described by our tuned \pythia spectra.
The tuned parameter values are consistent to about $\approx 1\sigma$ based on the uncertainties obtained using the method detailed in Appendix~\ref{app:errors}.
The precision on alphaSvalue is 0.0002, while the pTminChgQ confidence interval covers most of the its allowed range.
This difference reflects how well the event-shape distributions constrain each parameter.

\begin{figure}
  \centering
  \includegraphics[width=0.49\textwidth]{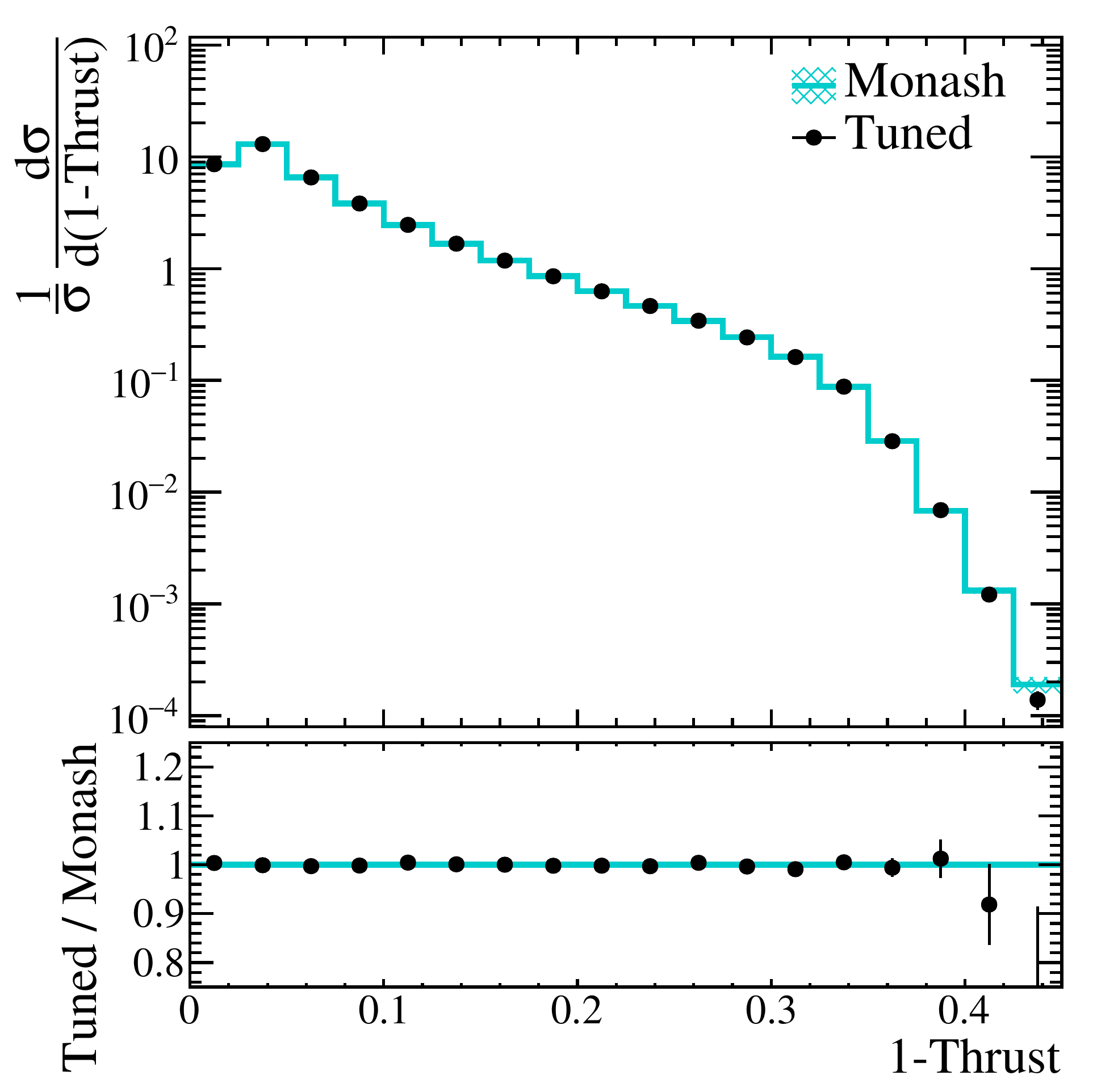}
  \includegraphics[width=0.49\textwidth]{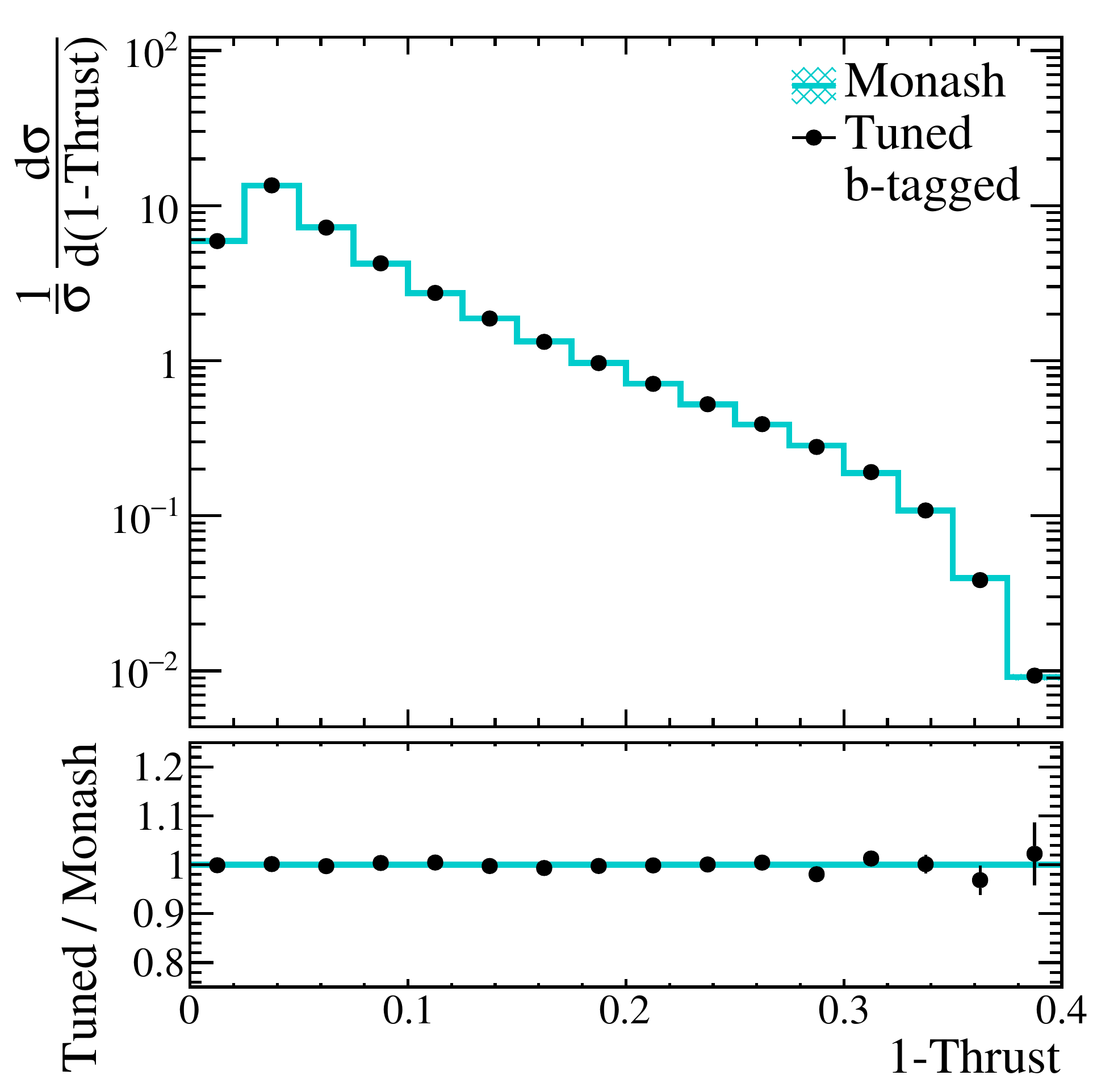}
  \includegraphics[width=0.49\textwidth]{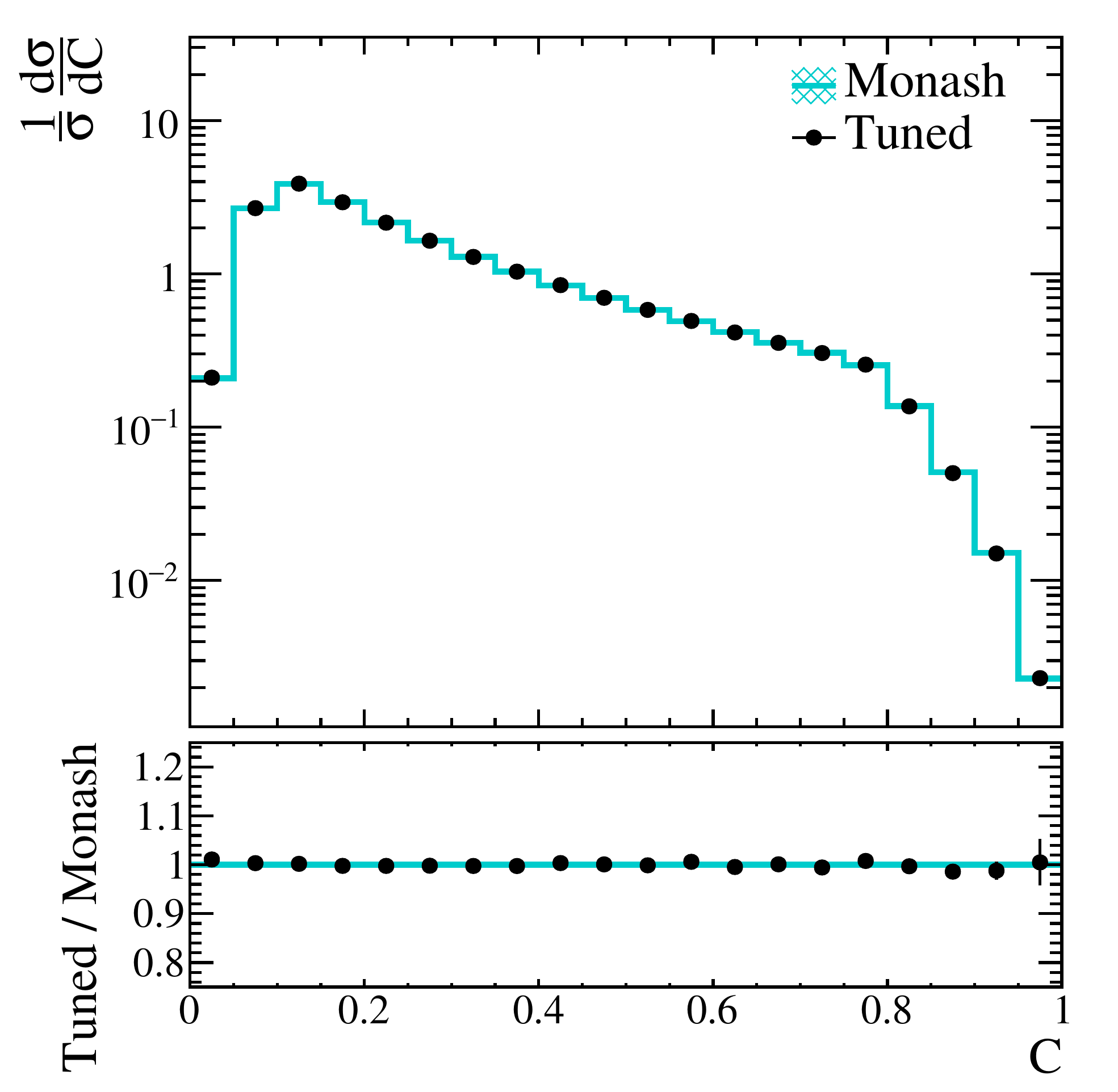}
  \includegraphics[width=0.49\textwidth]{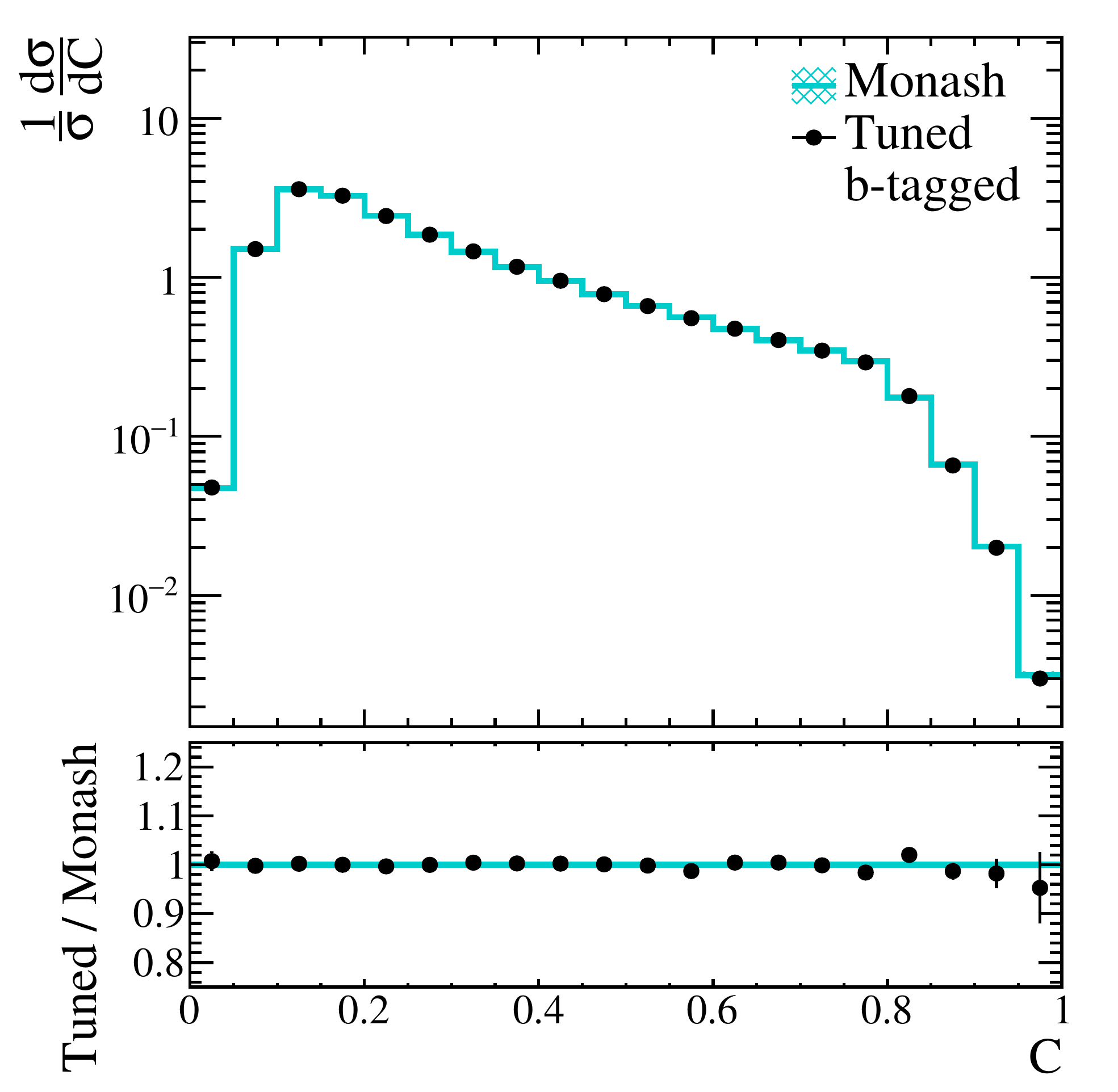}
  \includegraphics[width=0.49\textwidth]{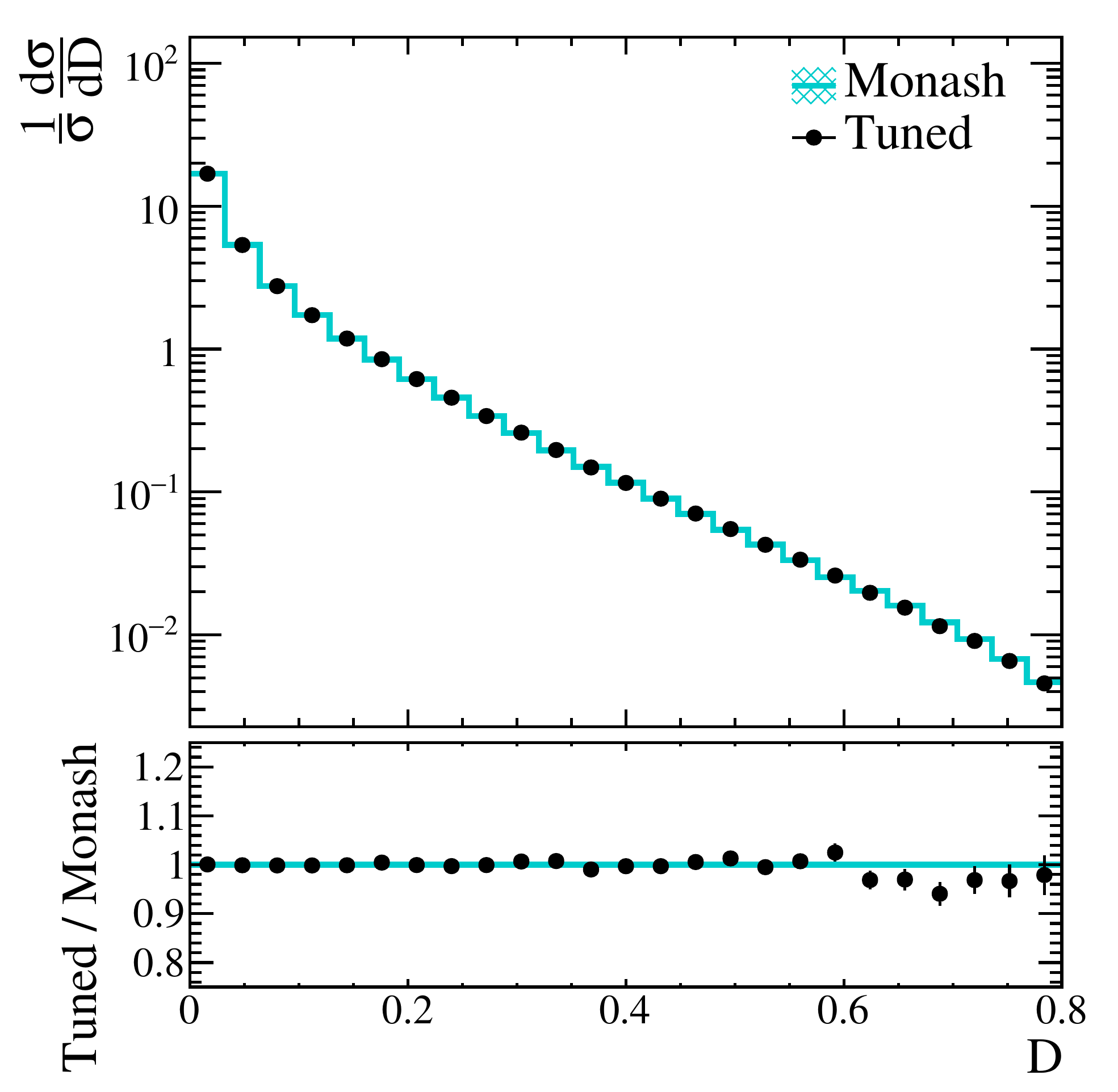}
  \includegraphics[width=0.49\textwidth]{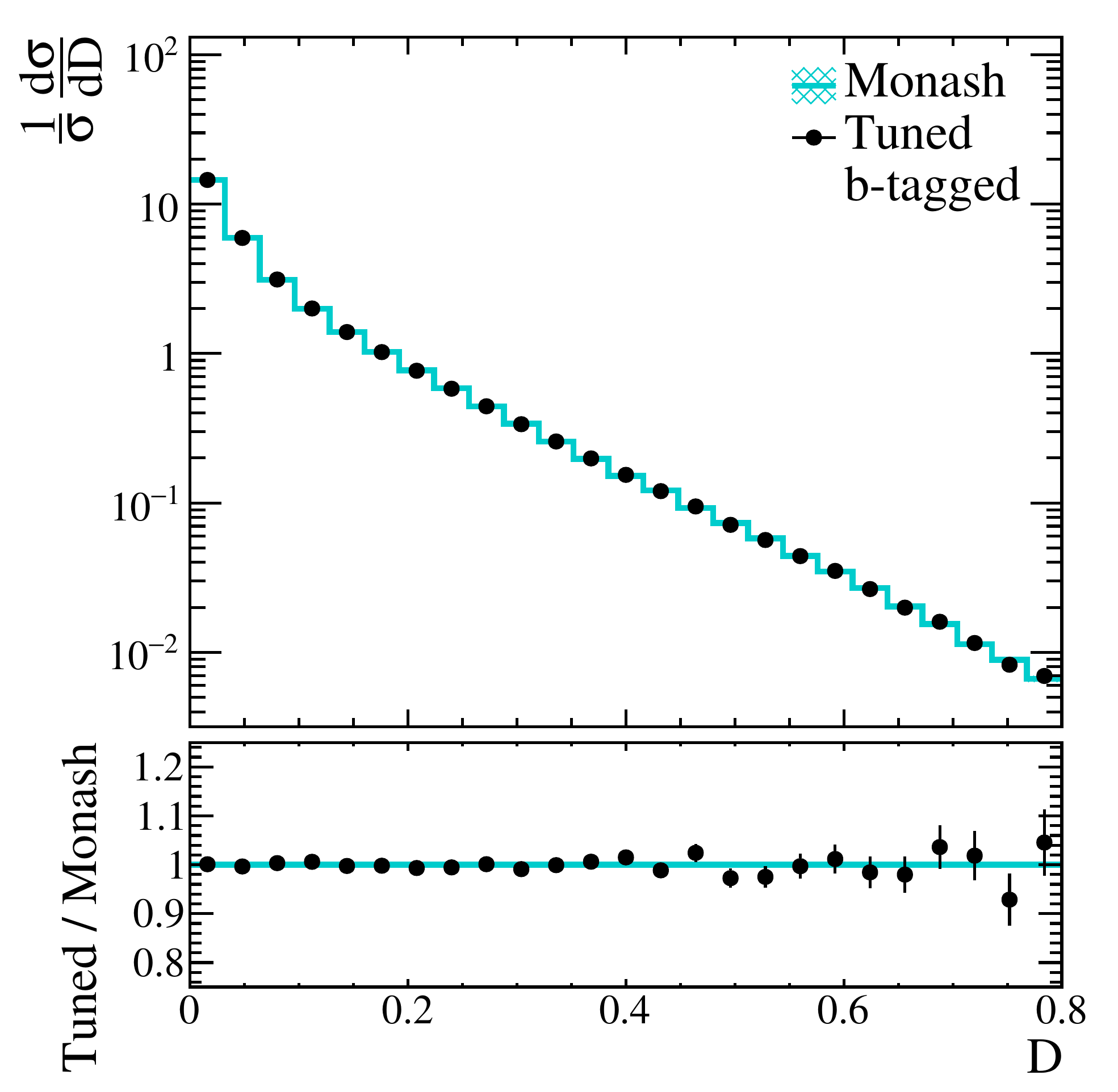}
  \caption{Event-shape distributions obtained from the Monash data sample compared to those obtained from our optimal tune of the parameters in block 1. Both samples used here have 10M events.
}
  \label{fig:block1_dists1}
\end{figure}

\begin{figure}
  \centering
   \includegraphics[width=0.49\textwidth]{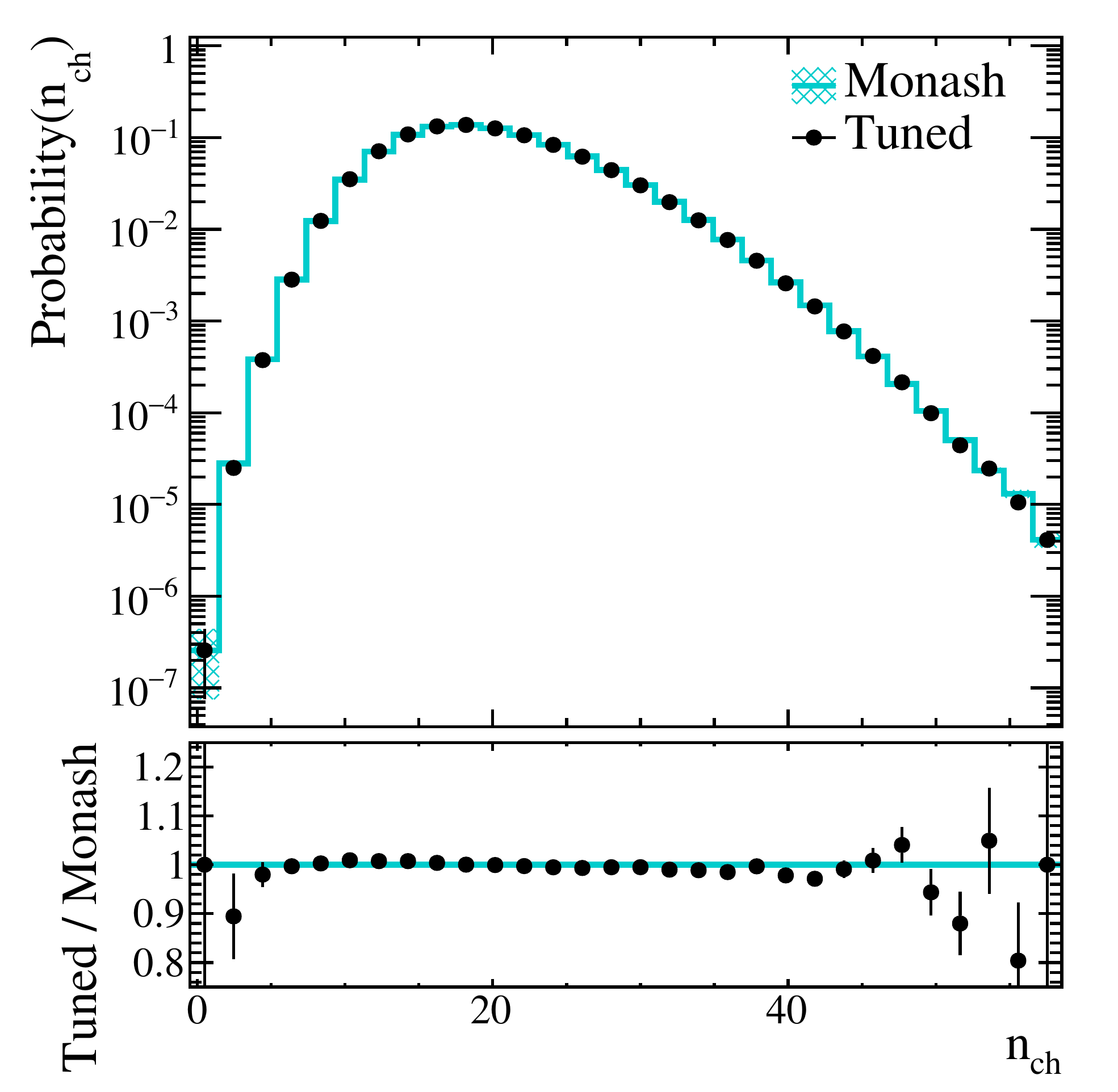}
   \includegraphics[width=0.49\textwidth]{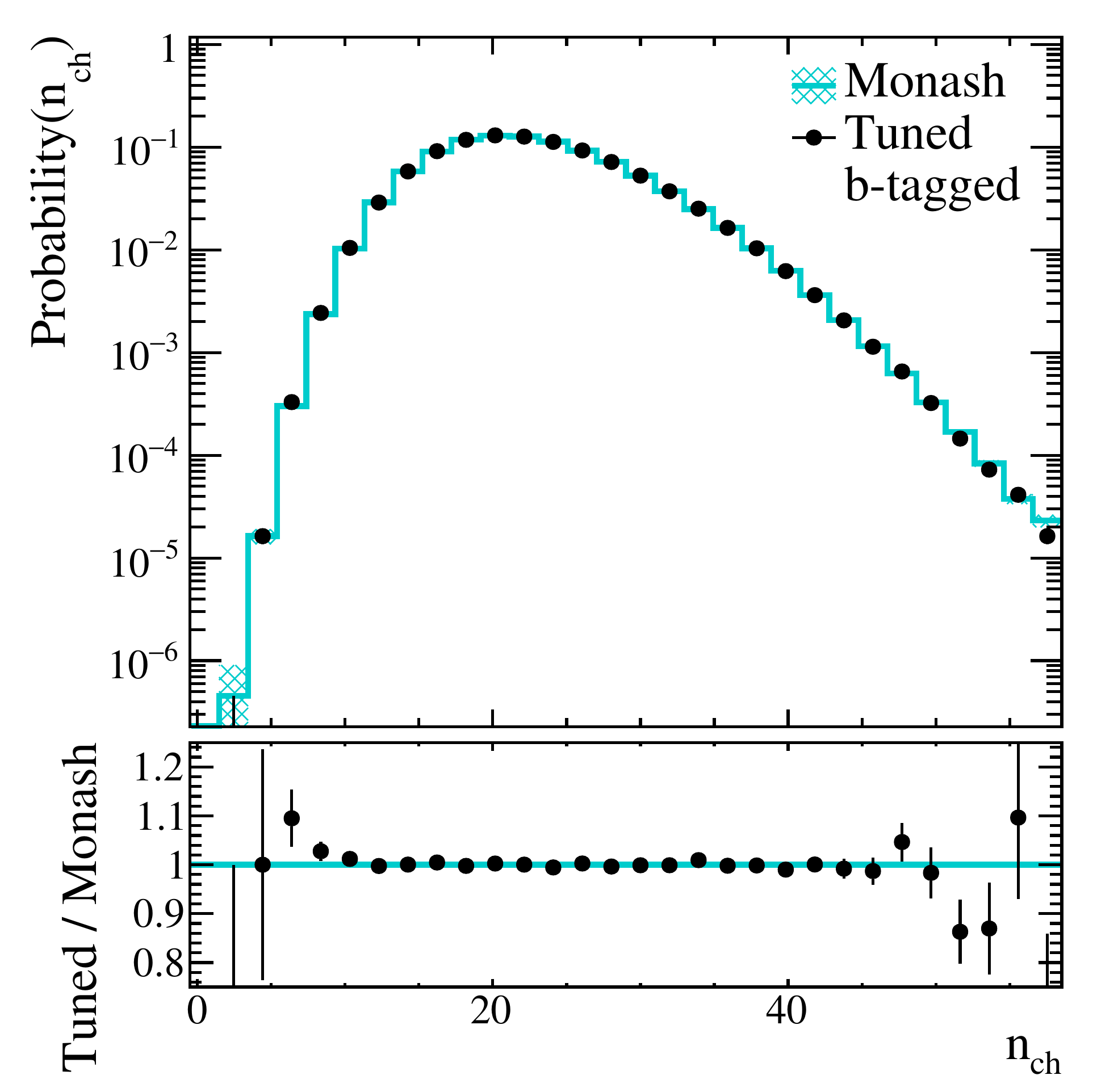}
   \includegraphics[width=0.49\textwidth]{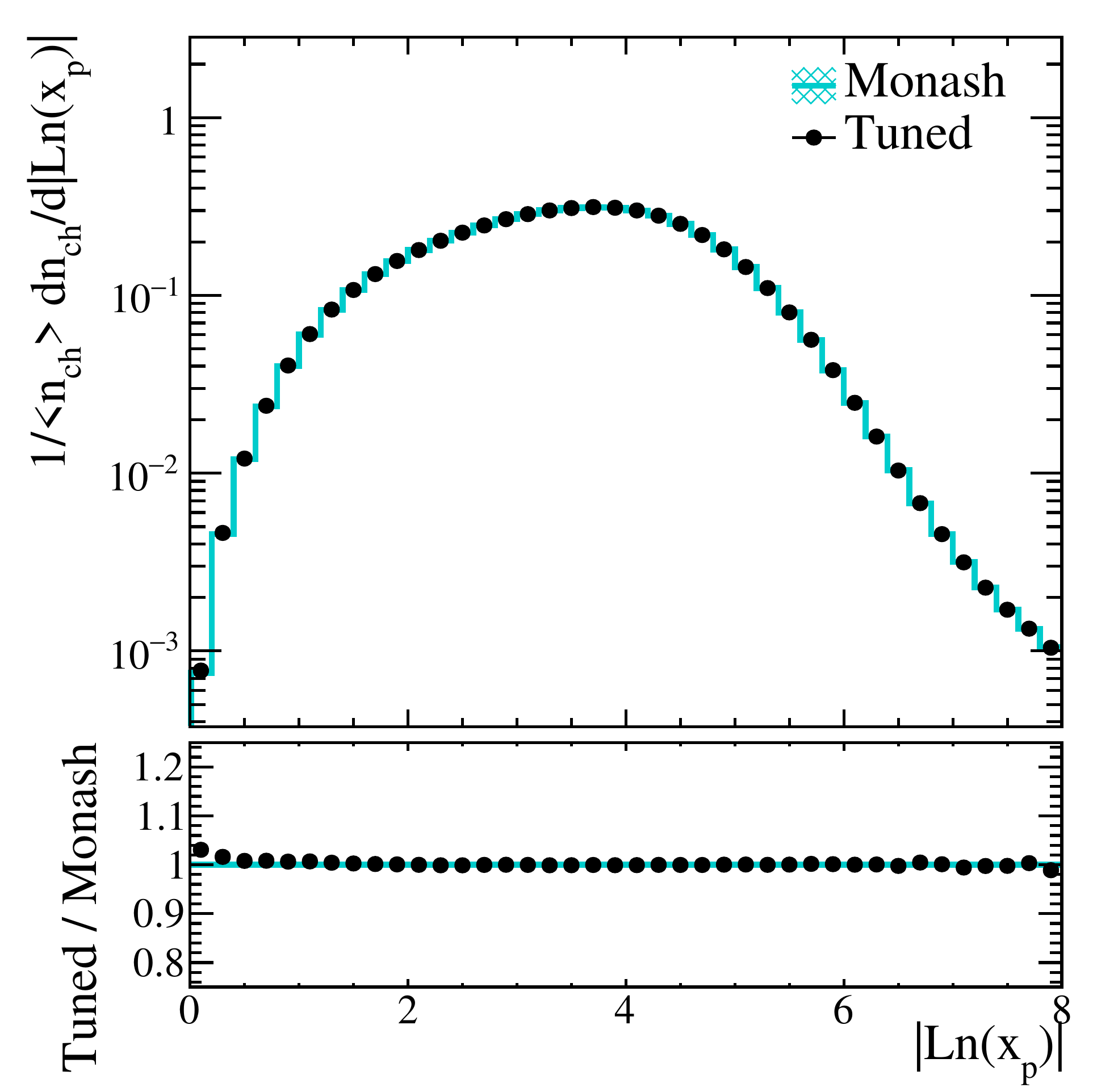}
   \includegraphics[width=0.49\textwidth]{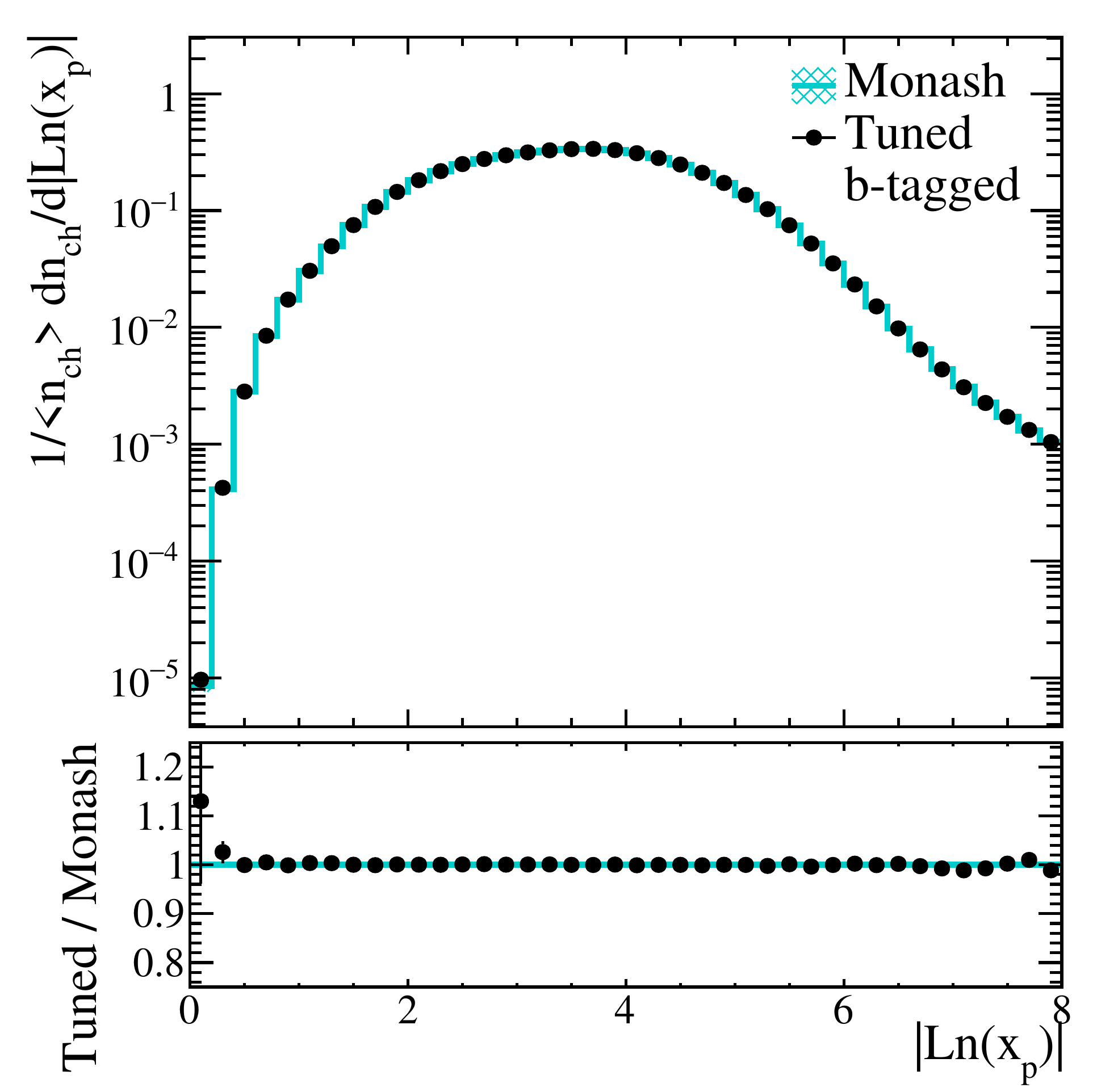}
  \includegraphics[width=0.49\textwidth]{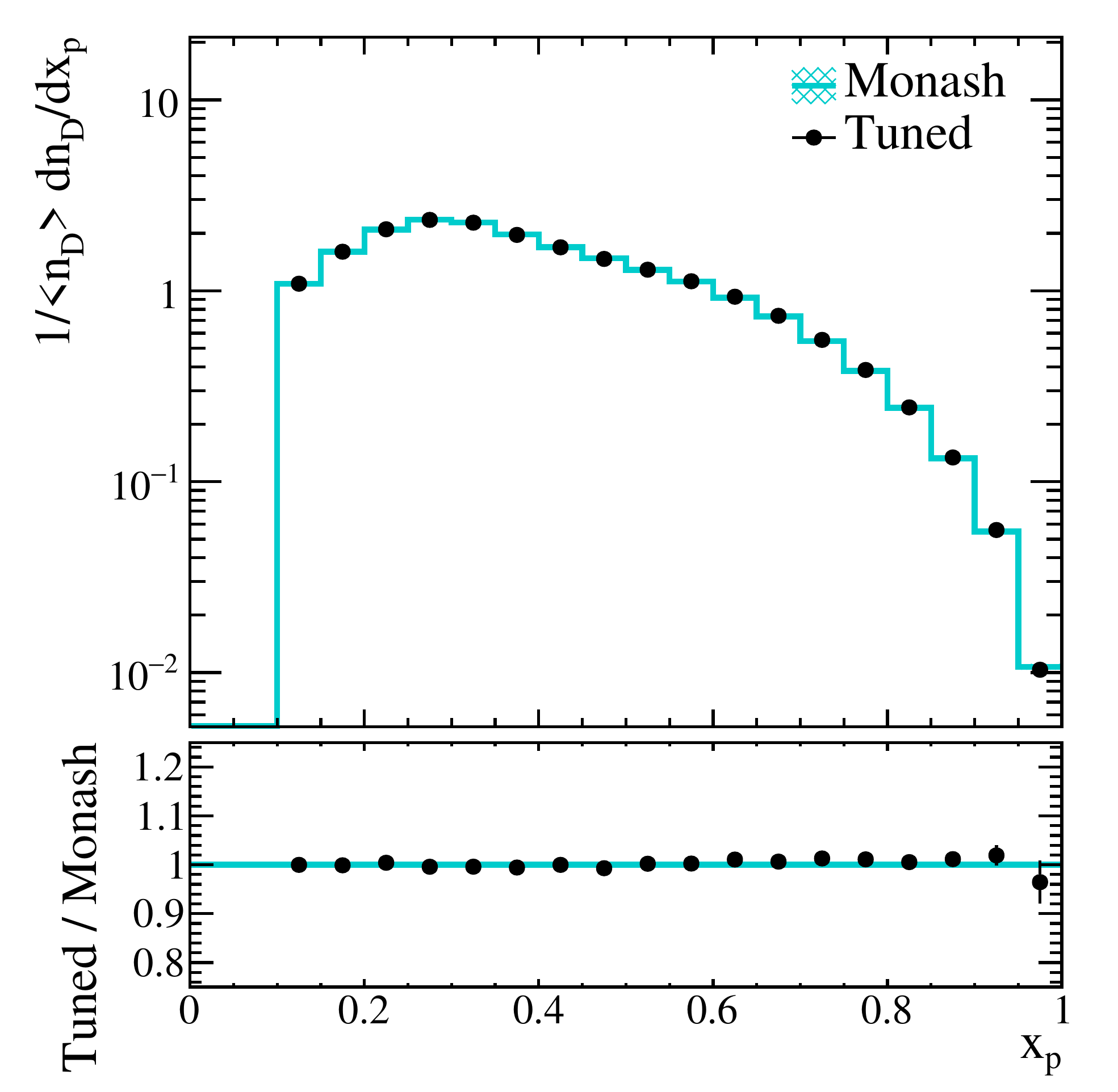}
  \includegraphics[width=0.49\textwidth]{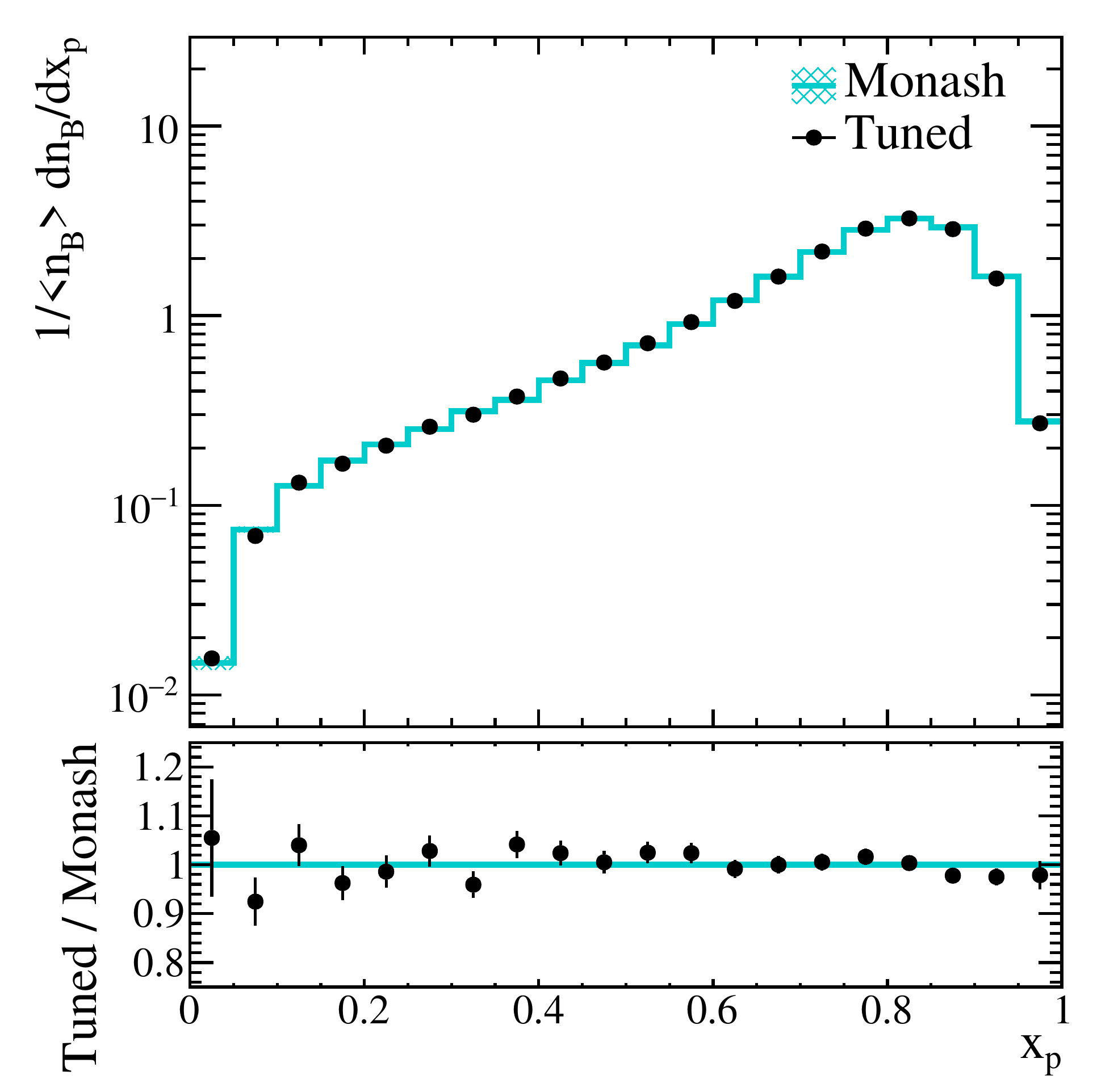}
  \caption{Fragmentation distributions obtained from the Monash data sample compared to those obtained from our optimal tune of the parameters in block 2. Both samples used here have 10M events.
}
  \label{fig:block2_dists}
\end{figure}

\begin{figure}
  \centering
  \includegraphics[width=0.99\textwidth]{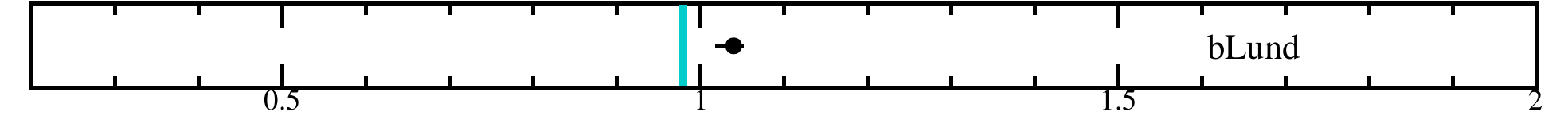}
  \includegraphics[width=0.99\textwidth]{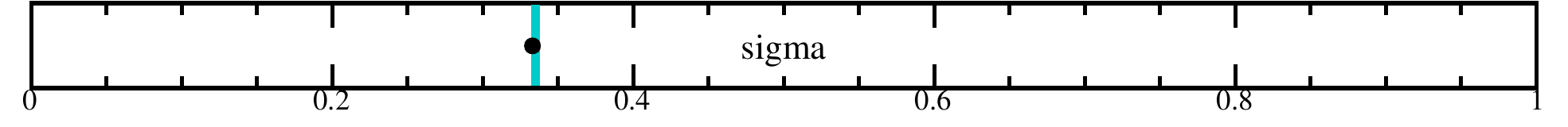}
  \includegraphics[width=0.99\textwidth]{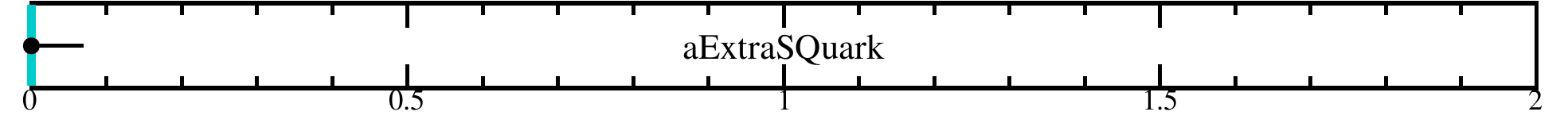}
  \includegraphics[width=0.99\textwidth]{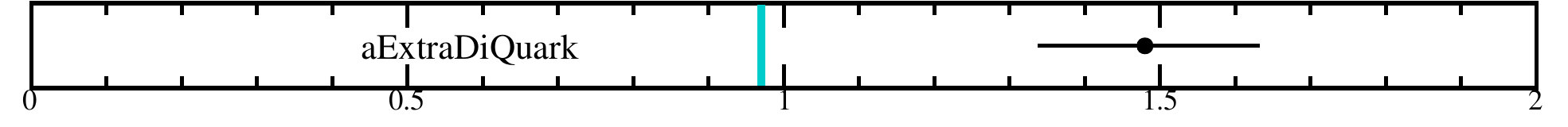}
  \includegraphics[width=0.99\textwidth]{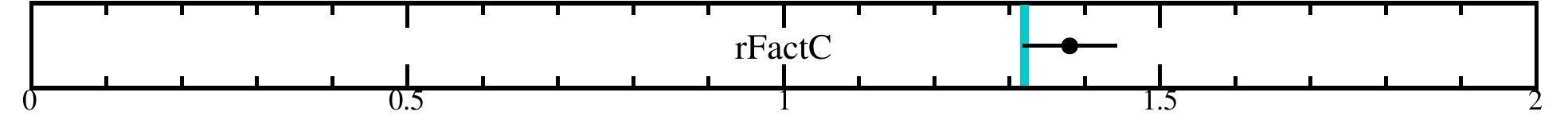}
  \includegraphics[width=0.99\textwidth]{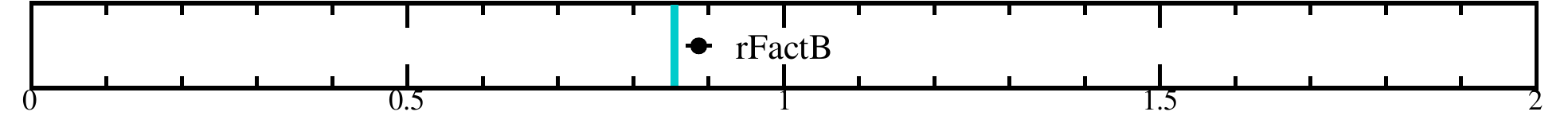}
  \caption{(black points) Block 2 parameters from our optimal tune compared to their (vertical cyan lines) Monash values. The horizontal-axis ranges are the regions considered by \spearmint during tuning.
}
  \label{fig:block2_pars}
\end{figure}

The results of tuning the block~2 parameters on the fragmentation distributions are presented in Figs.~\ref{fig:block2_dists} and \ref{fig:block2_pars}, and in Table~\ref{tab:block2_pars}.
We again find good agreement between the Monash and tuned spectra.
Several of the parameter uncertainties undercover in this case due to sizable correlations, {\em e.g.}, the aExtraDiQuark parameter is about $3\sigma$ from its Monash value.
This parameter is not well constrained by the distributions used and its impact on the $\chi^2$ is highly correlated with that of bLund.
A proper error bar can be assigned to this parameter by scanning the \spearmint model in 2-D; however, the simplified uncertainties assigned here, which ignore parameter correlations for the sake of saving CPU, are sufficient to convey how well each parameter is constrained.
The parameter aExtraSQuark is only constrained to be $\lesssim 0.1$ by the spectra used in the tune, and in such cases we find that \spearmint tends to select the edge of the region considered, {\em e.g.}, aExtraSQuark is tuned to its minimum allowed value of 0.

\begin{figure}
  \centering
 \includegraphics[width=0.49\textwidth]{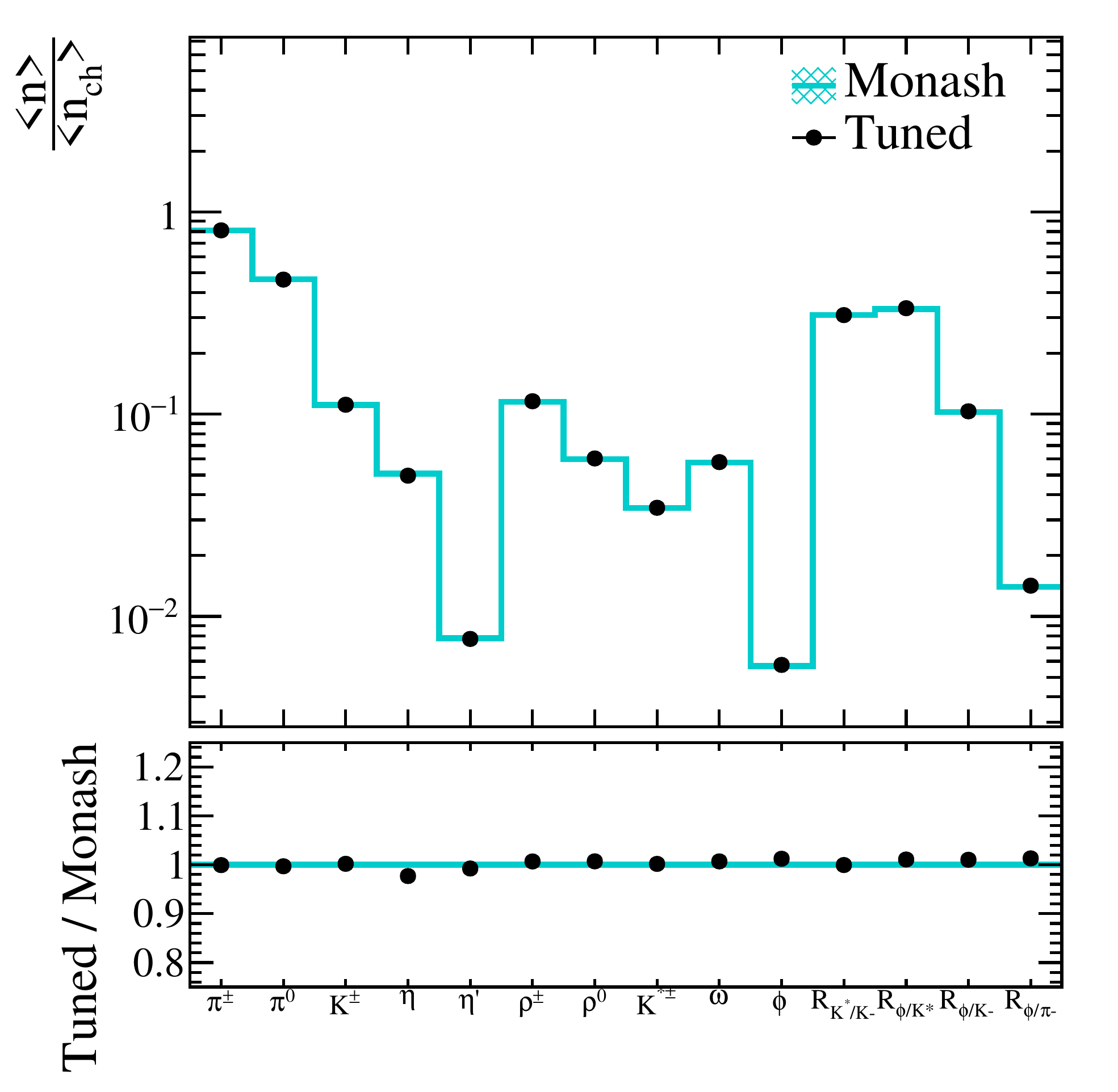}
  \includegraphics[width=0.49\textwidth]{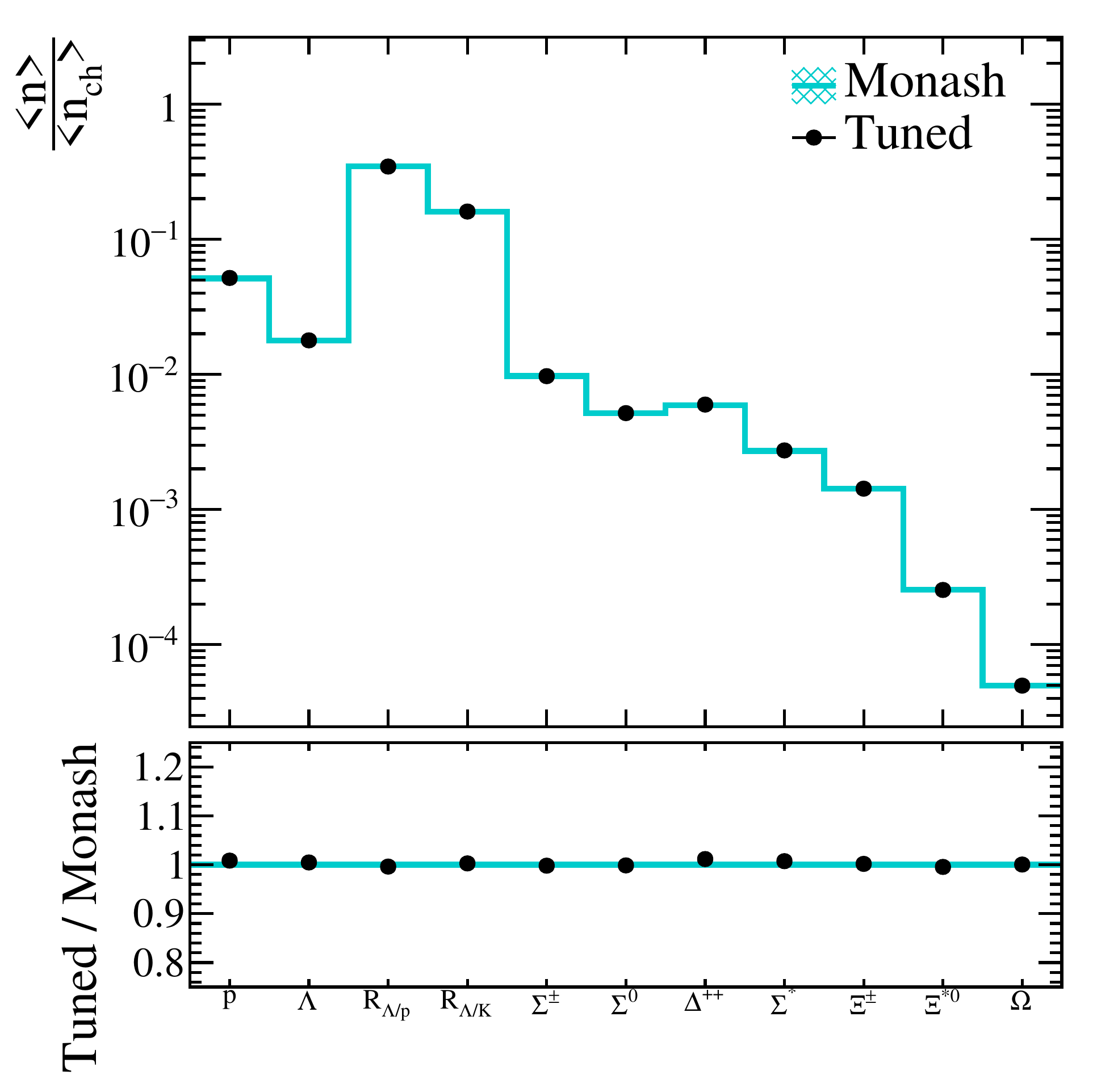}
  \includegraphics[width=0.49\textwidth]{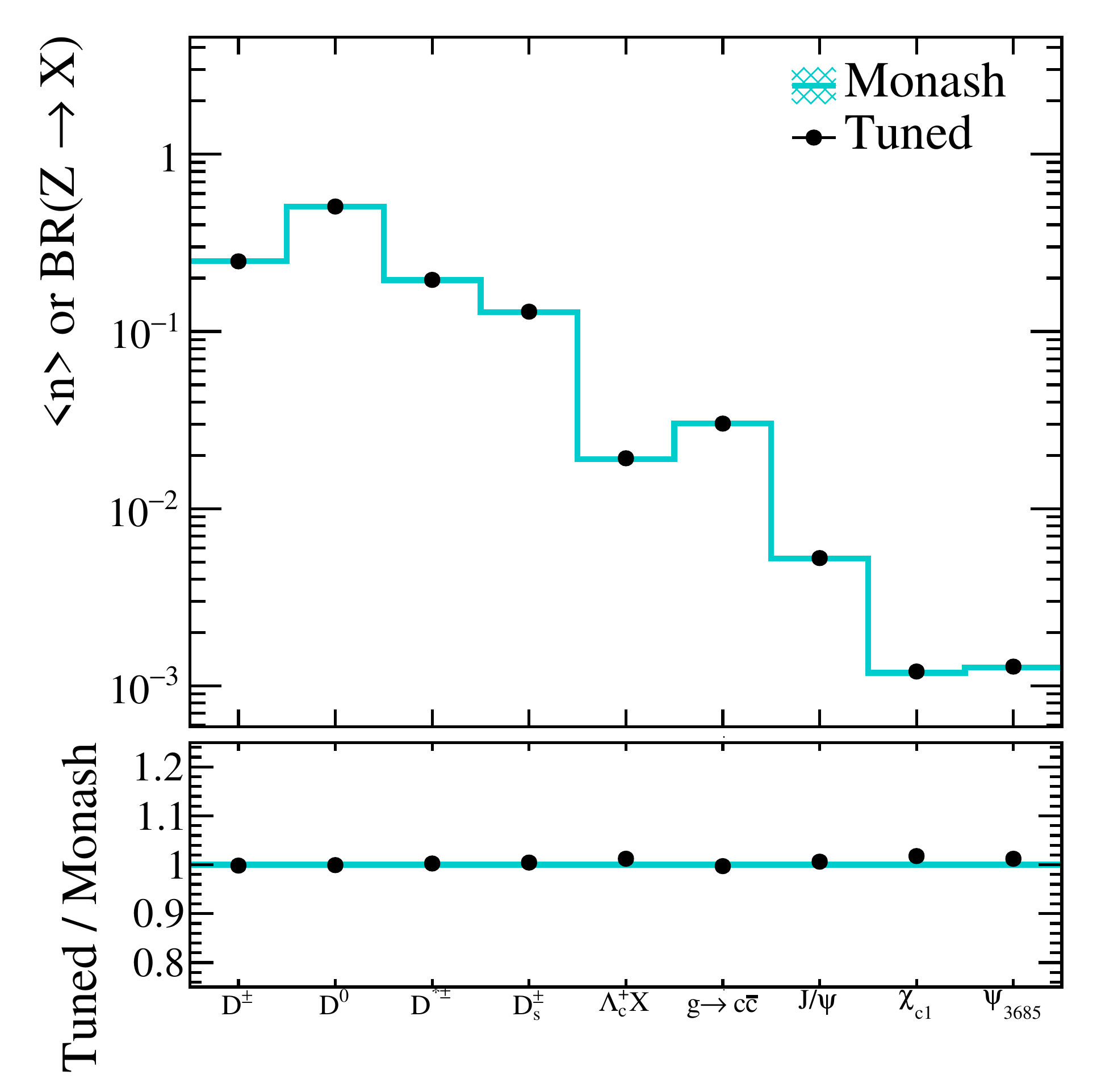}
  \includegraphics[width=0.49\textwidth]{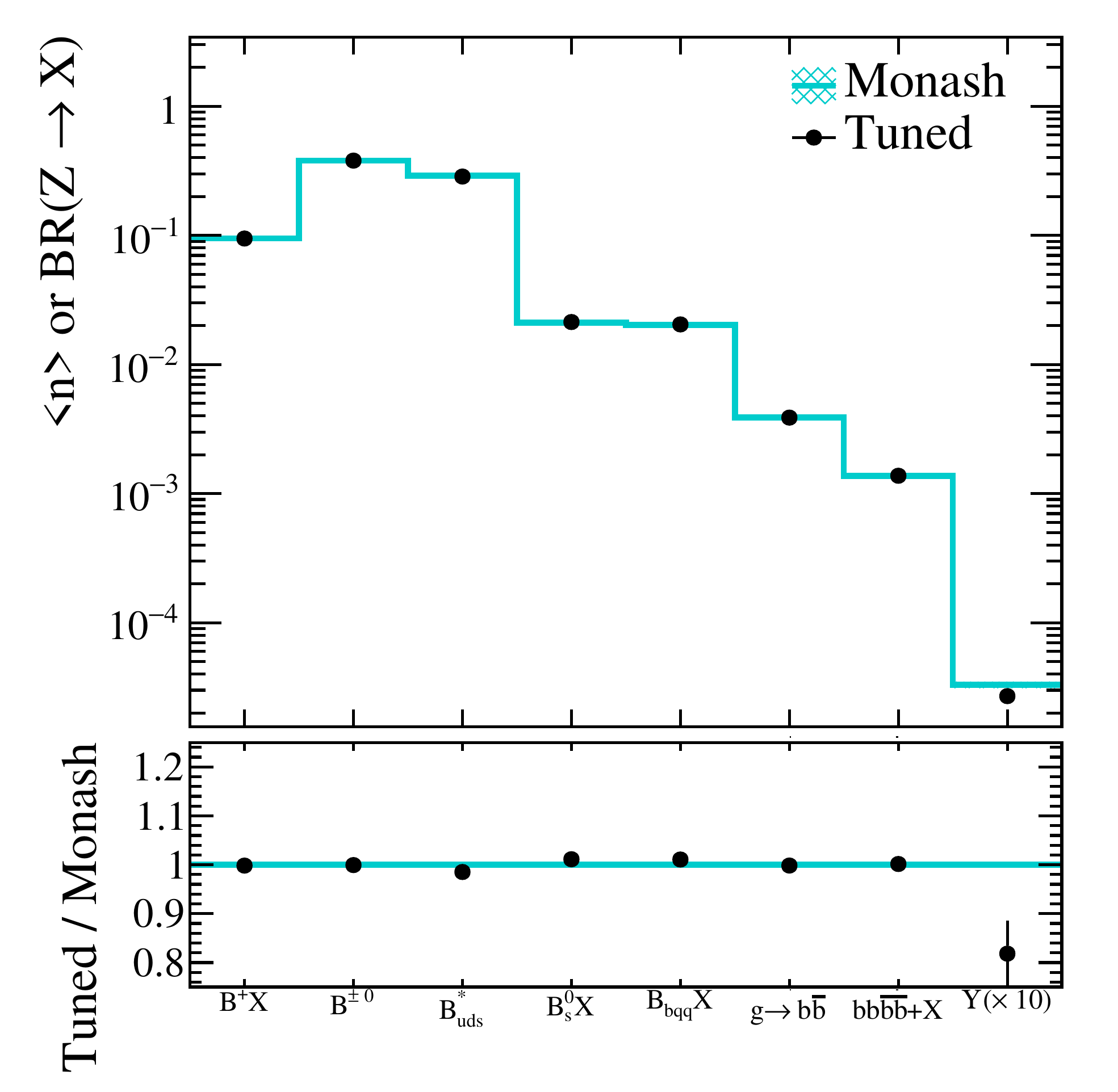}
  \caption{Hadron-type distributions obtained from the Monash data sample compared to those obtained from our optimal tune of the parameters in block 3.   Both samples used here have 10M events.
}
  \label{fig:block3_dists}
\end{figure}

\begin{figure}
  \centering
  \includegraphics[width=0.99\textwidth]{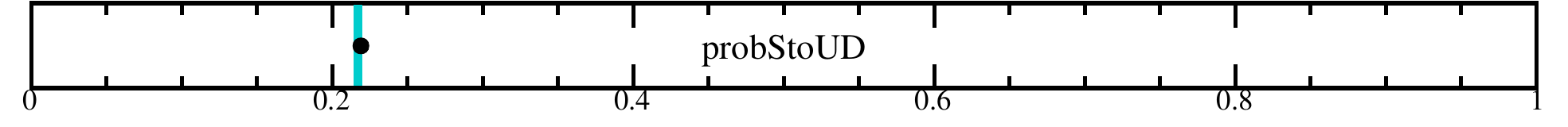}
  \includegraphics[width=0.99\textwidth]{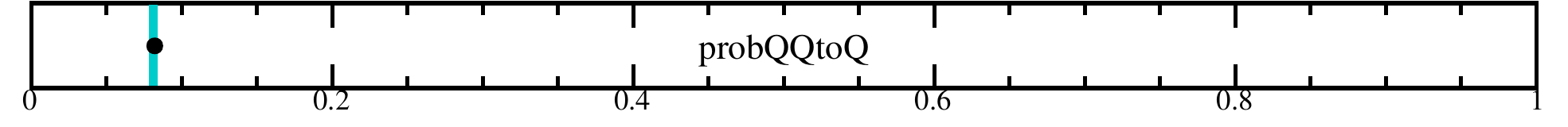}
  \includegraphics[width=0.99\textwidth]{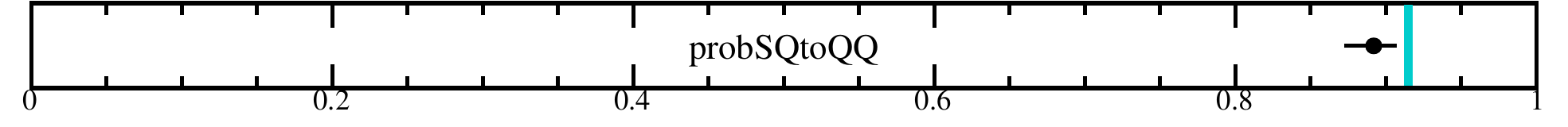}
  \includegraphics[width=0.99\textwidth]{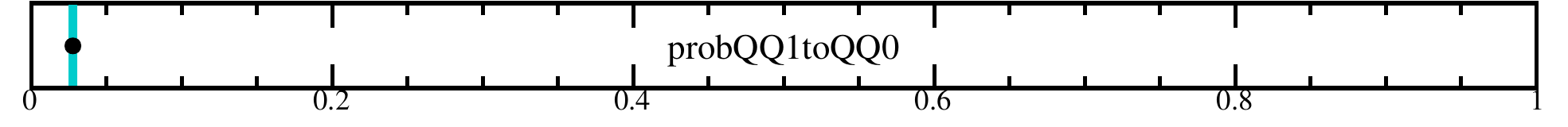}
  \includegraphics[width=0.99\textwidth]{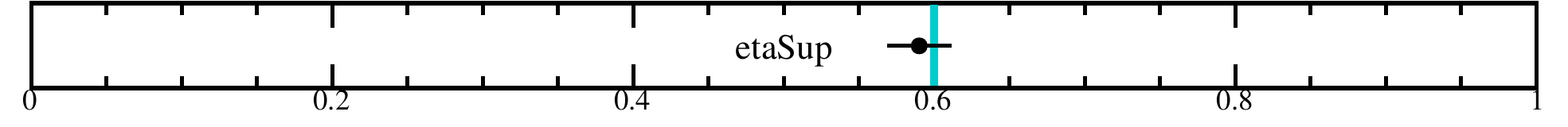}
  \includegraphics[width=0.99\textwidth]{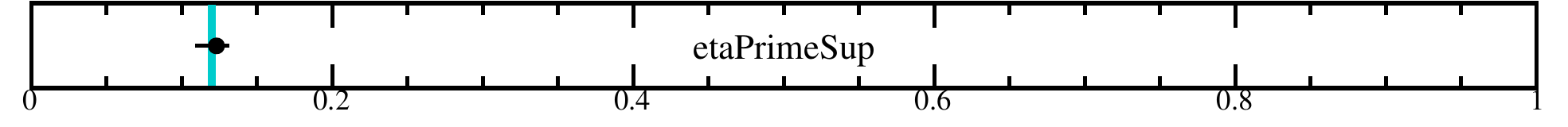}
  \includegraphics[width=0.99\textwidth]{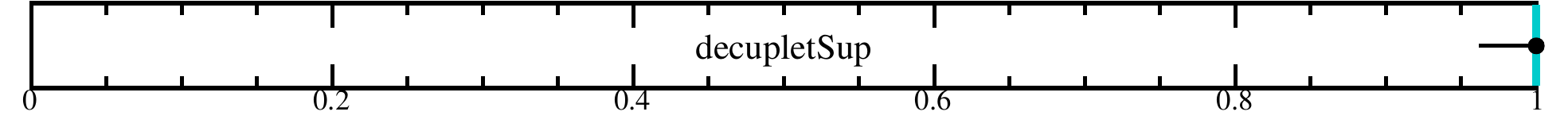}
  \includegraphics[width=0.99\textwidth]{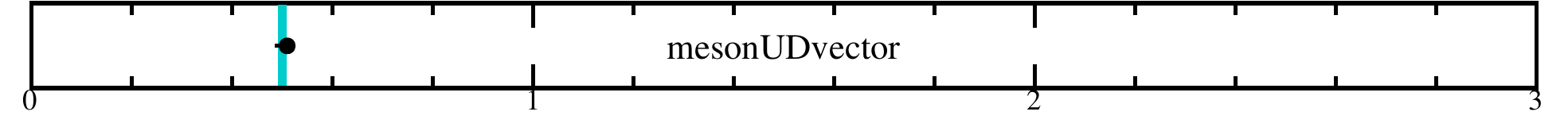}
  \includegraphics[width=0.99\textwidth]{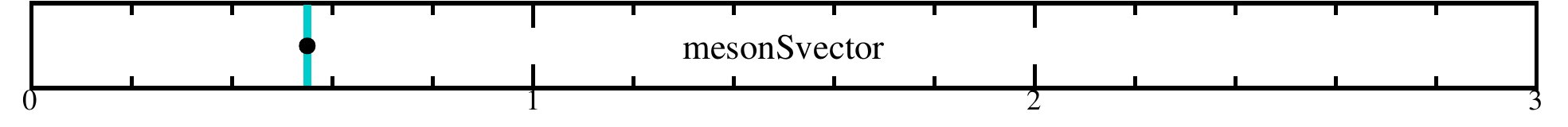}
  \includegraphics[width=0.99\textwidth]{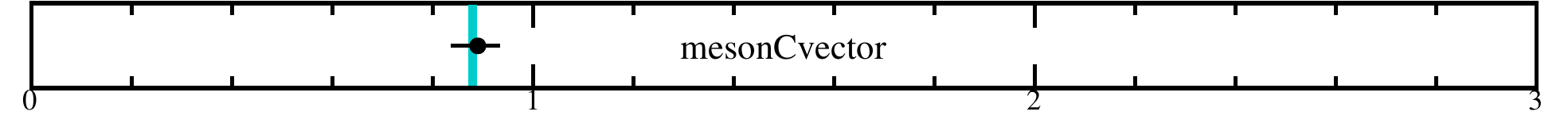}
  \includegraphics[width=0.99\textwidth]{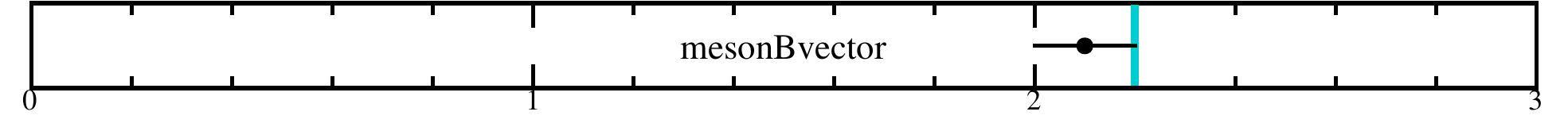}
  \caption{
(black points) Block 3 parameters from our optimal tune compared to their (vertical cyan lines) Monash values. The horizontal-axis ranges are the regions considered by \spearmint during tuning.
}
  \label{fig:block3_pars}
\end{figure}

The results of tuning the block~3 parameters on the hadron-type distributions are presented in Figs.~\ref{fig:block3_dists} and \ref{fig:block3_pars}, and in Table~\ref{tab:block3_pars}.
The block 3 tune involves 11 parameters, and we again find that all data distributions and \pythia parameter values are consistent with Monash.
Based on these results, we conclude that \spearmint has successfully performed a 20 parameter tune of \pythia~8 using the block-diagonal strategy that was also employed in the Monash tune~\cite{Skands:2014pea}.

\begin{figure}
   \includegraphics[width=0.99\textwidth]{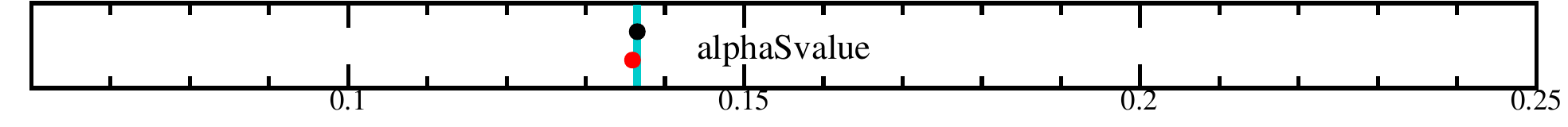}
  \includegraphics[width=0.99\textwidth]{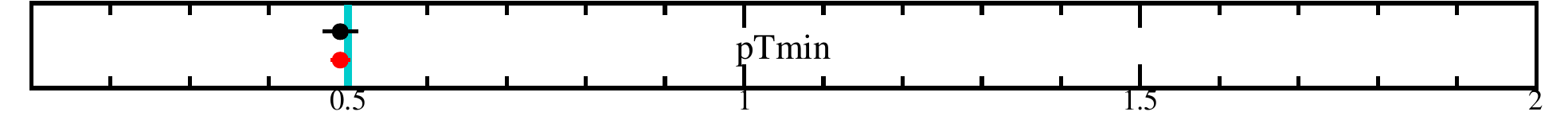}
  \vspace{-0.17in}\\
  \includegraphics[width=0.99\textwidth]{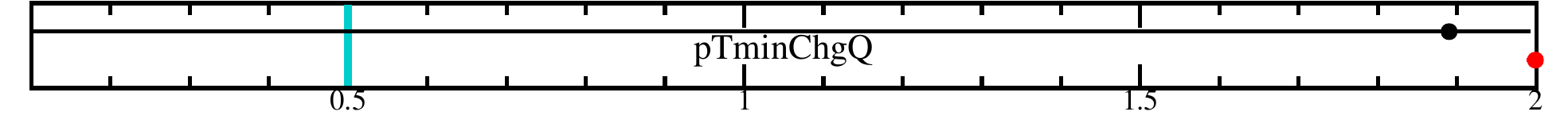}
  \includegraphics[width=0.99\textwidth]{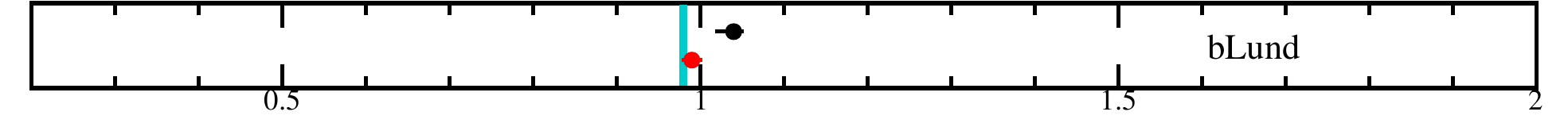}
  \includegraphics[width=0.99\textwidth]{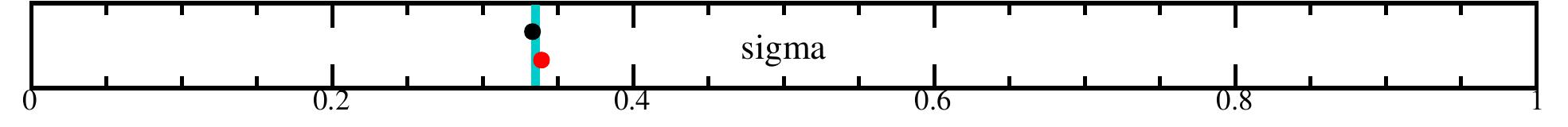}
  \vspace{-0.17in}\\
  \includegraphics[width=0.99\textwidth]{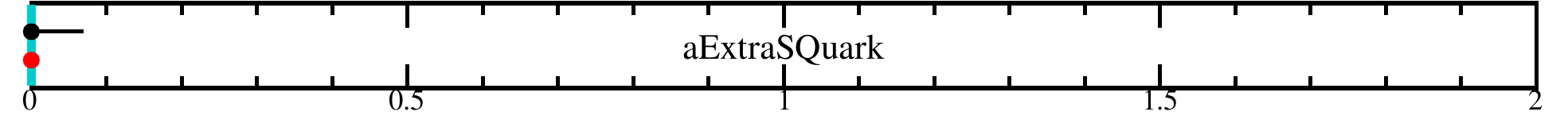}
  \vspace{-0.17in}\\
  \includegraphics[width=0.99\textwidth]{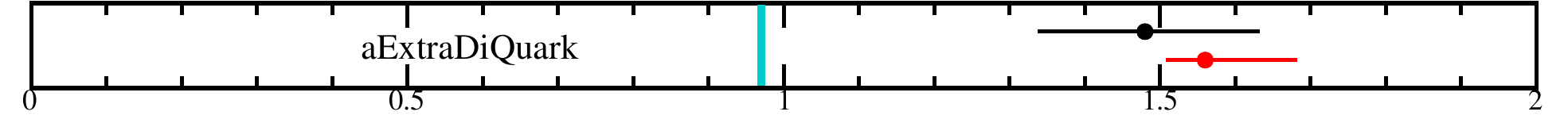}
  \vspace{-0.17in}\\
  \includegraphics[width=0.99\textwidth]{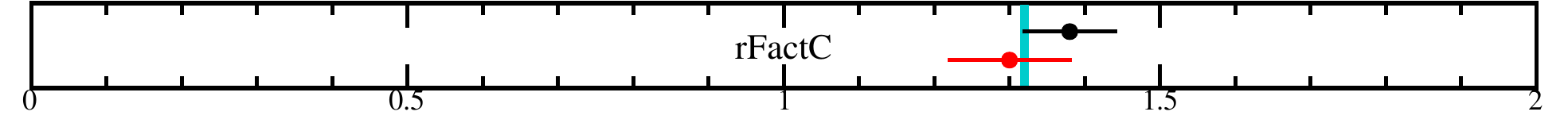}
  \vspace{-0.17in}\\
  \includegraphics[width=0.99\textwidth]{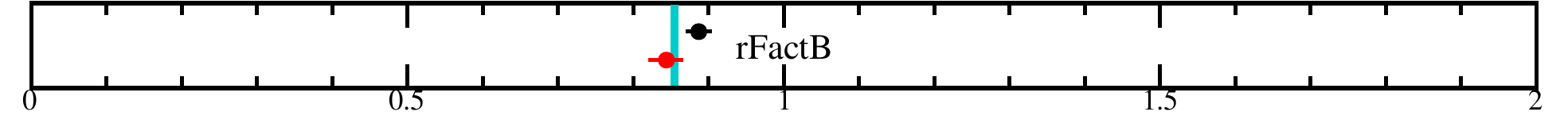}
   \includegraphics[width=0.99\textwidth]{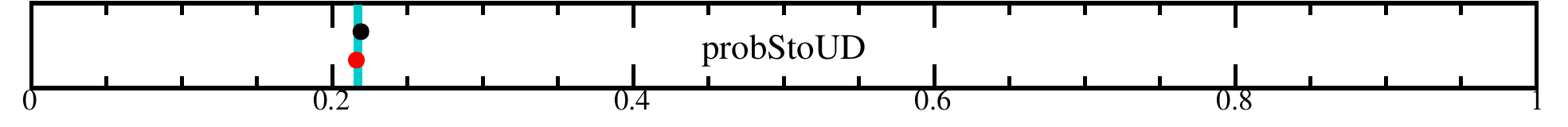}
  \vspace{-0.17in}\\
  \includegraphics[width=0.99\textwidth]{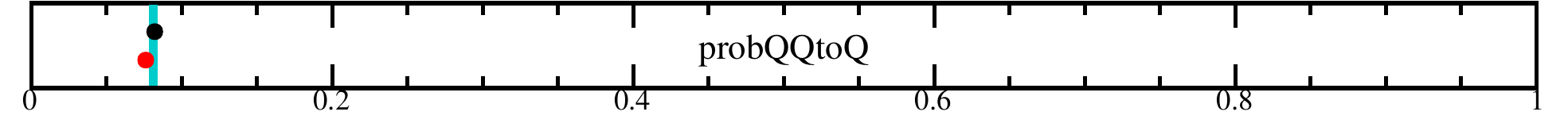}
  \vspace{-0.17in}\\
  \includegraphics[width=0.99\textwidth]{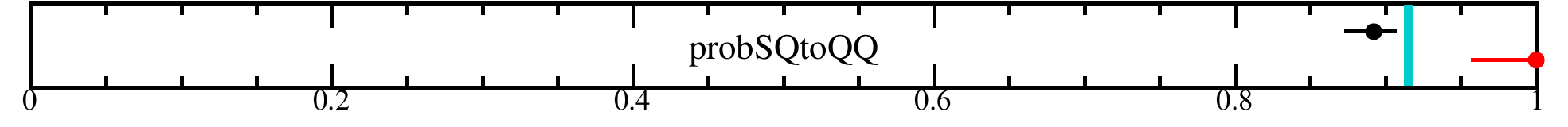}
  \vspace{-0.17in}\\
  \includegraphics[width=0.99\textwidth]{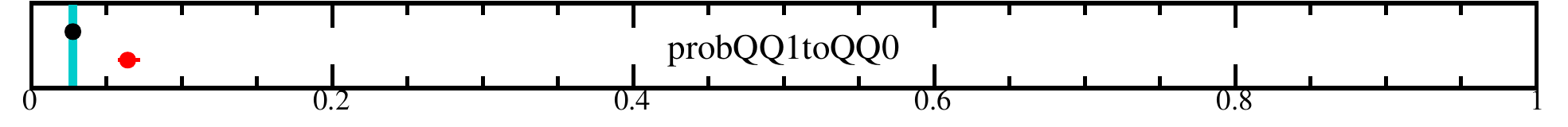}
  \vspace{-0.17in}\\
  \includegraphics[width=0.99\textwidth]{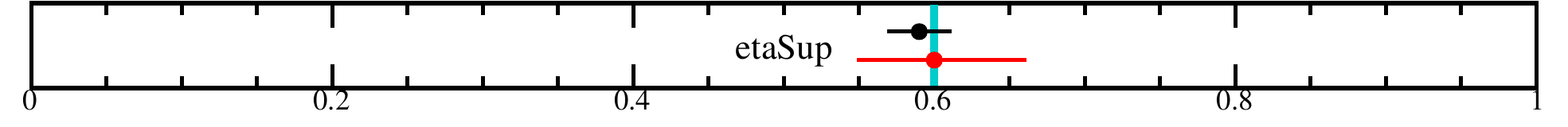}
  \vspace{-0.17in}\\
  \includegraphics[width=0.99\textwidth]{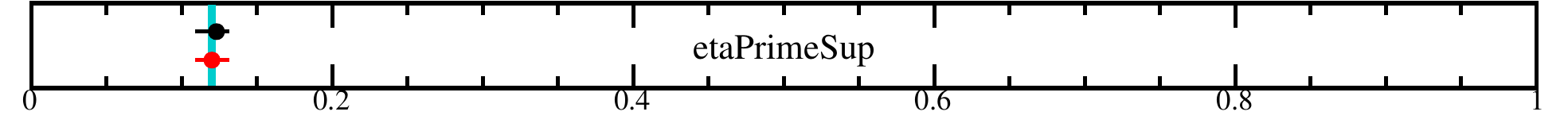}
  \vspace{-0.17in}\\
  \includegraphics[width=0.99\textwidth]{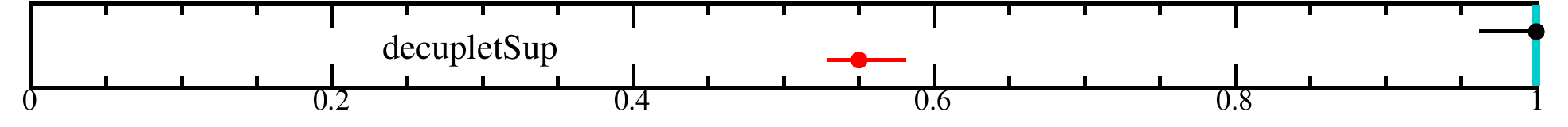}
 \includegraphics[width=0.99\textwidth]{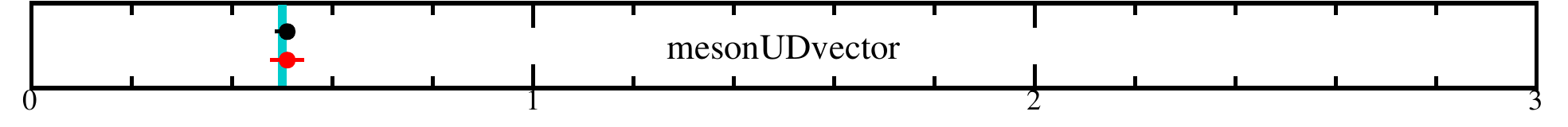}
  \vspace{-0.17in}\\
  \includegraphics[width=0.99\textwidth]{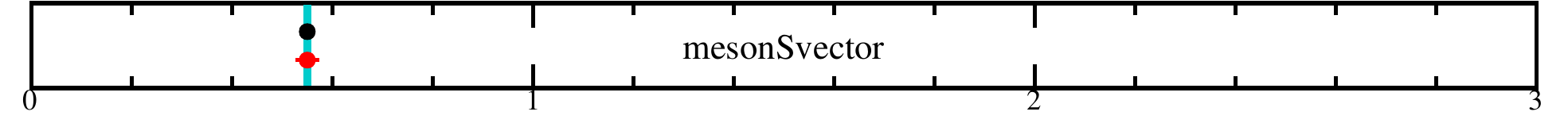}
  \vspace{-0.17in}\\
  \includegraphics[width=0.99\textwidth]{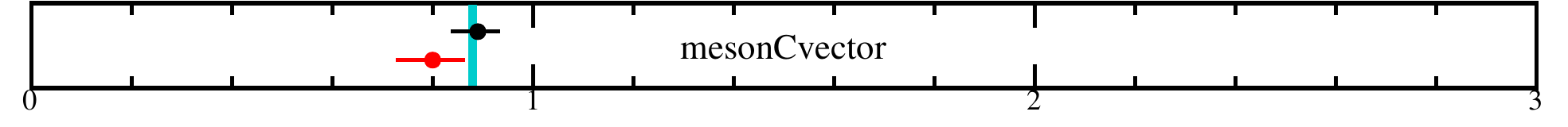}
  \vspace{-0.17in}\\
  \includegraphics[width=0.99\textwidth]{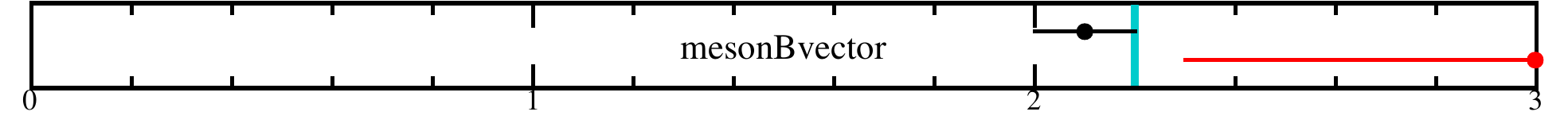}
  \caption{
(red points) Global tune of the block 1, 2, and 3 parameters compared to their (vertical cyan lines) Monash values. The horizontal-axis ranges are the regions considered by \spearmint during tuning. (black points) The block-diagonal results for each parameter are also shown.
}
  \label{fig:global_pars}
\end{figure}

A novel aspect of using the Bayesian optimization framework is the possibility to perform a global tune of all 20 parameters using all 20 distributions --- on a laptop.
Since most of the 20 parameters considered in our $e^+e^-$ tune are only strongly constrained by the distributions in a single block, we do not expect a global tune to improve the precision of the tuned parameter values.\footnote{If a parameter is only constrained by the distributions in a single block, the global tune should provide worse precision for the parameter due to the increased sample-to-sample variation in the global $\chi^2$ compared to that of the individual block $\chi^2$ values.}
That said, demonstrating that a 20 parameter tune is possible is of interest regardless of its utility in this specific example.
Figure~\ref{fig:global_pars} shows the optimal parameter values obtained from the global $e^+e^-$ tune, compared to those obtained from the tunes of the three blocks.
As expected, the global tune does not improve the precision;\footnote{The uncertainties are underestimated for a few of the parameters in the global tune, {\em e.g.}, pTminChgQ.  This is likely due to the global tune being terminated before it had fully converged.  Since the global tune is only presented here to demonstrate proof of principle, we have not investigated whether this issue  goes away with additional queries.} however, the fact that such a tune can be run on a laptop is an exciting result.
This capability may prove useful in future tunes of Monte Carlo event generators.

%\clearpage

%\clearpage

The results of our closure test of applying the Bayesian optimization framework to Monte Carlo event generator tuning are summarized as follows:
\begin{itemize}
\item The true Monash parameter values are accurately and precisely determined for all parameter blocks listed in Table~\ref{tab:pars}, even when a minimal amount of expert knowledge about each parameter is used.
\item The three blocks contain 3, 6, and 11 parameters, and no decrease in performance is observed in the 11 parameter tune relative to the others.  Furthermore, inclusion of a few poorly constrained parameters does not greatly impact the tuning process.
\item The uncertainties placed on the tuned parameters using the method described in Appendix~\ref{app:errors} provide reasonable coverage properties in the absence of sizable correlations, and in all cases are sufficient for estimating how well a parameter is constrained by the experimental data distributions.
\item A 20 parameter global tune converges to a consistent set of parameter values, except for a few poorly constrained parameters.  While the global approach provides no improvement for this particular example, which is expected, the fact that performing a 20 parameter tune is possible is both novel and exciting.
\end{itemize}
Therefore, this closure test demonstrates that Bayesian optimization is a viable method for tuning Monte Carlo event generators.  The real value in its usage for this task will be shown in Sec.~\ref{sec:cpu}, where we demonstrate the small CPU required to perform the tunes presented in this section.

\subsection{Possible Improvements}
\label{sec:improve}

An obvious question arises from the study above: is it possible to transfer knowledge gained from the block-diagonal tunes to the global tune?
This would permit the global tune to focus on regions of the 20-dimensional parameter space that are known to be promising, rather than simply starting over and ignoring the block-diagonal results.
Related applications include transferring knowledge from a tune at one beam energy to another, or from a tune to $e^+e^-$ data to one on proton-proton data.
An interesting area of research on this topic is {\em multi-task} Bayesian optimization~\cite{NIPS2013_5086}, which could prove to be useful for tuning Monte Carlo event generators.
Another way of potentially speeding up the tuning process would be to first consider the $\chi^2$ obtained using a smaller generated sample, and then decide whether or not to generate the 1M-event sample based on the $\chi^2$ value~\cite{NIPS2013_5086}.

\clearpage

\subsection{Tuning Discrete Parameters}
\label{sec:discrete}

As noted above, the Gaussian process framework is not suited for tuning discrete parameters; however, other automated optimization procedures do handle discrete parameters well.
For example, tree-based optimizers handle both discrete and continuous parameters naturally.
Such optimizers are available in open-source packages, {\em e.g.}, the \skopt package~\cite{skopt}.
To study this, we perform a modified tune of block 1 using \skopt, where the under-constrained parameter pTminChgQ is left fixed to its Monash value, and the discrete parameter MEcorrections is added to the tuning process.\footnote{Technically, MEcorrections here refers to two \pythia parameters: TimeShower:MEcorrections and SpaceShower:MEcorrections. In this tune, we turn both of them either on or off, so effectively there is one binary parameter.}
MEcorrections is a binary parameter in \pythia that turns on or off matrix-element corrections in the parton shower.
Within about 20 queries, \skopt chooses MEcorrections to be on (the same as its Monash setting), while also tuning alphaSvalue and pTmin to within about 10\% of their Monash values.
This is an encouraging result, and demonstrates the potential to automatically tune discrete parameters; however, we find that the precision achieved on the continuous parameters is worse when using a tree-based optimizer, which is not surprising given that the tree-based approach does not try and model the dependence of the $\chi^2$ on the parameters.
There is also no obvious way of assigning uncertainties to the continuous parameters using this approach that itself does not require substantial CPU resources.
Therefore, some combination of tree-based and Gaussian-process-based optimization may be desirable, if discrete parameters are also to be tuned.
The tree-based approach can be used first to fix any discrete parameters, and to determine a smaller region to explore for each continuous parameter. Next, the Gaussian process framework can be employed to precisely determine the continuous parameters, and assign uncertainties to them.

\section{CPU Usage}
\label{sec:cpu}

\begin{figure}[t]
  \centering
  \includegraphics[width=0.32\textwidth]{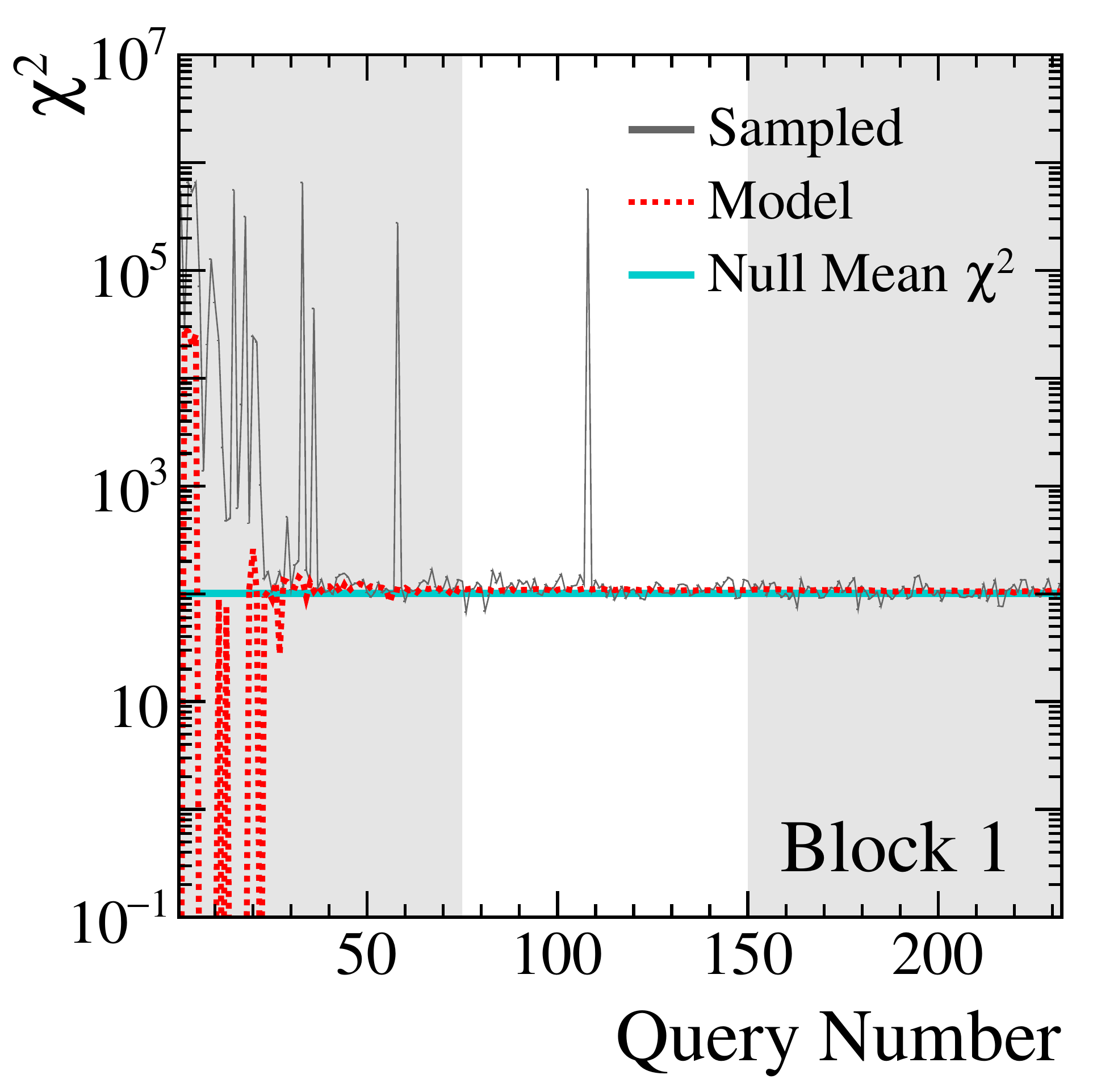}
  \includegraphics[width=0.32\textwidth]{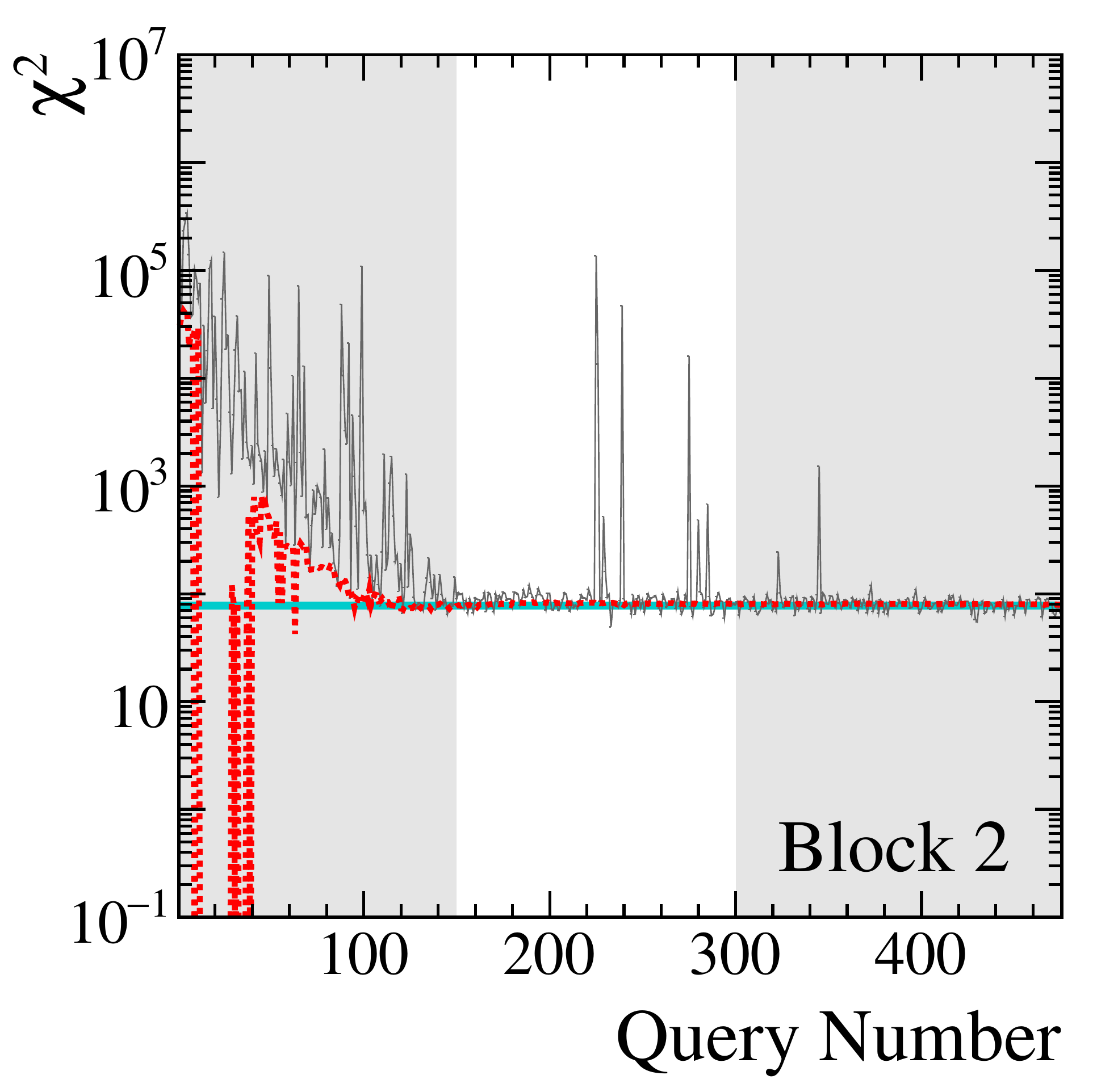}
  \includegraphics[width=0.32\textwidth]{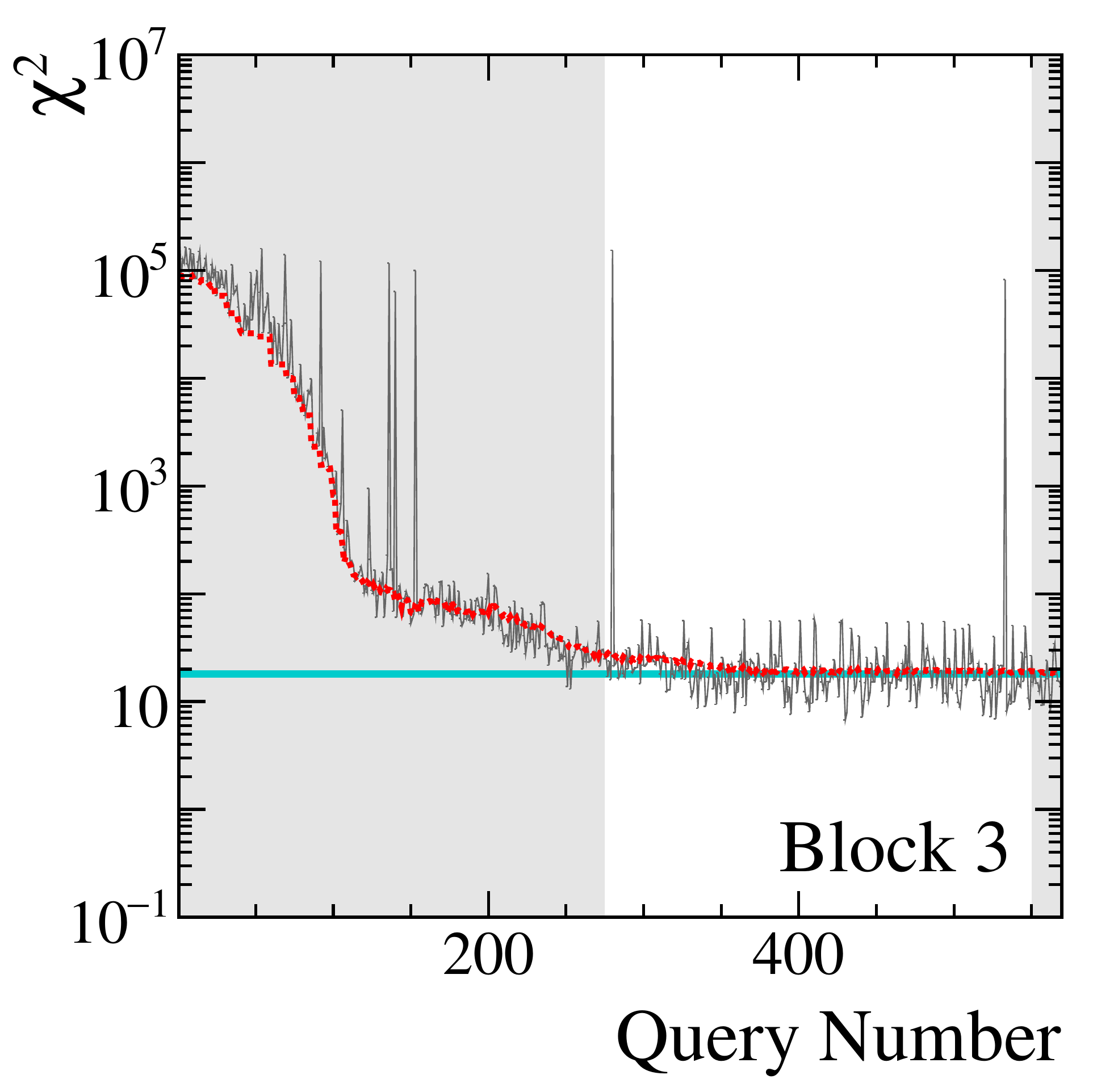}
  \caption{The $\chi^2$ value for each \spearmint query obtained using (black) the \pythia sample produced for the current query and (red) the \spearmint $\chi^2$ model.
The white regions show $25\cdot n({\rm par})$--$50\cdot n({\rm par})$ queries for each block.
}
  \label{fig:chi2vcalls}
\end{figure}

The CPU cost of performing these tunes depends on how many queries are made by \spearmint; therefore, determining when to terminate the optimization process governs how much total CPU is required.
Figure~\ref{fig:chi2vcalls} presents the evolution of the \spearmint model $\chi^2$ value versus query number for each tuning block.
In each case, the \spearmint model $\chi^2$ converges to a value close to the mean $\chi^2$ value under the null hypothesis (see Appendix~\ref{app:errors} for details on how the mean value is obtained); {\em i.e.}, the \spearmint model $\chi^2$ converges to the mean $\chi^2$ value expected using the true parameter values.
Therefore, by first computing the null mean $\chi^2$ using a Monte Carlo data sample constructed to have per-bin errors that match the experimental data distributions --- where the true parameters are known --- it is possible to obtain an estimate of what the \spearmint model $\chi^2$ value should converge to.

Figure~\ref{fig:chi2vcalls} shows that for each tuning block considered in our study, the \spearmint model $\chi^2$ value is unstable until about $25\cdot n({\rm par})$ queries are made, and that each block has fully stabilized by $50\cdot n({\rm par})$ queries ($n({\rm par})$ denotes the number of parameters being tuned).
Figures~\ref{fig:block1_parsvcalls}-\ref{fig:block3_parsvcalls} show how the optimal \spearmint model parameter values evolve with the number of queries.
The optimal parameter values also begin to stabilize around $25\cdot n({\rm par})$ queries, and are found to vary by negligible amounts relative to their uncertainties beyond $50\cdot n({\rm par})$ queries.
 All tuning results shown in the previous section are obtained using $50\cdot n({\rm par})$ queries.
Based on the results of the tunes performed in this study, some possible stopping criteria are:
\begin{itemize}
\item the number of queries reaching a maximum value, {\em e.g.}, $50\cdot n({\rm par})$ for the case where each parameter is allowed to vary freely within a large region (less queries are required if smaller regions are explored, see below);
\item comparing the \spearmint model $\chi^2$ value after each query to the expected null mean $\chi^2$ value, and terminating the tuning process when these values converge, {\em e.g.}, if they differ by less than a few $\chi^2$ units;
\item terminating once
the stability of the \spearmint model $\chi^2$ over the previous $\approx 5\cdot n({\rm par})$ queries is better than a few $\chi^2$ units.
\end{itemize}
Applying any of these criteria to our tunes results in stopping after roughly the same number of queries, with negligible differences in the tuned parameter values.
{\em N.b.}, it is possible to restart a tune in \spearmint after it has been terminated, even if the termination was executed via a SIGINT call, {\em e.g.}, CTRL-c.

The wall time required to perform these tunes on a quad-core i7 2.8~GHz 2015 Macbook Pro laptop are about 6, 14, and 25 hours for blocks 1, 2, and 3, respectively.\footnote{This laptop has 8 virtual cores. We run \spearmint on one core, and \pythia event generation is performed in parallel on the remaining 7 cores.}
The CPU per parameter increases roughly linearly going from 3 to 11 parameters.
In total, 45 hours of wall time is required to perform the full 20 parameter block-diagonal $e^+e^-$ tune; therefore, a full  $e^+e^-$ tune of 20 \pythia parameters can be performed on a laptop in less than 2 days using \spearmint.
The event-generation processes dominate the total CPU required to tune each block.
Since event generation is trivial to do in parallel, the tunes of each block could be performed much faster using more computing power.

Bayesian optimization implementations like \spearmint are not designed for the case where $n({\rm par}) \gg 10$ parameters; however, as shown above, a global tune of all 20  $e^+e^-$  parameters does converge to a consistent set of optimal parameter values, and it does so using the same total number of queries, {\em i.e.}\
our global tune was also terminated after $50\cdot n({\rm par})$ queries.
Despite this, the global tune takes about 3 times more wall time to run because \spearmint uses more CPU to determine the next set of parameters to query when there are 20 parameters rather than when there are $\lesssim 11$.
This results in a sizable increase in the wall time required to perform the global tune because \spearmint runs on a single core.
For $n({\rm par}) \approx 20$, this likely could be sufficiently mitigated by parallelizing \spearmint to permit running on a few cores.
To use Bayesian optimization to perform a tune with a much larger number of parameters, changes to the algorithm are likely required (see, {\em e.g.}, Ref.~\cite{ML2015}).
%Cite 1503.01673 here.

Finally, we note that a much smaller number of queries is required to reach convergence if the parameters are restricted to localized regions around their true values.
For example, restricting each well-constrained parameter to vary within a $\pm10\sigma$ region around its Monash value, where $\sigma$ denotes the quoted uncertainty on each parameter in our block-diagonal tunes, results in all three tuning blocks converging in $\lesssim 10\cdot n({\rm par})$ queries.
This is similar to the number of queries required by parametric tuning methods~\cite{Buckley:2009bj}, where such restricted regions must be used to ensure that the generator response is well approximated by a low-order parametric function.
Which approach should be employed in a real-world tune depends on what knowledge exists about the parameters.
If it is known that the optimal parameter values must be within a small region, and that within this region the generator response is well approximated by a low-order parametric function, then the parametric approach is likely the best option.
Conversely, if little is known about the optimal parameter values, then Bayesian optimization would be preferable.
One could also consider employing \spearmint to determine the parameter values used by a subsequent parametric tune.

\section{Towards a Real-World Tune}
\label{sec:real}

There are a number of issues that arise in a real-world tune (see, {\em e.g.}, Refs.~\cite{ATLAS:2012uec,Khachatryan:2015pea,Belyaev:2010yga}) that are absent from the closure test presented above; however, our view is that each of these factorizes from the process of efficiently exploring the parameter space.
\begin{itemize}
  \item The goal of the tuning closure test presented above was to demonstrate the power of the Bayesian optimization method; therefore,  we chose to use minimal expert knowledge by placing a uniform prior over each parameter within a large specified range.
  In a real-word tune, expert knowledge could be used here by assigning non-uniform priors to the parameters that capture the physics belief about their behavior, which could include expected correlations between parameters.
  \item In a real-world tune, it may be desirable to weight the various bin contributions in the $\chi^2$ definition, rather than treating them all equally as we did. This is trivial to implement as it only requires changing the $\chi^2$ value reported to the optimizer.
  \item Our tuning closure test only involved $e^+e^-$ collisions at a single energy, whereas many real-world tuning applications involve several beam types at multiple energies. From the perspective of the optimizer, this situation is no different than the simple case of one beam type and energy. The optimizer provides the set of parameter values for the next query, then waits to receive the $\chi^2$ value; it does not need to know how the $\chi^2$ is obtained. We presented the CPU requirements in the previous section in terms of the number of queries because this is more universally applicable than CPU time; {\em i.e.}\ the CPU per query depends greatly on the beam types and energies to be generated, but in all cases the approach that minimizes the number of queries should also require the least CPU resources.
  \item In a real-world tune, even once the optimal parameter values are found, one expects discrepancies between the Monte Carlo and data will remain. Discovering such situations as quickly as possible should be viewed as one of the goals of the parameter-optimization process.
\end{itemize}
As an example, rather than treating the Monte Carlo generated using \pythia and the Monash parameters as experimental data, we performed a tune of block 1 using actual experimental data~\cite{Achard:2004sv}. The tune converges in about the same number of queries as in the closure test, and the $\chi^2$ value obtained is almost two times better than $\chi^2$ obtained using the Monash parameter values.
While this result demonstrates successful application of the optimization process to experimental data, some care is needed to interpret this result.
Since Monash is meant to be a global tune of \pythia, the selection of the block 1 parameters includes expert-level physics knowledge beyond the input distributions used to construct the $\chi^2$.
Such knowledge could be included in the Bayesian optimization process by placing non-uniform priors over the parameters.
Alternatively, since a model of parameter dependence of the $\chi^2$ is built during the tuning process, the expert could choose to perform various block-specific tunes first using uniform priors. Next, the expert could study how each $\chi^2$ value depends on the parameters and choose how to combine these results using their knowledge to obtain the optimal result --- or to decide where to improve the Monte Carlo generator.
Regardless, by combining the Bayesian optimization approach with expert-level knowledge, it should be possible to produce better tunes in the future by making it much faster and easier to both optimize the generator parameters and to study discrepancies between Monte Carlo and experimental data.

\section{Summary}
\label{sec:sum}

Monte Carlo event generators contain a large number of parameters that must be determined by comparing the output of the generator with experimental data.
Generating enough events with a fixed set of parameter values to enable making such a comparison is extremely CPU intensive.
All available tunes provided with the \pythia~8 package were obtained either manually or parametrically. %, and each of these approaches has both merit and limitations.
In this article, we proposed to instead treat Monte Carlo event generator tuning as a black-box optimization problem and  addressed it using the framework of Bayesian optimization.
We showed that Monte Carlo generator parameters can be accurately obtained using Bayesian optimization and minimal expert-level physics knowledge.
Using this approach, a tune of \pythia~8 using $e^+e^-$ data, where 20 parameters were optimized, was run on a laptop in just two days.
Finally, we believe that combining the Bayesian optimization approach with expert-level knowledge should enable producing better tunes in the future, by making it faster and easier to study discrepancies between Monte Carlo and experimental data. The code used in this study is available at Ref.~\cite{TuneMC}.

\acknowledgments

This work was supported by DOE grants DE-SC0010497 and DE-FG02-94ER40818.
We thank A.~Buckley, T.~Head, G.~Louppe, and T.~Sj\"{o}strand for useful discussions and feedback.

\appendix

\section{Parameter Uncertainties}
\label{app:errors}

\begin{figure}[t]
  \centering
  \includegraphics[width=0.32\textwidth]{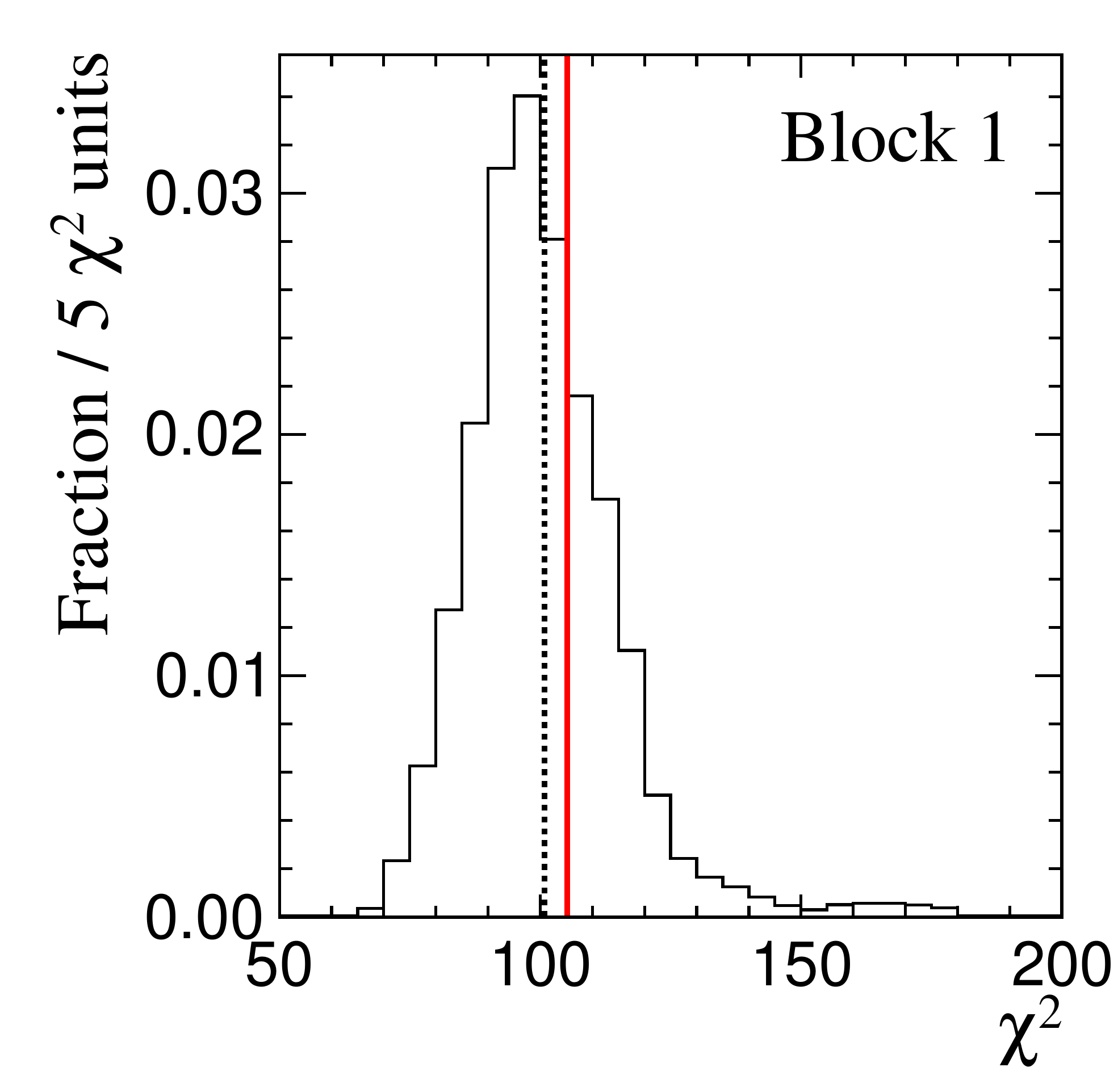}
  \includegraphics[width=0.32\textwidth]{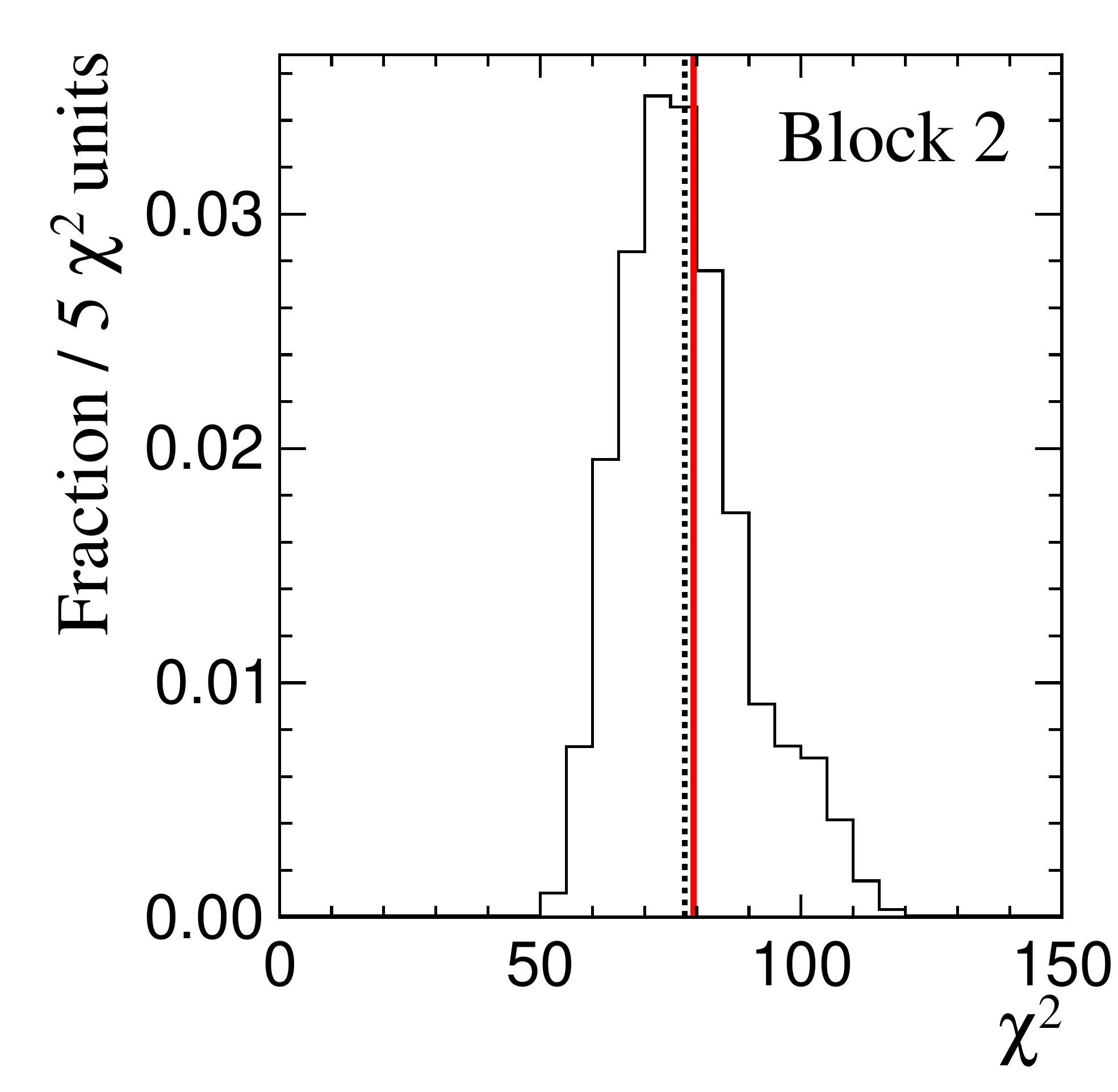}
  \includegraphics[width=0.32\textwidth]{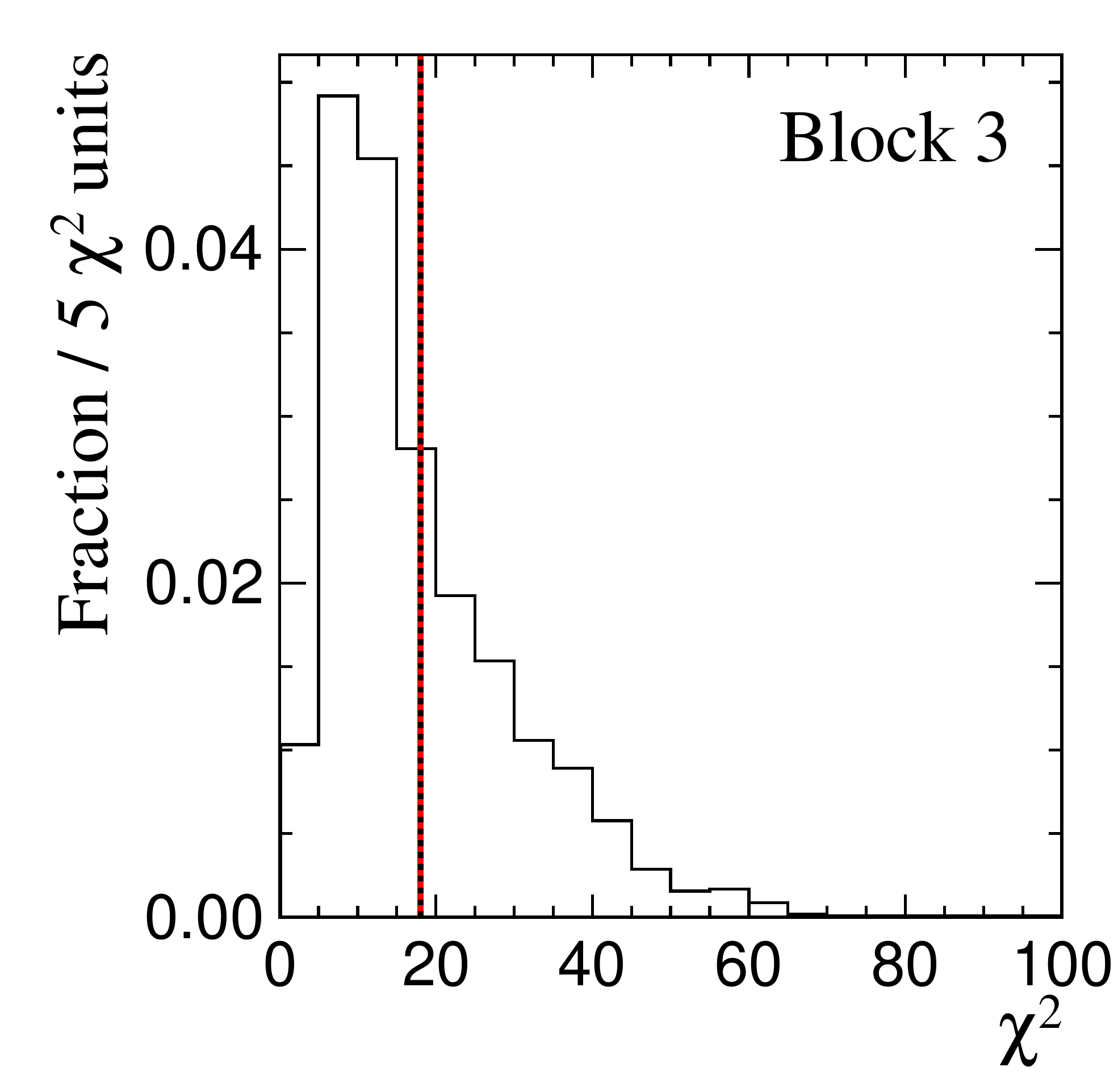}
  \caption{$\chi^2$ distributions for each tuning block obtained using the true (Monash) parameter values.
The vertical lines show (dashed black) the mean $\chi^2$ value and (solid red) the final \spearmint model of the $\chi^2$ value.
}
  \label{fig:chi2_null}
\end{figure}

The \spearmint model of how the objective function depends on the parameters is conceptually different than a $\chi^2$ statistic obtained from a single two-sample test; therefore, we cannot simply use $\Delta\chi^2=1$, or similar criteria, to estimate the uncertainty on the parameters.
Figure~\ref{fig:chi2_null} shows the distribution of $\chi^2$ values obtained from an ensemble of two-sample tests in each parameter block under the null hypothesis.
These distributions are obtained by performing a large number of two-sample comparisons using the $\chi^2$ in Eq.~\ref{eq:chi2}, where all samples are generated using the Monash parameters, one sample contains 1M events, and the other 10M events.
Figure~\ref{fig:chi2_null} shows that the \spearmint model accurately predicts the mean value of each $\chi^2$ distribution.

As an {\em ad hoc} method for assigning uncertainties to tuning parameters, we scan the \spearmint model $\chi^2$ value while varying each parameter independently and holding all other parameters fixed to their tuned values (often referred to as the plugin method).
The $1\sigma$ confidence interval for each parameter is taken to include all parameter values for which  the \spearmint model $\chi^2$ is less than $\chi^2(p=0.32)$, where $\chi^2(p=0.32)$ is the $\chi^2$ value corresponding to a $p$-value of 0.32 for the case where the number of degrees of freedom equals the mean $\chi^2$ value predicted by the \spearmint model.
While this approach is certainly {\em ad hoc}, it produces confidence intervals with reasonable coverage properties in each of the tunes presented above, and at minimal CPU cost.

\section{Parameter Results}

\begin{figure}[p]
  \centering
  \includegraphics[width=0.32\textwidth]{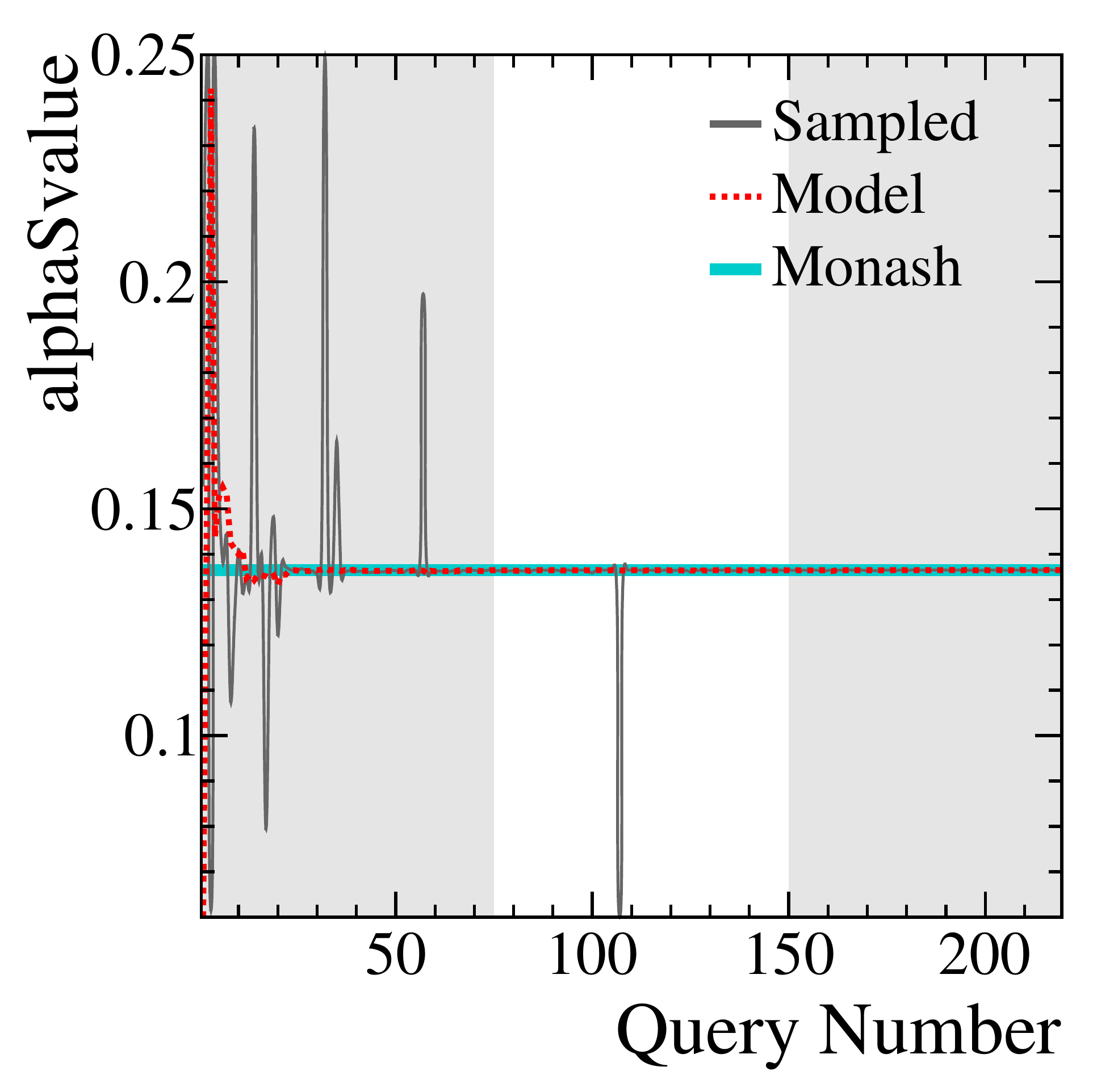}
  \includegraphics[width=0.32\textwidth]{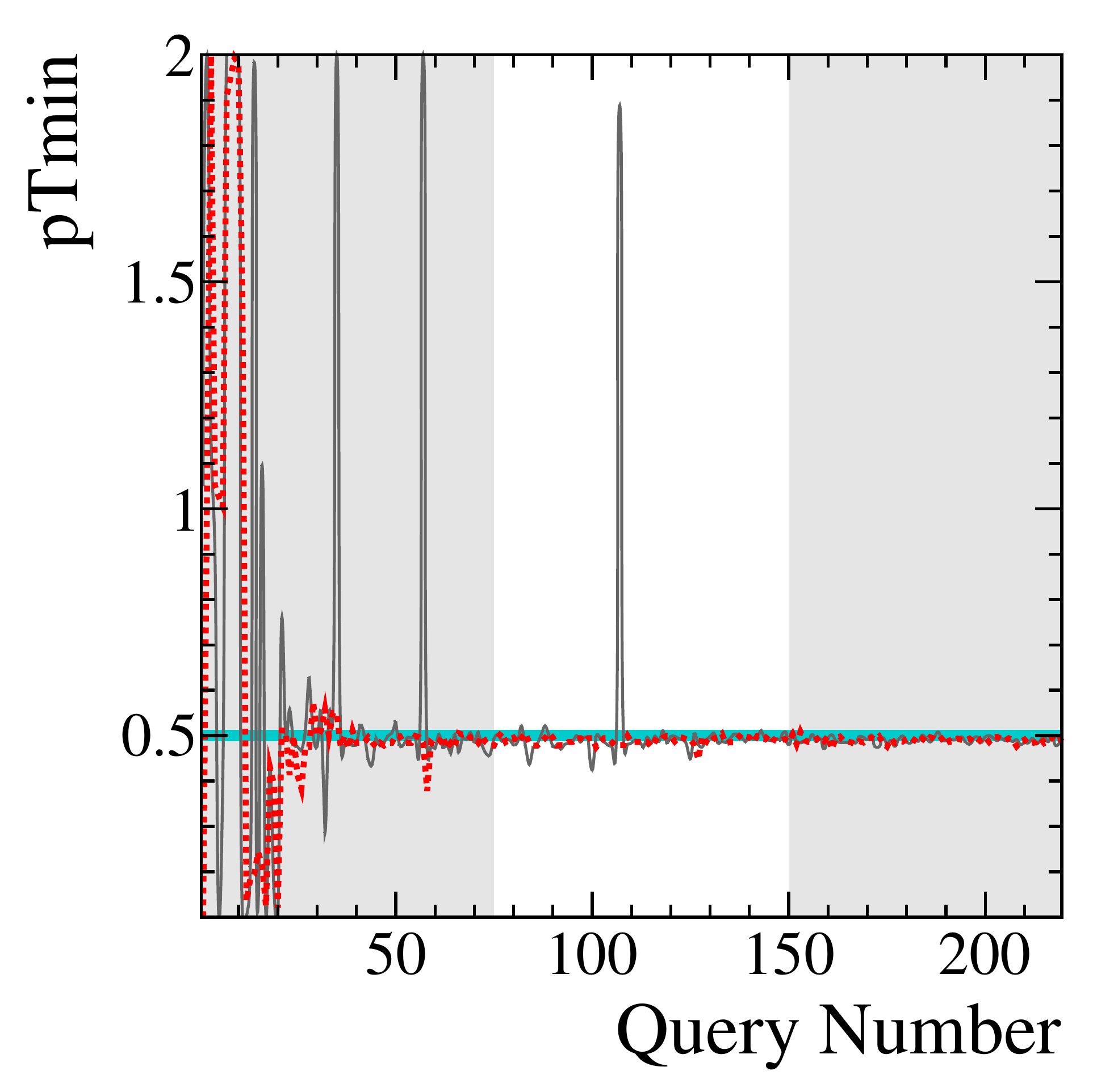}
  \includegraphics[width=0.32\textwidth]{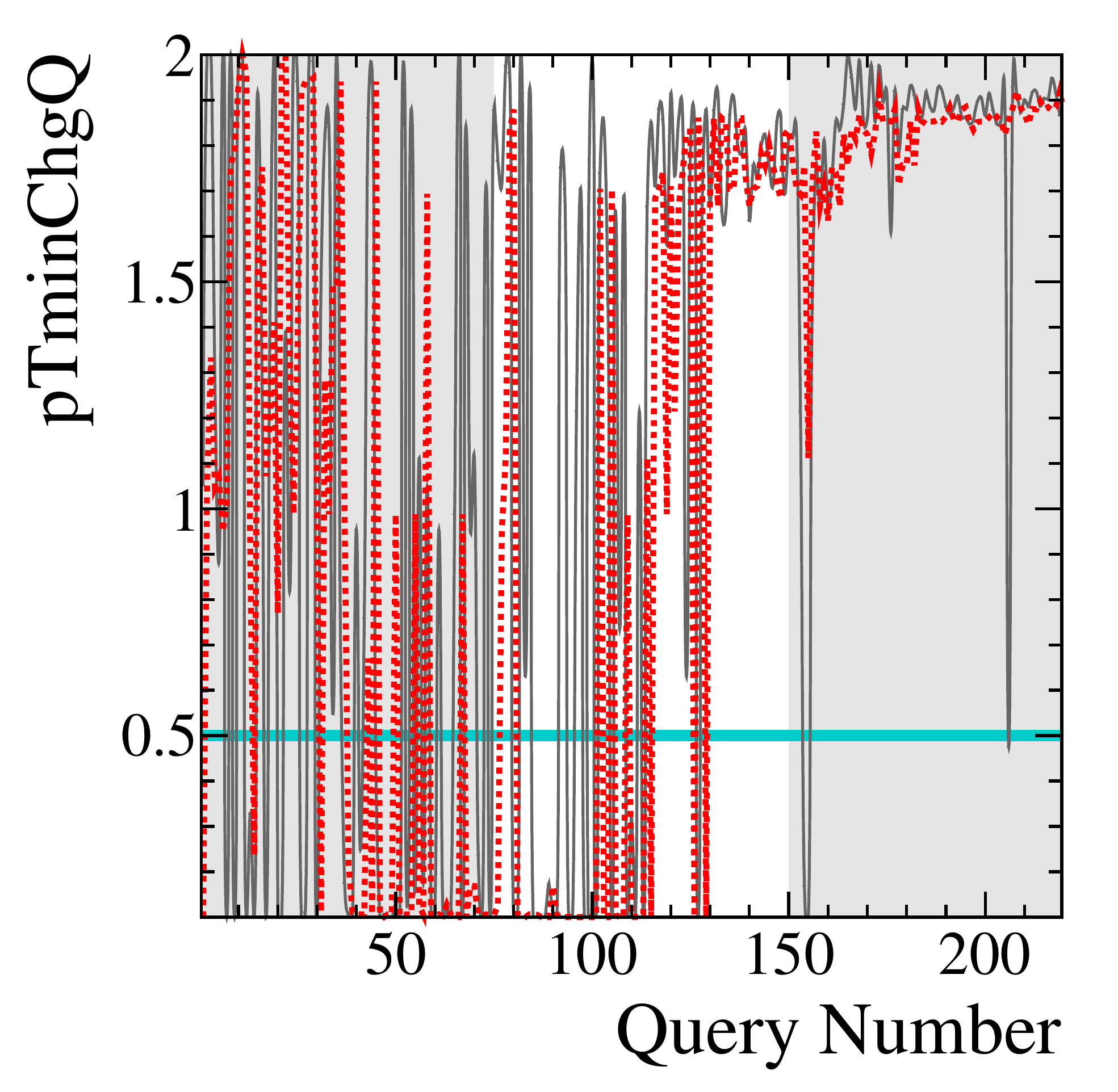}
  \caption{
  Evolution of the parameter values in block 1 during the tuning process.
  The white regions are $25\cdot n({\rm par})$--$50\cdot n({\rm par})$ queries.
}
  \label{fig:block1_parsvcalls}
\end{figure}

\begin{figure}[p]
  \centering
  \includegraphics[width=0.32\textwidth]{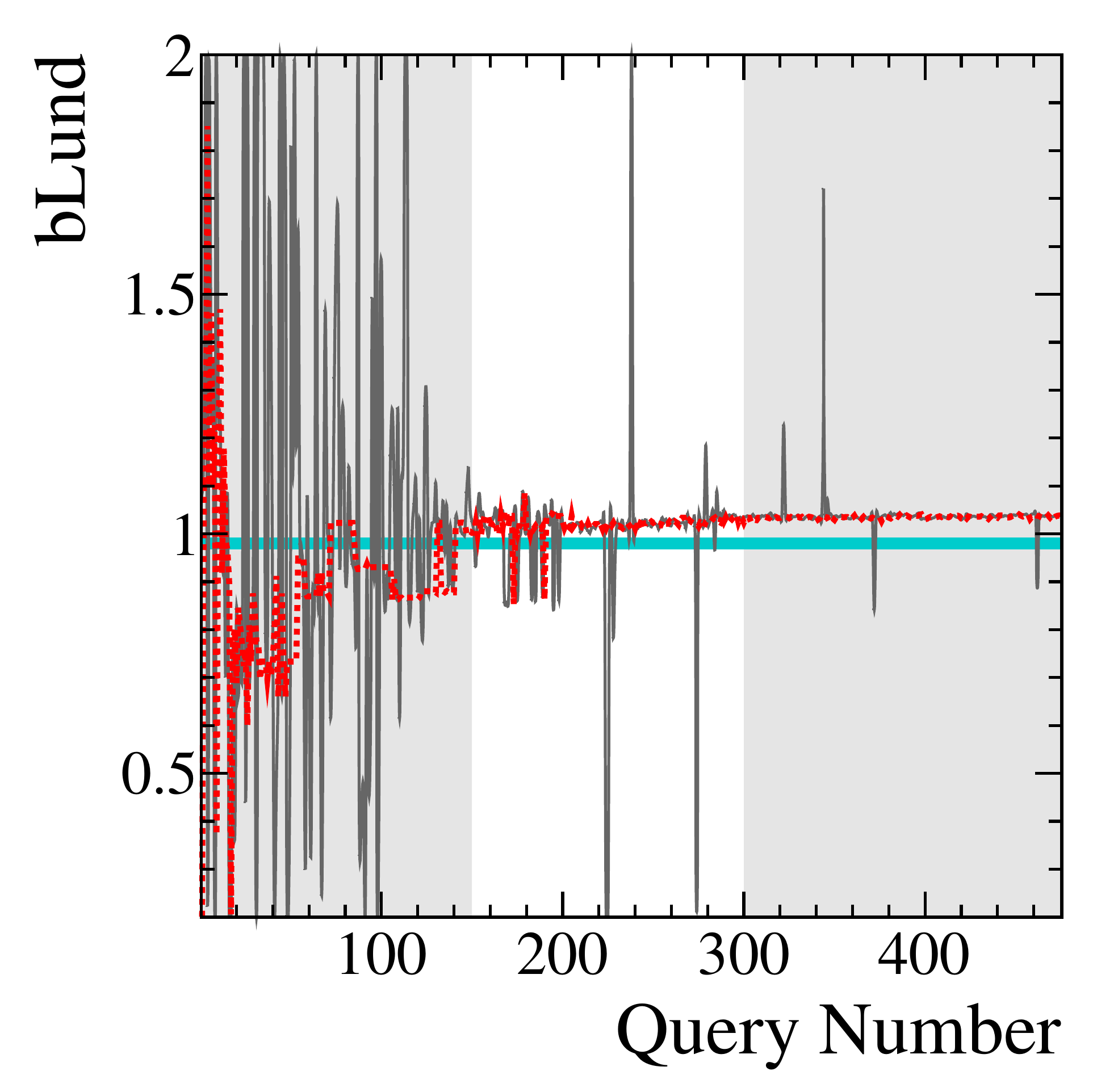}
  \includegraphics[width=0.32\textwidth]{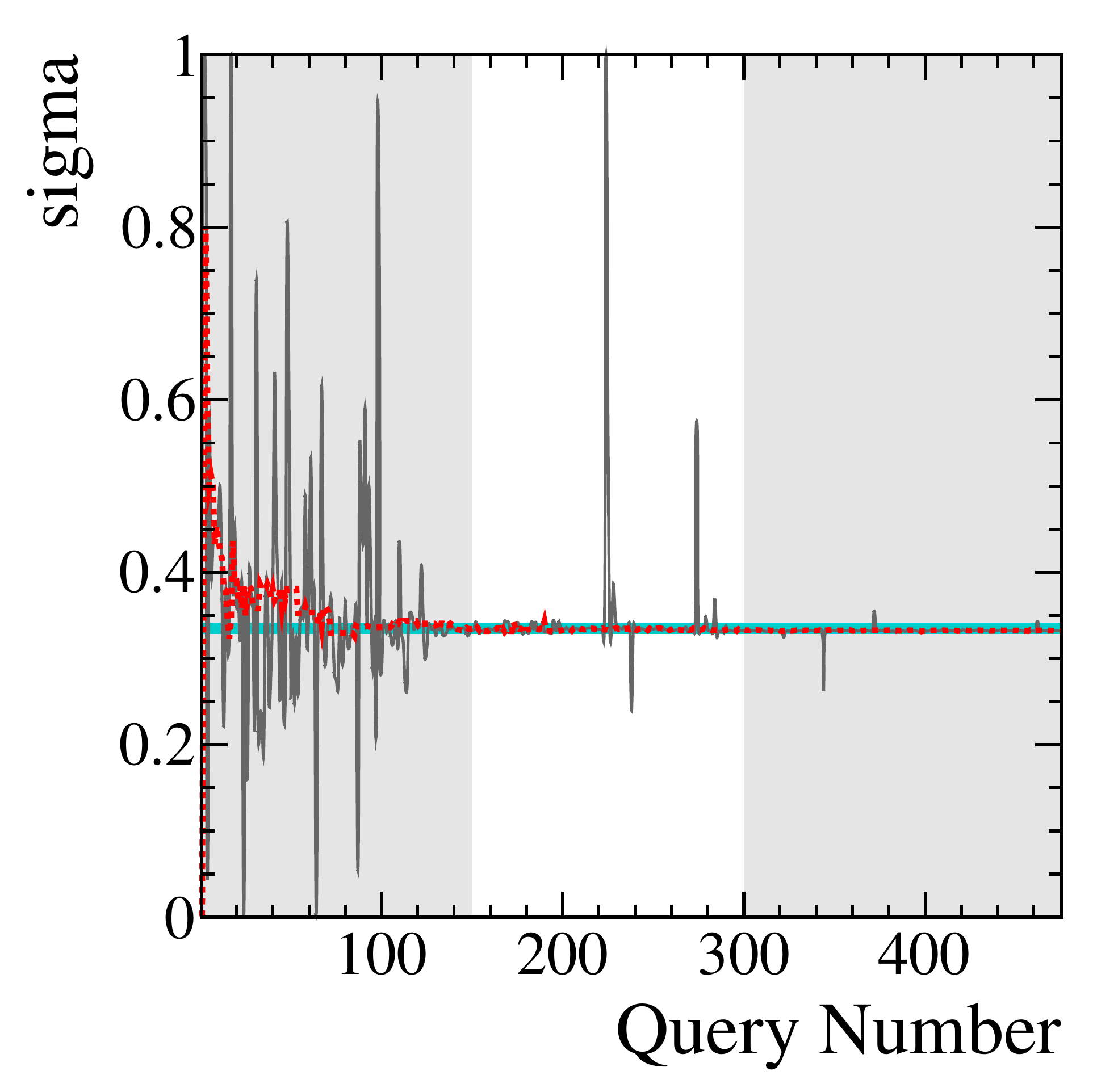}
  \includegraphics[width=0.32\textwidth]{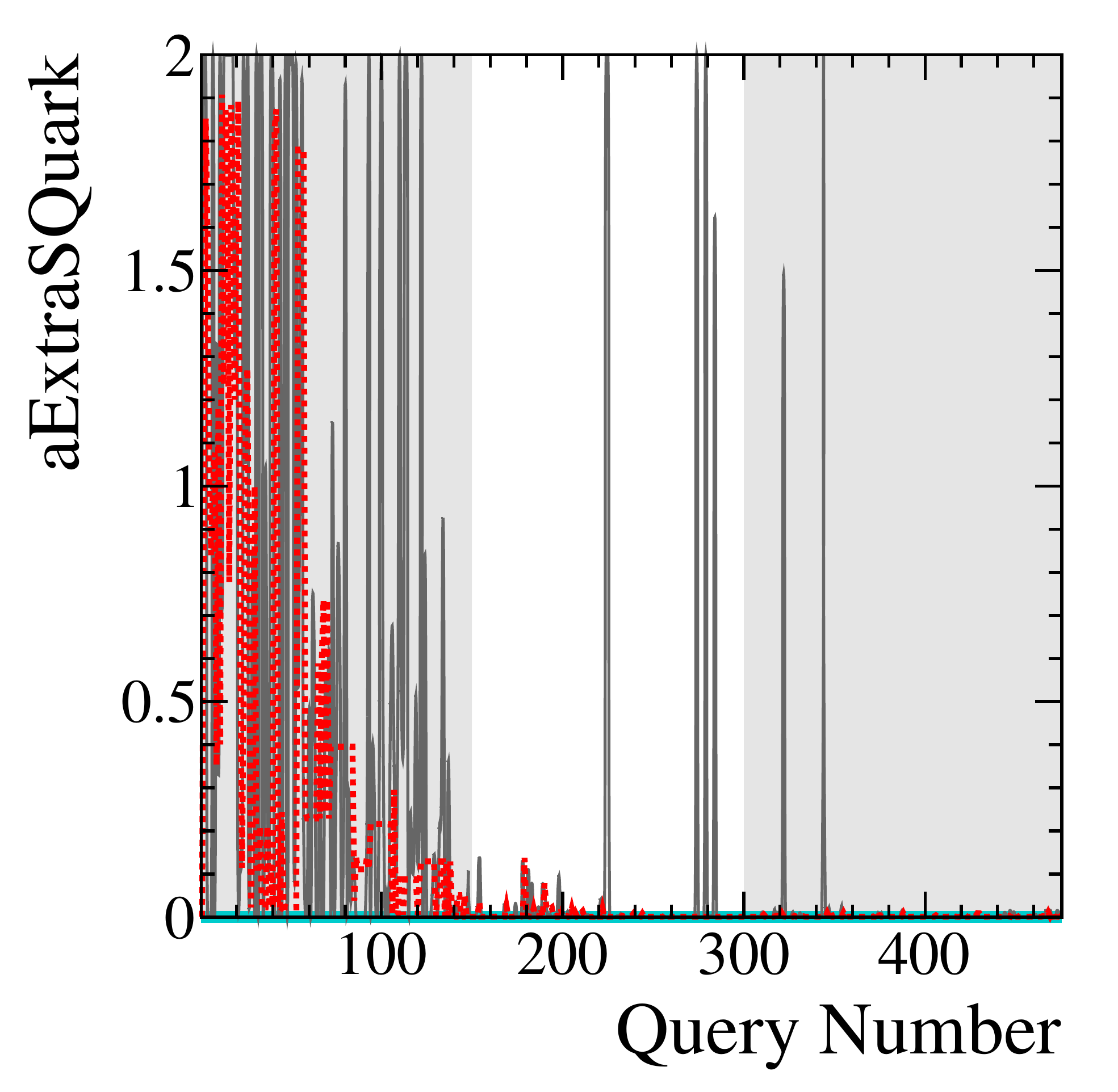}
  \includegraphics[width=0.32\textwidth]{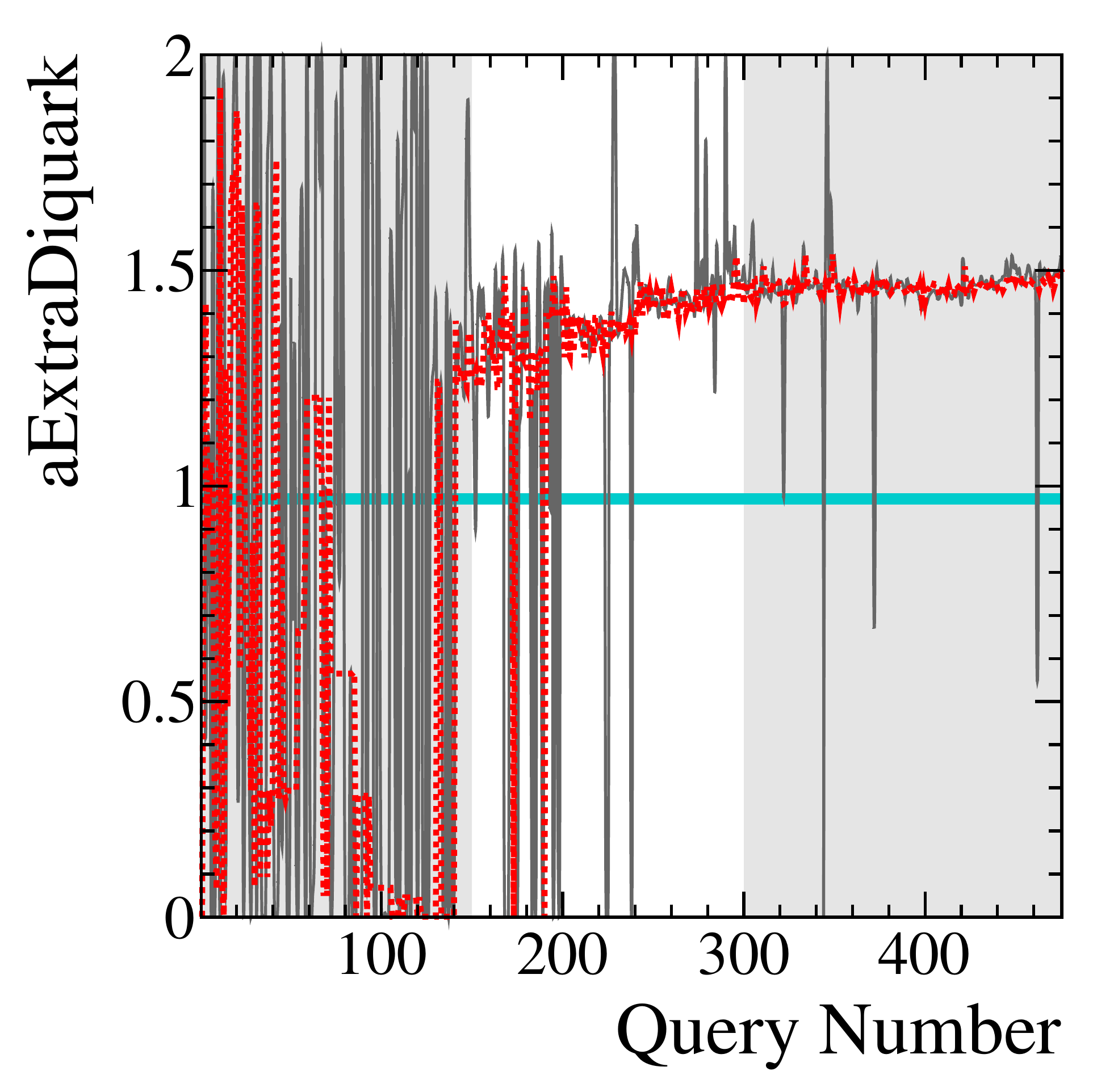}
  \includegraphics[width=0.32\textwidth]{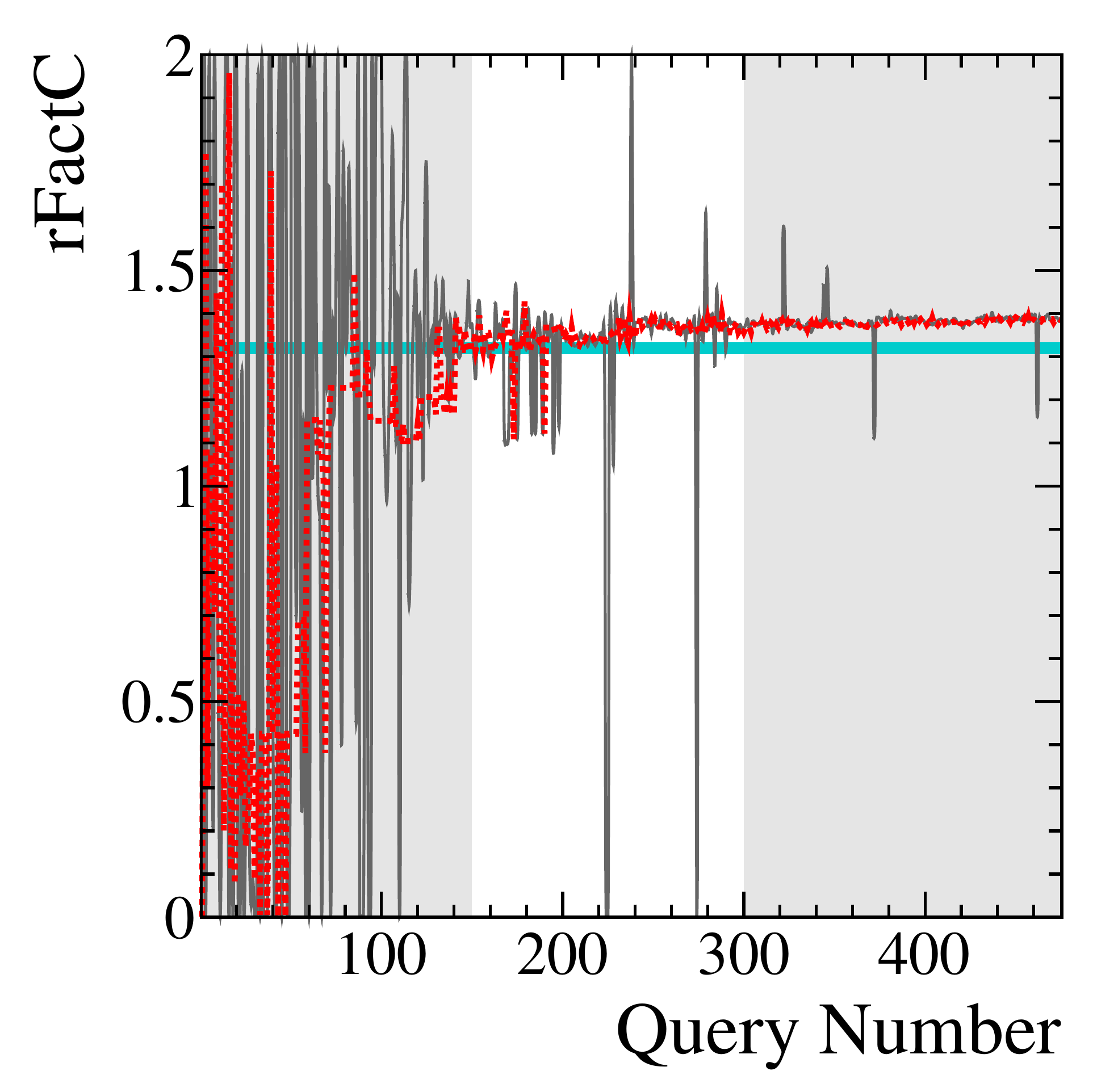}
  \includegraphics[width=0.32\textwidth]{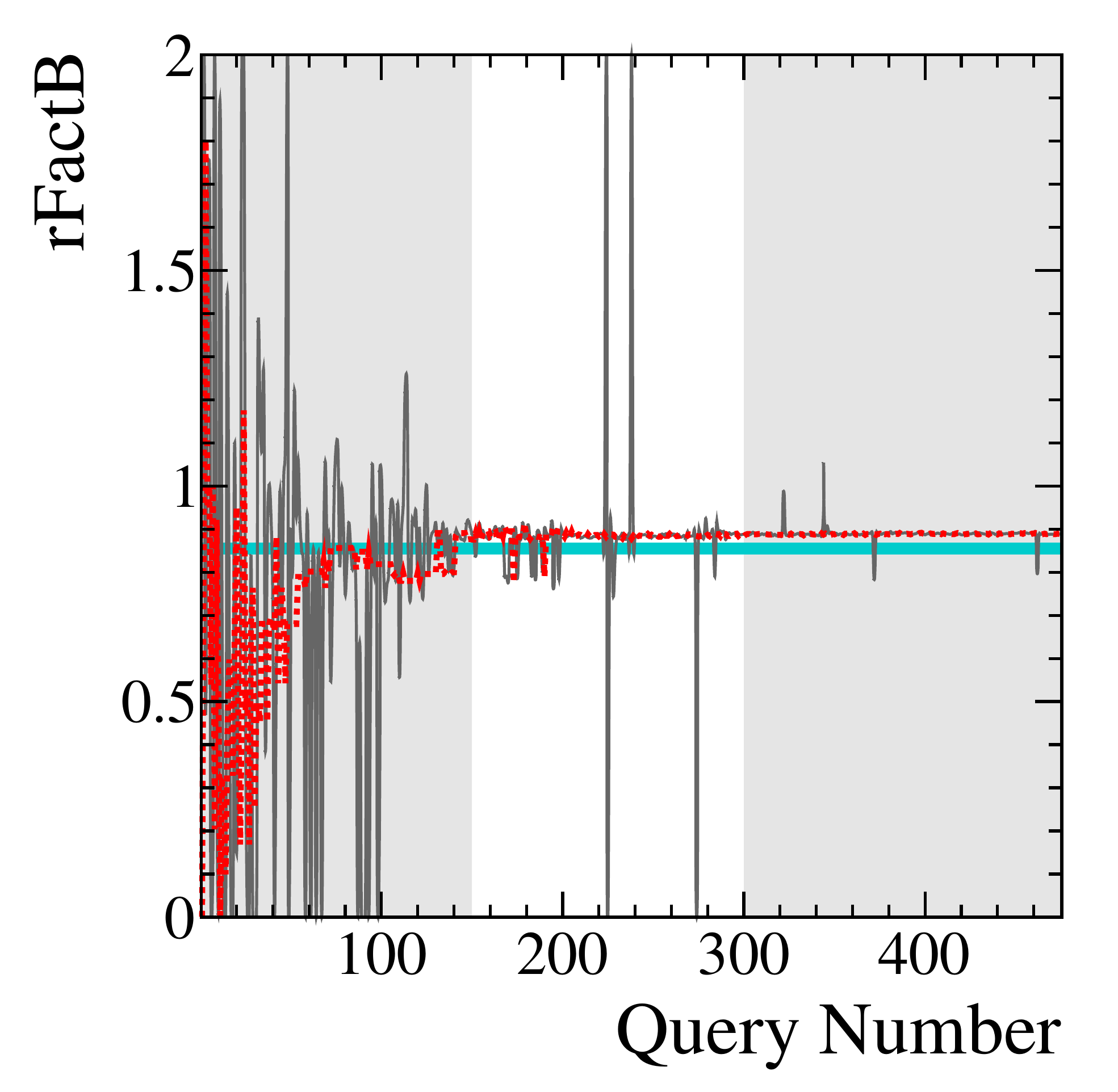}
  \caption{
  Evolution of the parameter values in block 2 during the tuning process.
The white regions are $25\cdot n({\rm par})$--$50\cdot n({\rm par})$ queries.
}
  \label{fig:block2_parsvcalls}
\end{figure}

\begin{figure}[p]
  \centering
  \includegraphics[width=0.32\textwidth]{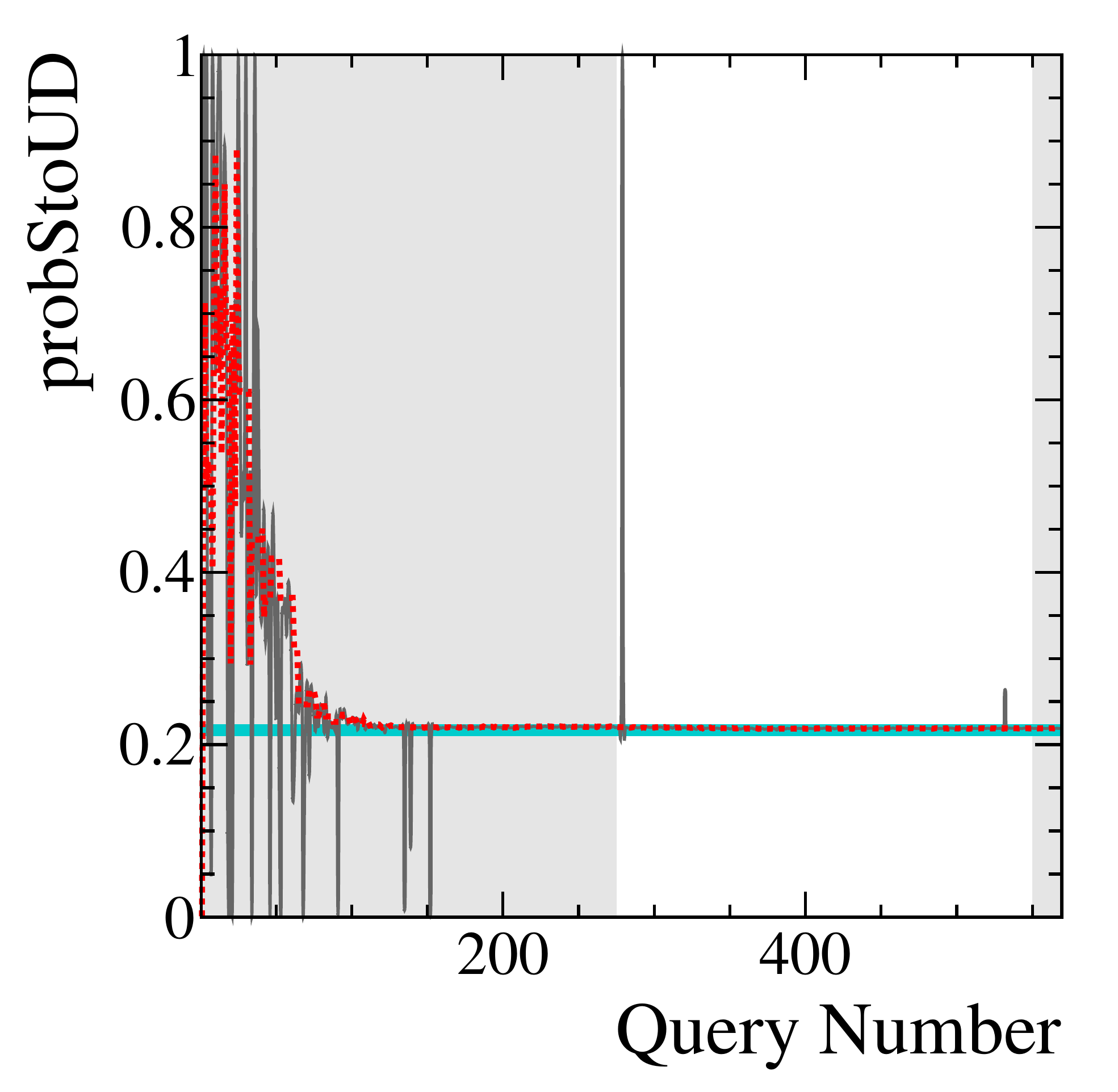}
  \includegraphics[width=0.32\textwidth]{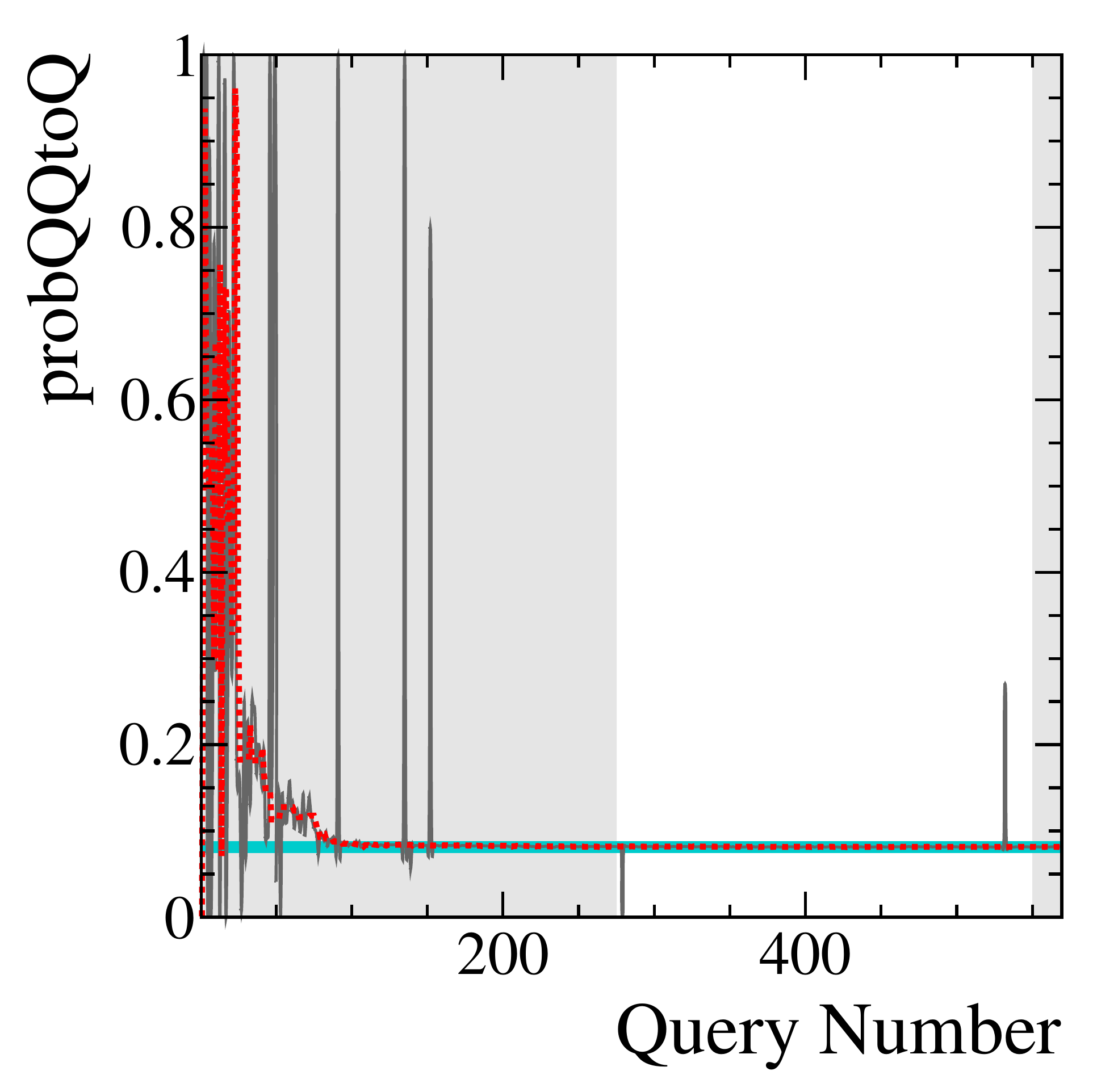}
  \includegraphics[width=0.32\textwidth]{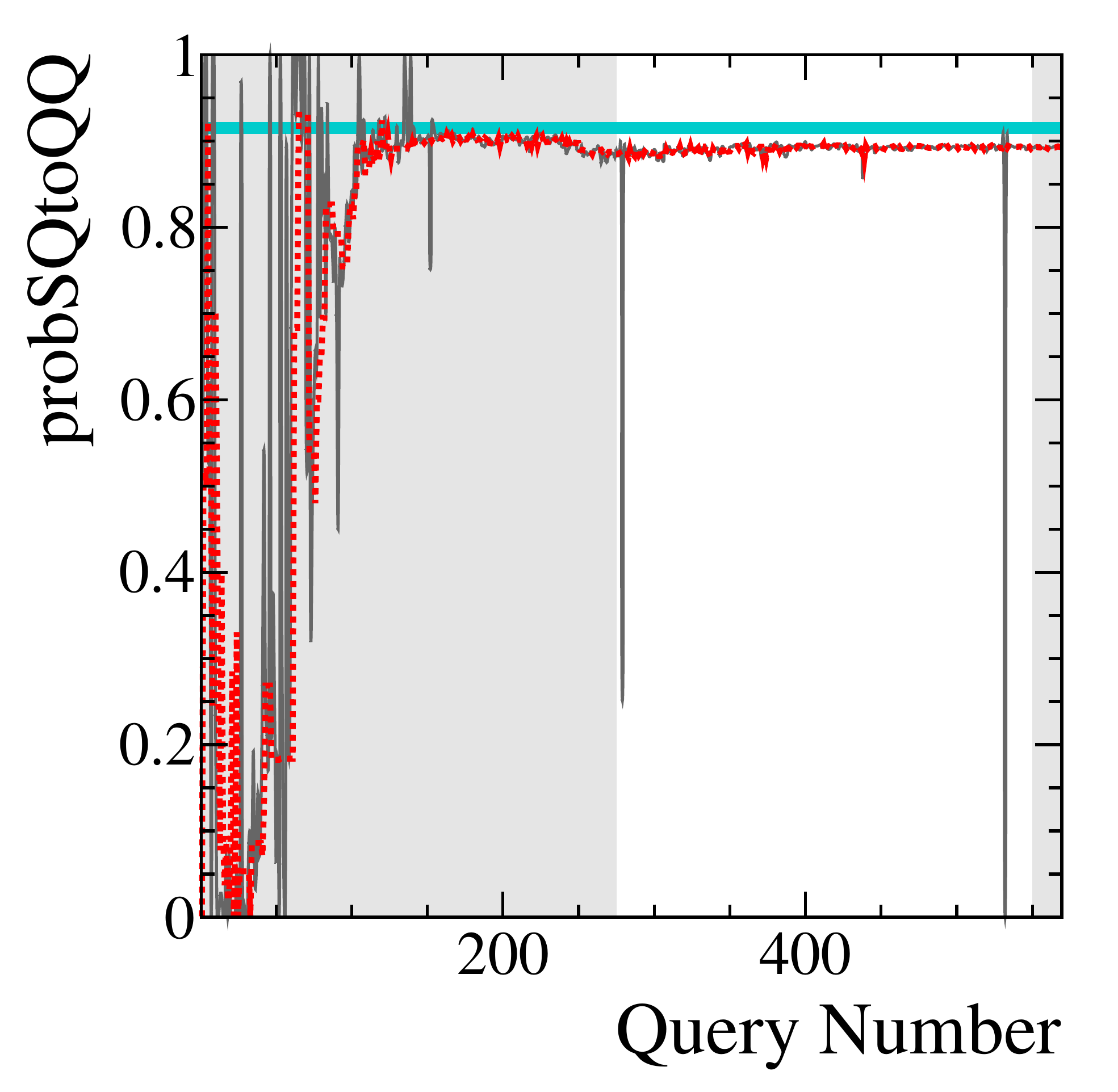}
  \includegraphics[width=0.32\textwidth]{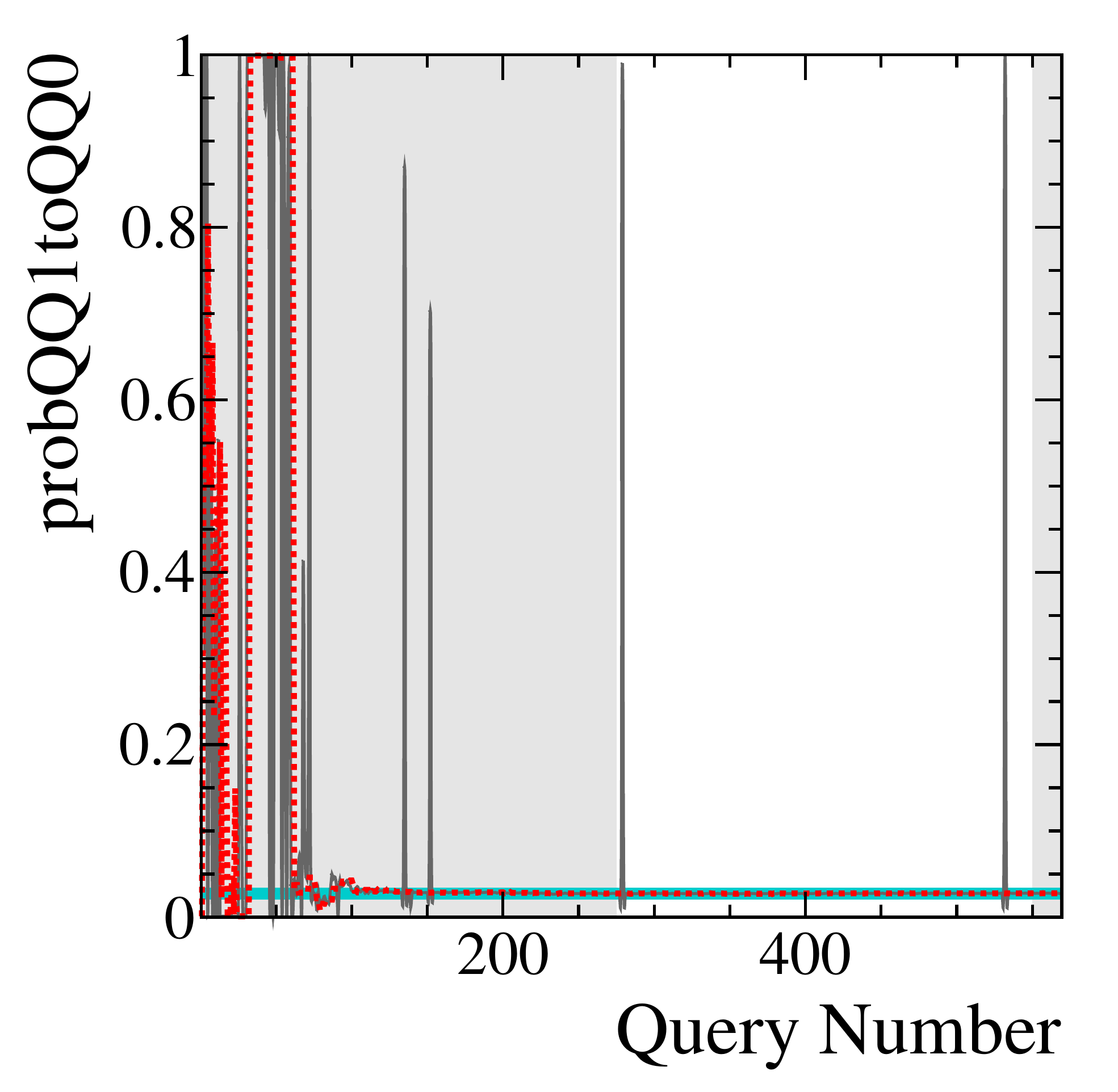}
  \includegraphics[width=0.32\textwidth]{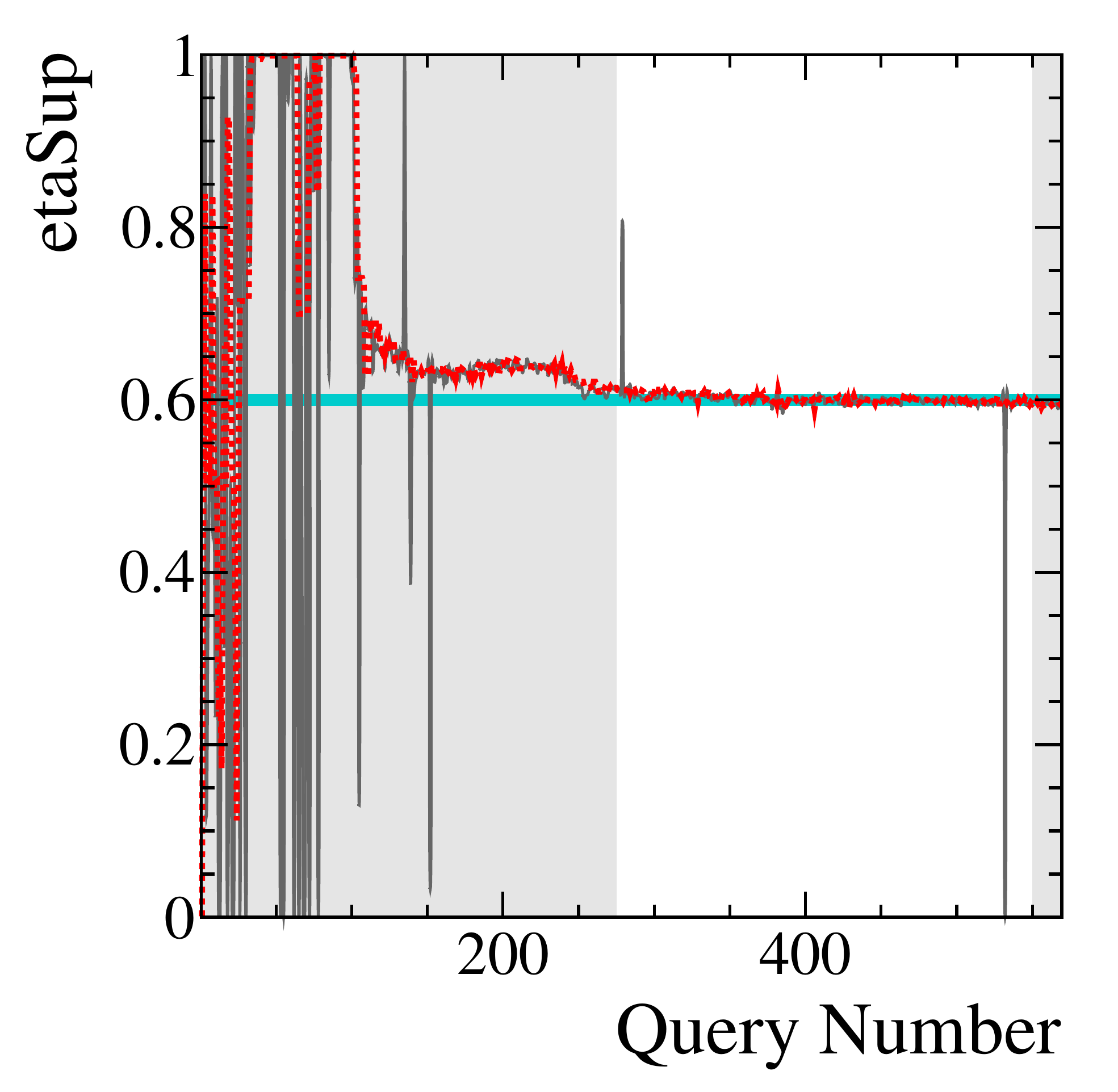}
  \includegraphics[width=0.32\textwidth]{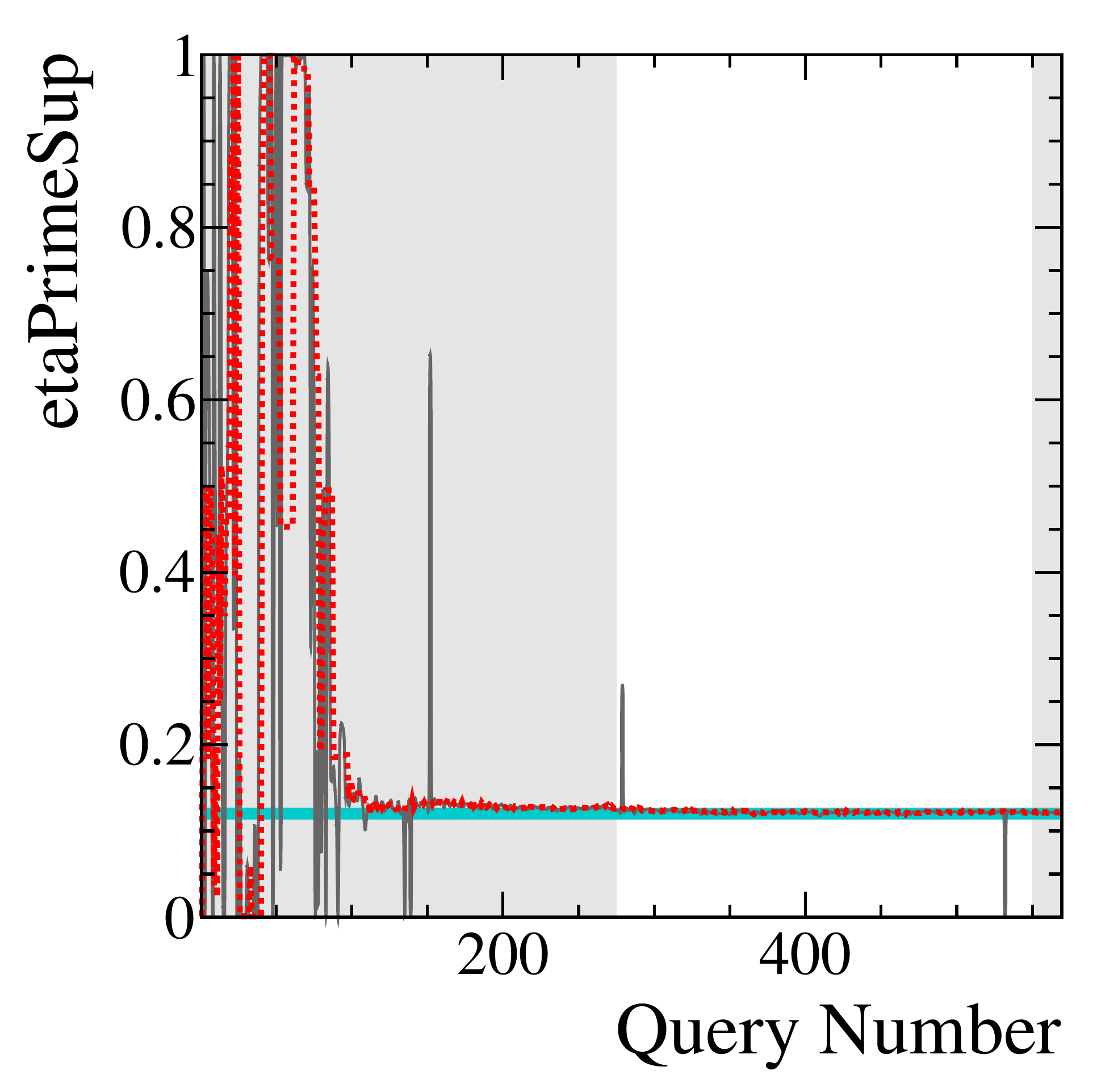}
  \includegraphics[width=0.32\textwidth]{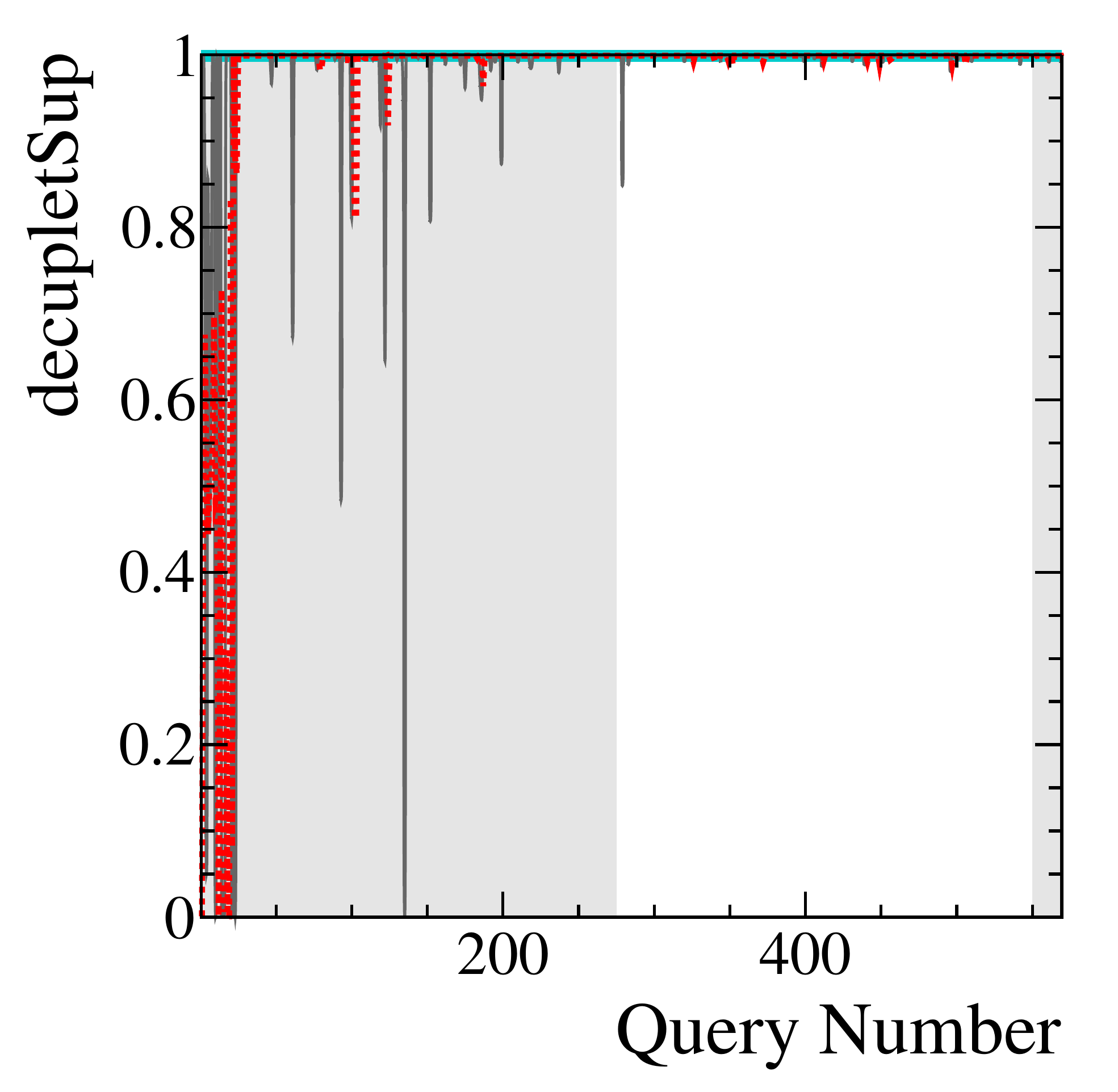}
  \includegraphics[width=0.32\textwidth]{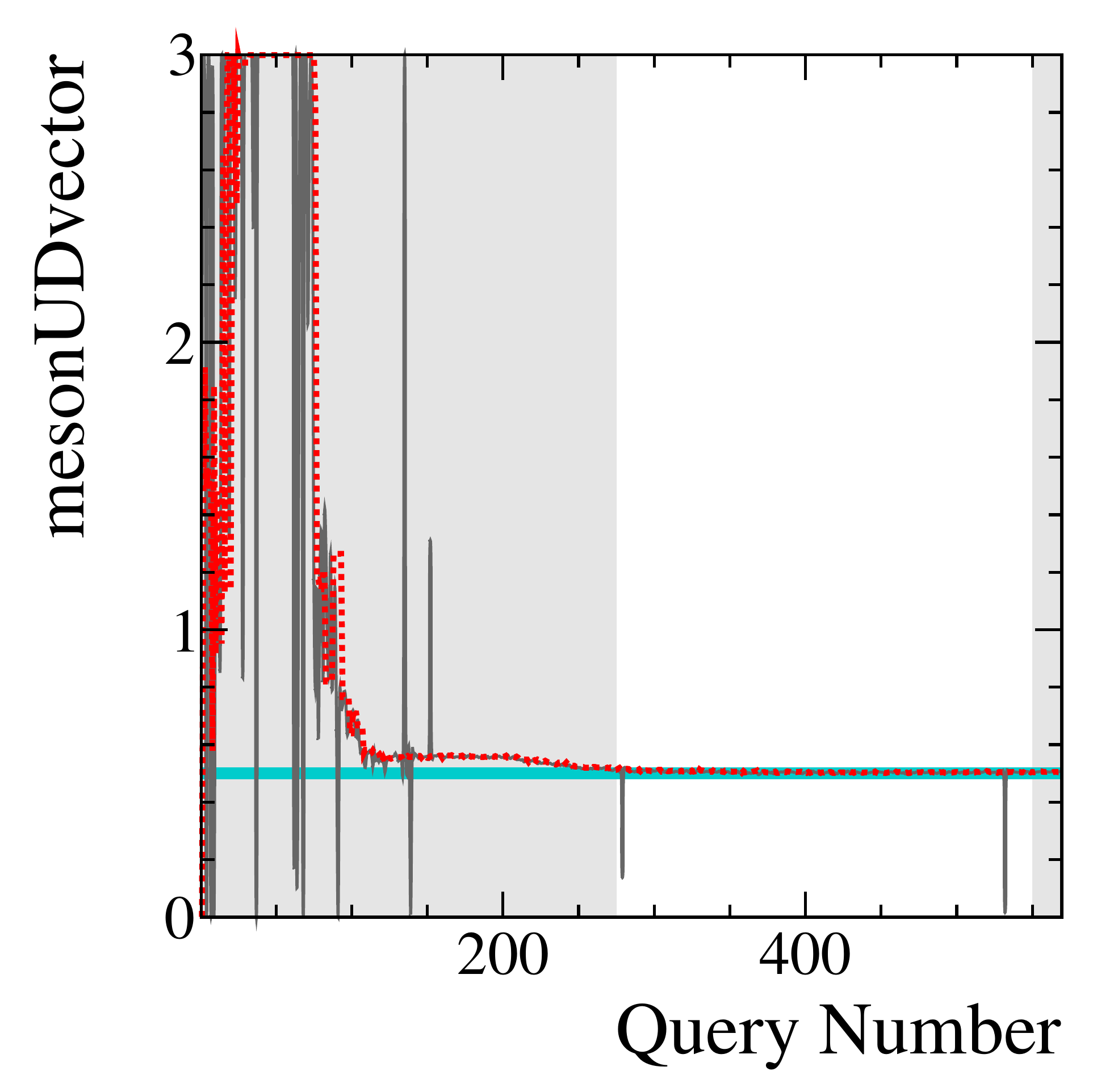}
  \includegraphics[width=0.32\textwidth]{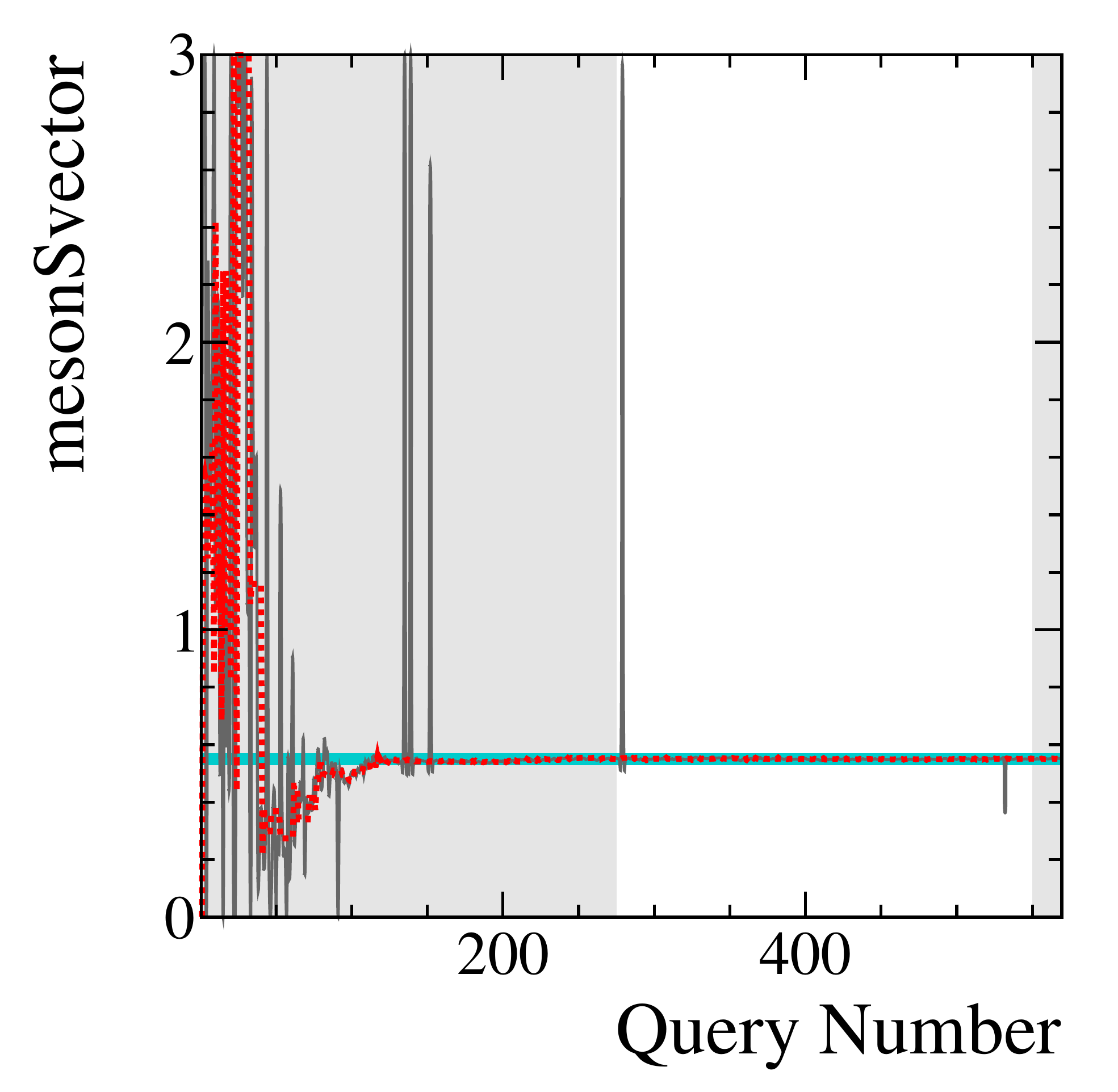}
  \includegraphics[width=0.32\textwidth]{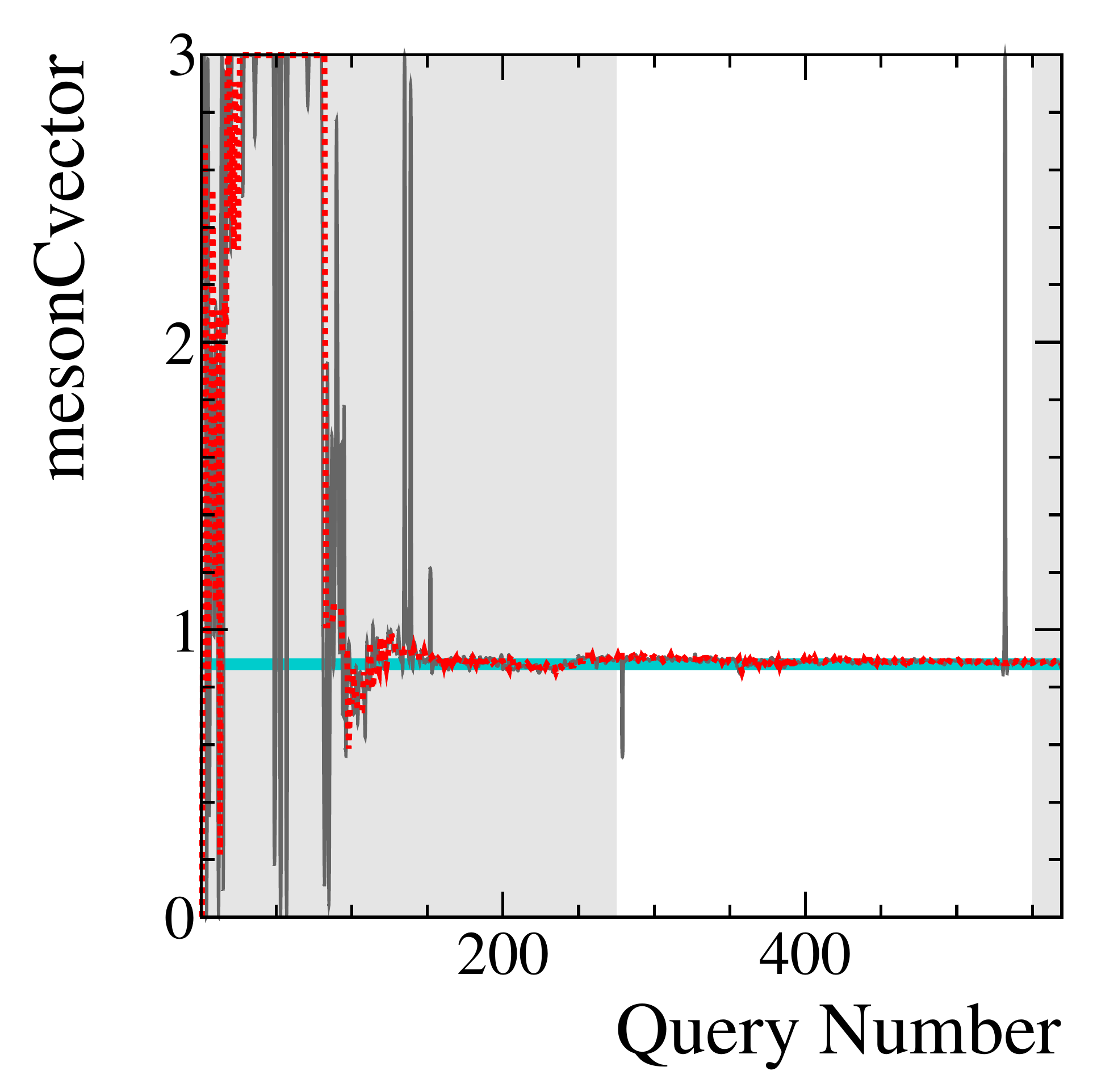}
  \includegraphics[width=0.32\textwidth]{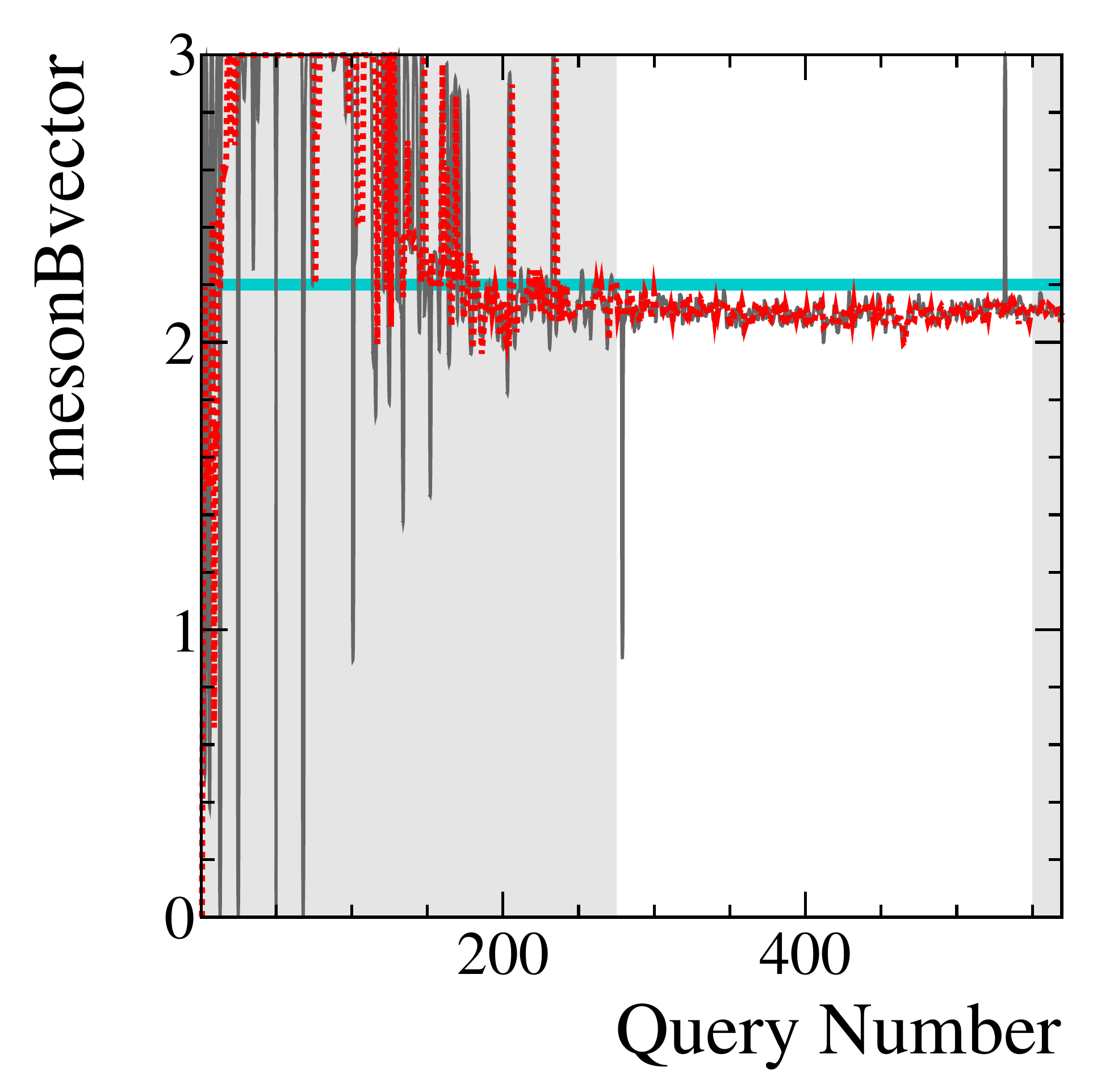}
  \caption{
  Evolution of the parameter values in block 3 during the tuning process.
The white regions are $25\cdot n({\rm par})$--$50\cdot n({\rm par})$ queries.
}
  \label{fig:block3_parsvcalls}
\end{figure}

The tuning results for each parameter for the three blocks are given in Tables~\ref{tab:block1_pars}--\ref{tab:block3_pars}. The evolution of the parameter values during the tuning processes are shown in Figs.~\ref{fig:block1_parsvcalls}--\ref{fig:block3_parsvcalls}.

\begin{table}[h]
  \begin{center}
    \caption{\label{tab:block1_pars}Tuning results for block 1.}
      \begin{tabular}{cccc}
        \hline
        Parameter & Monash Value & Tune Value & Range Considered   \\
        \hline
         alphaSvalue & 0.1365 & $0.1365\pm0.0002$ & $[0.06,0.25]$\\
         pTmin  & 0.5 & $0.49\pm0.02$ & $[0.1,2.0]$ \\
         pTminChgQ & 0.5 & $1.89^{+0.10}_{-1.79}$ & $[0.1,2.0]$ \\
        \hline
      \end{tabular}
  \end{center}
\end{table}

\begin{table}[h]
  \begin{center}
    \caption{\label{tab:block2_pars}Tuning results for block 2.}
      \begin{tabular}{cccc}
        \hline
        Parameter & Monash Value & Tune Value & Range Considered \\
        \hline
        sigma & 0.335 & $0.333_{-0.002}^{+0.001}$ & $[0,1]$ \\
        bLund & 0.98 & $1.04^{+0.01}_{-0.02}$ & $[0.2,2]$ \\
        aExtraSQuark & 0 & $0^{+0.07}_{-0}$ & $[0,2]$ \\
        aExtraDiQuark & 0.97 & $1.48^{+0.15}_{-0.14}$ & $[0,2]$ \\
        rFactC & 1.32 & $1.38\pm0.06$ & $[0,2]$ \\
        rFactB & 0.855 & $0.887\pm0.015$ & $[0,2]$ \\
        \hline
      \end{tabular}
  \end{center}
\end{table}

\begin{table}
  \begin{center}
    \caption{\label{tab:block3_pars}Tuning results for block 3.}
      \begin{tabular}{cccc}
        \hline
        Parameter & Monash Value & Tune Value & Range Considered \\
        \hline
        probStoUD & 0.217 & $0.219^{+0.001}_{-0.002}$ & $[0,1]$ \\
        probQQtoQ & 0.081 & $0.082\pm0.01$ & $[0,1]$ \\
        probSQtoQQ & 0.915 & $0.892^{+0.014}_{-0.018}$ & $[0,1]$ \\
        probQQ1toQQ0 & 0.0275 & $0.0276\pm0.0009$ & $[0,1]$ \\
        etaSup & 0.6 & $0.59\pm0.02$ & $[0,1]$ \\
        etaPrimeSup & 0.12 & $0.12\pm0.01$ & $[0,1]$ \\
        decupletSup & 1 & $1^{+0}_{-0.04}$ & $[0,1]$ \\
        mesonUDvector & 0.5 & $0.51^{+0.01}_{-0.02}$ & $[0,3]$ \\
        mesonSvector & 0.55 & $0.55\pm0.01$ & $[0,3]$ \\
        mesonCvector & 0.88 & $0.89^{+0.04}_{-0.05}$ & $[0,3]$ \\
        mesonBvector & 2.2 & $2.1\pm0.1$ & $[0,3]$ \\
        \hline
      \end{tabular}
  \end{center}
\end{table}

\section{Tuning Procedure}
\label{app:procedure}
In this section, we provide more details regarding the technical aspects of the tuning procedure. 
All of the code required to reproduce the studies presented in this work is available at the GitHub repository \github~\cite{TuneMC}, and fully documented there (including installation and usage instructions).
Schematically, the tuning proceeds as follows:
\begin{itemize}
\item a large (10M-event) $e^+e^-$ data sample is generated using \pythia~8 with its default parameter values, collectively referred to as the {\em Monash} tune~\cite{Skands:2014pea}, and various observable distributions are built from the Monash simulated data sample and treated as experimental data (in a real-world tuning application, this step would be replaced by the use of experimental distributions);
\item a set of parameters in \pythia is chosen for tuning (we chose to tune 20 parameters), and for each parameter a range of values to consider is defined; 
\item for each query, the Bayesian optimization package \spearmint provides a set of parameter values which are passed to \pythia~8 and used to generate 1M events (we chose to use a fixed number of events per sample though, as discussed in the text, one could consider using a variable number~\cite{NIPS2013_5086} which may improve the CPU usage);
\item once the \pythia sample is generated for each parameter set, the observable distributions are constructed and used to calculate  the objective function value according to Eq.~\ref{eq:chi2}, which is provided to \spearmint and used to update its internal model from which the next set of parameters to query is determined and passed to \pythia;
\item the query steps are repeated until some chosen stopping criterion (see Sec.~\ref{sec:cpu}) is met.
\end{itemize}
{\em N.b.}, by default \spearmint provides the suggested parameter sets in series, though it is possible for these to be provided in parallel. However, since the CPU usage is dominated by \pythia event generation---and because we performed this study on a small number of CPU cores---we chose to parallelize the event-generation processes for each sample, rather than generate multiple samples in parallel. Specifically, we split up the generation of each 1M-event sample into multiple jobs run in parallel, but ran \spearmint in series mode. For a larger-scale tuning application, it may also be advantageous to parallelize the queries.

%\clearpage

\bibliographystyle{h-physrev}
\bibliography{refs}

\end{document}